\RequirePackage[2024-06-01]{latexrelease}

\documentclass[reprint,
superscriptaddress,
 amsmath,
 amssymb,
 aps,
pra,
]{revtex4-2}

\usepackage[hyperindex,breaklinks]{hyperref}

\usepackage[dvipsnames,svgnames,x11names,hyperref]{xcolor}
\usepackage[caption=false]{subfig}
\usepackage{graphicx,dcolumn,bm,tikz}
\usetikzlibrary{external}
\usetikzlibrary{decorations.markings}
\usetikzlibrary{shapes.geometric}
\usetikzlibrary{arrows,calc}
\usetikzlibrary{quantikz2}
\usetikzlibrary{backgrounds}
\usepackage{adjustbox}
\usepackage{dsfont}
\usepackage[normalem]{ulem}

\definecolor{NodeGray}{RGB}{134,134,134} 
\definecolor{EdgeGray}{RGB}{185,185,185} 
\definecolor{HighGray}{RGB}{230,230,230} 
\definecolor{LGreen}{RGB}{50,250,40}     

\definecolor{SpinBlue}{HTML}{348AD8} 
\definecolor{SpinRed}{HTML}{C2361A}
\definecolor{NodePurple}{HTML}{983DA2} 
\definecolor{EdgePurple}{HTML}{983DA2}
\definecolor{GreenCell}{HTML}{5DA631}

\def\RectSize{0.4}
\def\RectRadius{8pt}
\def\cnotRadius{0.35}

\newcommand{\cnot}[4]{\fill[black]( (#1,#2) circle (5pt);
    \draw[black,line width=1pt] (#3,#4) circle[radius=\cnotRadius];
    \draw[-,line width=1pt](#1,#2)--(#3,#4);
    \draw[-,line width=1pt] (#3-\cnotRadius,#4)--(#3+\cnotRadius,#4);
    \draw[-,line width=1pt] (#3,#4-\cnotRadius)--(#3,#4+\cnotRadius); }

\newcommand{\dcnot}[4]{\fill[black]( (#1,#2) circle (5pt);
    \draw[black,line width=1pt] (#3,#4) circle[radius=\cnotRadius];
    \draw[-,line width=1pt](#1,#2)--(#3,#4);
    \draw[-,line width=1pt,opacity=1.0] (#3-\cnotRadius/1.414,#4-\cnotRadius/1.414)--(#3+\cnotRadius/1.414,#4+\cnotRadius/1.414);
    \draw[-,line width=1pt,opacity=1.0] (#3+\cnotRadius/1.414,#4-\cnotRadius/1.414)--(#3-\cnotRadius/1.414,#4+\cnotRadius/1.414); }

\makeatletter
\DeclareRobustCommand{\rvdots}{%
  \vbox{
    \baselineskip4\p@\lineskiplimit\z@
    \kern-\p@
    \hbox{.}\hbox{.}\hbox{.}
  }}
\makeatother

\definecolor{CustomLink}{RGB}{0,118,138}
\hypersetup{colorlinks=true, linkcolor=CustomLink, citecolor=CustomLink}

\newcommand{\adam}[1]{{\color{red}[AS] #1}}  
\newcommand{\arash}[1]{{\color{blue}[AJ] #1}}

\newcommand{\ceil}[1]{\left\lceil #1 \right\rceil}

\begin{document}

\title{A recipe for local simulation of strongly-correlated fermionic matter on quantum computers: the 2D Fermi-Hubbard model}

\author{Arash Jafarizadeh}
\affiliation{School of Physics and Astronomy, University of Nottingham, Nottingham, NG7 2RD, UK}
\affiliation{Centre for the Mathematics and Theoretical Physics of Quantum Non-Equilibrium Systems, University of Nottingham, Nottingham, NG7 2RD, UK}
\author{Frank Pollmann}
\affiliation{Department of Physics, TFK, Technische Universität München, James-Franck-Straße 1, D-85748 Garching, Germany}
\affiliation{Munich Center for Quantum Science and Technology (MCQST), Schellingstr. 4, 80799 München, Germany}
\author{Adam Gammon-Smith}
\affiliation{School of Physics and Astronomy, University of Nottingham, Nottingham, NG7 2RD, UK}
\affiliation{Centre for the Mathematics and Theoretical Physics of Quantum Non-Equilibrium Systems, University of Nottingham, Nottingham, NG7 2RD, UK}

\date{\today}

\begin{abstract}
The simulation of quantum many-body systems, relevant for quantum chemistry and condensed matter physics, is one of the most promising applications of near-term quantum computers before fault-tolerance. However, since the vast majority of quantum computing technologies are built around qubits and discrete gate-based operations, the translation of the physical problem into this framework is a crucial step. This translation will often be device specific, and a suboptimal implementation will be punished by the exponential compounding of errors on real devices. The importance of an efficient mapping is already revealed for models of spinful fermions in two or three dimensions, which naturally arise when the relevant physics relates to electrons. Using the most direct and well-known mapping, the Jordan-Wigner transformation, leads to a non-local representation of local degrees of freedom, and necessities efficient decompositions of non-local unitary gates into a sequence of hardware accessible local gates. In this paper, we provide a step-by-step recipe for simulating the paradigmatic two-dimensional Fermi-Hubbard model on a quantum computer using only local operations. To provide the ingredients for such a recipe, we briefly review the plethora of different approaches that have emerged recently but focus on the Derby-Klassen compact fermion mapping in order to make our discussion concrete. We provide a detailed recipe for an end-to-end simulation including embedding on a physical device, preparing initial states such as ground states, simulation of unitary time evolution, and measurement of observables and spectral functions. We explicitly compute the resource requirements for simulating a global quantum quench and conclude by discussing the challenges and future directions for simulating strongly-correlated fermionic matter on quantum computers.
\end{abstract}

\maketitle

\section*{Introduction}

Quantum computers have been now been realized using a variety of different technologies. While their ultimate promise and impact on society is potentially dramatic, near-term devices are not fault-tolerant and are subject to significant amount of noise and errors. Whether these noisy quantum devices can perform useful tasks that are beyond current analytical and numerical methods remains a subtle and unsettled question~\cite{Arute2019,Pednault2019,Wu2021,Daley2022,Dalzell2023,Kim2023,Tindall2024}. 

One of the most promising near-term applications of quantum computers---and quantum simulators~\cite{Cirac2012,Altman2021,Georgescu2014}---is the study of complex quantum many-body systems~\cite{Feynman1982,Fauseweh2024,Daley2022}, relevant for quantum chemistry~\cite{Bauer2020,Lee2023,Santagati2024} and condensed matter physics~\cite{Smith2019,Fauseweh2024,Daley2022,Arute2020,Chiaro2022,Mi2022}. These systems are particularly inaccessible with current methods due to the exponential explosion of Hilbert space dimension with system size, and the generic ballistic growth of entanglement entropy under unitary dynamics~\cite{Nahum2017,vonKeyserlingk2018}. While significant progress can be made for the ground states of one-dimensional quantum systems due to the entanglement area-law~\cite{Verstraete2006,Hastings2007,Eisert2010,Arad2013,Brandao2015}, typical ground states in two dimensions or higher still have extensive entanglement entropy in the boundary of the partition. 

Quantum computers may provide a natural approach for extending the set of quantum states that we can study. Naturally there is a trade-off in the resources and expressibility of the methods we use. While classical numerical methods based on tensor networks are limited by entanglement~\cite{White1992,Vidal2004,Orus2014,Cirac2021}, quantum circuits can efficiently generate entanglement and the relevant limitation becomes complexity (e.g. circuit depth)~\cite{Haferkamp2022}. For example, simulating unitary time evolution using a Trotter decomposition requires a circuit depth that scales polynomially with the simulation time~\cite{Trotter1959,Lloyd1996}, in contrast to the generic exponential scaling of the bond-dimension for matrix product state simulations. There is also hope that quantum circuits may provide efficient representations of the ground state of quantum many-body systems, particularly in higher dimensions.

When simulating quantum many-body systems, the first consideration is about what the relevant effective degrees of freedom are. In the majority of condensed matter or quantum chemistry settings, the electronic degrees of freedom are most relevant, and are described by spinful fermions. Since quantum computers are built around qubits, for these fermionic systems---and indeed any effective model not simply composed of spin-1/2 degrees of freedom---we require a mapping in order to utilize the quantum computer. Such a mapping is not unique, and the choice of mapping will depend on the specific quantum computer architecture and problem being studied. For fermions, the simplest and most direct mapping would be the Jordan-Wigner transformation~\cite{JordanWigner}. However, this leads to a non-local representation of the fermionic operators. In two-dimensions or higher, local fermionic Hamiltonians will be mapped to non-local and multi-body qubit Hamiltonians, and their simulation on quantum computers will require the efficient decomposition of non-local unitary gates into a sequence of hardware accessible local gates.

A classic example of such a model for describing the relevant physics of a wide range of materials is the Fermi-Hubbard model~\cite{Hubbard1963,Gutzwiller1963,Junjiro1963}. The model is deceptively simple and takes a tight-binding approximation for the itinerant electrons, treats only the dominant electronic orbital from each atom, and approximates the screened Coulomb interaction by an on-site interaction~\cite{Ashcroft1976,Simon2013}. Despite its simplicity, the Fermi-Hubbard model exhibits a rich phase diagram, including a metallic phase, a Mott insulating phase, and a variety of magnetic phases. It is also believed that variants of the Fermi-Hubbard model can capture the essential physics of high-temperature superconductivity. However, beyond one-dimension~\cite{Essler2005}, there is no known analytic solution, and understanding the phase diagram is subject of intense numerical study~\cite{LeBlanc2015,Zheng2017}. 

Given the importance of fermionic models such as the Fermi-Hubbard model, combined with the wide range of recent advances in quantum simulation using quantum computers that are scattered in the literature, in this paper we review the current state of play and present a step-by-step recipe for simulating the Fermi-Hubbard model on a quantum computer. As with any good recipe, we have structured the paper into the ingredients, recipe, and finishes touches (plating up, if you will). One of the main ingredients is a mapping from fermions to qubits. Here we focus on local fermion mappings that make use of ancillary qubits, and particularly discuss the Derby-Klassen compact fermion mapping~\cite{Derby2021} to provide a concrete choice. The recipe essentially consists of three main steps: (i) State preparation, (ii) Unitary time evolution, and (iii) Measurement of observables. In the finishing touches, we discuss practical details of running on a device, such as the use of error mitigation. Finally, we explicitly compute the resource requirements for simulating a concrete global quench protocol. Our work therefore provides a practical lower bound on the resource requirements to perform this type of fermionic simulation at a scale that would challenge classical numerical methods. A detailed table of contents is provided below.

\tableofcontents

\section{The Ingredients}

The first step in any good recipe is to set out the ingredients that we will be using. In this paper, this section on ingredients serves two purposes. First, we briefly review the available fermion-to-qubit mappings, and discuss the advantages and disadvantages of each. Second, we will make some concrete choices about the model we are studying and the mapping we will focus on. This allows the following recipe to be more prescriptive, but the reader should understand that the choices we make here are not unique, and the optimal choice will depend on the specific problem and available hardware, which we endeavour to point out at appropriate places throughout. The recipe should then be used as a template and adapted as necessary. 

\subsection{The Fermi-Hubbard Model}

In this work, we focus on the details of how to implement the Fermi-Hubbard model using the compact fermion mapping. The Fermi-Hubbard model is a model of spinful fermions hopping on a lattice with an on-site interaction~\cite{Hubbard1963,Gutzwiller1963,Junjiro1963}. It is described by the Hamiltonian
\begin{equation}\label{eq:Hubbard model}
H = -J \sum_{\langle i,j \rangle, \sigma} \left( c_{i,\sigma}^\dagger c^{}_{j,\sigma} + \text{H.c.} \right) + U \sum_i n_{i,\uparrow} n_{i,\downarrow},
\end{equation}
where in the sum $\langle i, j \rangle$ is over nearest-neighbour sites, and $n_{i, \sigma} = c^\dag_{i,\sigma} c_{i, \sigma}$ is the density operator for the fermion species $\sigma \in \{\uparrow, \downarrow\}$. In this paper we will consider only a square lattice, although the following can be readily generalised to more general lattices, see e.g. Ref.~\cite{Derby2021}.

\begin{figure}[t]
    \begin{center}
     \usetikzlibrary {shapes.geometric}
\begin{tikzpicture}[scale=1.47,>=angle 60,thick]
    \foreach \x in {1,3,5}{ 
	\foreach \y in {0,1,2,3}{
		\node at (\x,\y)[shape=dart,shape border rotate=270,fill=SpinRed,scale=0.5]{};
    }	}
    
    \foreach \x in {0,2,4}{ 
	\foreach \y in {0,1,2,3}{
		\node at (\x,\y)[shape=dart,shape border rotate=90,fill=SpinBlue,scale=0.5]{};
    }	}

    \def \padding {0.33}
    \foreach \x in {1,3,5}{
	\foreach \y in {0,1,2}{
		\draw[SpinRed,-] (\x,{\y+0.5-pow(-1,\x)*\padding})--(\x,{\y+0.5+pow(-1,\x)*\padding});
    }   }

    \foreach \x in {0,2,4}{
	\foreach \y in {0,1,2}{
		\draw[SpinBlue,-] (\x,{\y+0.5-pow(-1,\x)*\padding})--(\x,{\y+0.5+pow(-1,\x)*\padding});
    }   }
    \foreach \x in {0,1,2,3}{
	\foreach \y in {0,1,2,3,4}{
		\draw[black,-] ({\y+0.5-pow(-1,\x)*\padding},\x)--({\y+0.5+pow(-1,\x)*\padding},\x);
    }   }	

    \foreach \x in {0,2,4}{
	\foreach \y in {0,1,2,3}{
        \draw[GreenCell,rounded corners=\RectRadius*1.4, line width=0.35mm] (\x-\RectSize/1.5,\y-\RectSize/1.5) rectangle (\x+1+\RectSize/1.5,\y+\RectSize/1.5);
    }   }

    \draw [NodeGray,dashed,<->] (0,3+.2) to [out=60,in=120] (2,3+.2);
    \draw [NodeGray,dashed,<->] (0-.2,2) to [out=-170,in=170] (0-.2,1);
    \draw [NodeGray,dashed,<->] (3.2,1) to [out=-10,in=15] (3.2,0);
    \draw [NodeGray,dashed,<->] (3.1,2+.2) to [out=60,in=120] (4.9,2+.2);
    \draw [NodeGray,dotted,very thick,-] (0.1,0.1) to [out=20,in=160] (1-.1,0.1);
    \draw [NodeGray,dotted,very thick,-] (4.1,1-.1) to [out=-20,in=-160] (5-.1,1-.1);
\end{tikzpicture}
    \end{center}
    \caption{Spinful fermions are represented by a square lattice of spinless fermions. The blue vertices represent the spin-up fermions and the red vertices represent spin-down fermions. The green oval region is a unit cell with two fermions with opposite spins. The dashed arrows indicate the inter-cell hopping terms in the Fermi-Hubbard model in Eq.~\eqref{eq:Hubbard model}, and the dotted lines indicate the intra-cell interaction terms.}\label{fig:Sq_Spinfull}
\end{figure}
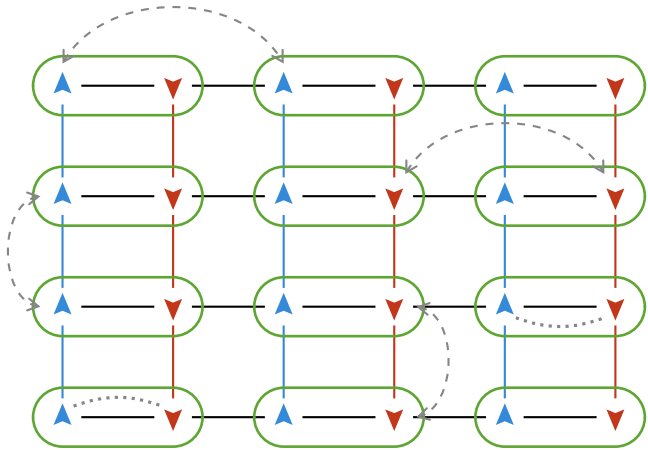

The fermion to qubit mapping allows us to directly map between qubits and \emph{spinless} fermions. To study the Fermi-Hubbard model we must first embed the spinful fermions on a square lattice into the spinless fermions. If we have access to all-to-all connectivity of the qubits, then we can separately map the up and down species of fermions. For local connectivity, the spinful fermions can be embedded in pairs, with one spin up and one spin down fermion in each unit cell, as illustrated in Fig.~\ref{fig:Sq_Spinfull}. The horizontal nearest-neighbour hopping terms then necessarily become next-nearest neighbour in terms of spinless fermions, and on-site interaction is implemented as a nearest-neighbour density interaction.

\subsection{From Fermions to Qubits}

Fermions, particularly electrons, are ultimately responsible for many material properties, as well as the structure and interactions of simple and complex molecules. Fermions are characterized by their statistics, which can most neatly be summarized by the anti-commutation relations for the creation/annihilation's operators,
\begin{equation}\label{eq:c-fermion_algebra}
    \{c^{\dagger}_{j},c^{}_{l}\}=\delta_{jl},\qquad \{c^{\dagger}_{j},c^{\dagger}_{l}\}=0,\qquad \{c^{}_{j},c^{}_{l}\}=0, 
\end{equation}
where the subscripts label the position of the fermion. These algebraic relations lead to the physical effects of Pauli-exclusion, and the anti-symmetry of wavefunctions under the exchange of two fermions. In contrast, nearly all quantum computers are based on qubits. Qubits are two-level systems, with raising and lowering operators, $\sigma^+_j$, $\sigma^-_k$, satisfying commutation relations,
\begin{equation}\label{eq:qubit_algebra}
    [\sigma^+_{j},\sigma^-_{l}]=\delta_{jl},\qquad (\sigma^+_j)^2 = (\sigma^-_j)^2 = 0  \qquad (\sigma^+_j)^\dagger = \sigma^-_j.
\end{equation}
Nonetheless, quantum computers based on qubits are in-principle universal, meaning that they can deal with Fermionic statistics. The typical approach to dealing with fermionic statistics is to find combinations of operators on qubits that satisfy the same commutation relations as the fermionic creation/annihilation operators, that is, a mapping from fermion to qubits.

\subsubsection{Review of fermion-to-qubit mappings}\label{sec:Fermion_to_Qubit_Mappings}

The most well known and widely used fermion to qubit mapping is the Jordan-Wigner transformation~\cite{JordanWigner},
\begin{equation}\label{eq:JW_Transformation_Sec1}
    c_l \rightarrow \frac{1}{2} \sigma^-_l\prod_{j<l} Z_j,
\end{equation}
where $X_j = \sigma^+_j + \sigma^-_j$, $Y_j = -i(\sigma^+_j - \sigma^-_j)$, and $Z_j = \sigma^+_j\sigma^-_j - \sigma^-_j \sigma^+_j$, are the standard Pauli operators. Importantly, the Jordan-Wigner transformation uses a linear but arbitrary, ordering of the sites $j<l$. The resulting non-local product of Pauli-$Z$ operators is known as a Jordan-Wigner string. The benefit of this type of mapping is that (up to subtleties with periodic boundary conditions) the Hilbert spaces are in one-to-one correspondence, with each real-space fermionic mode being mapped to a qubit. However, in higher than one-dimension, local fermion operators are mapped to non-local Pauli-strings that scale at least with the linear size of the system. The Jordan-Wigner transformation has proven to be a powerful theoretical tool, for instance revealing the free-fermion solvability of the transverse-field Ising model~\cite{Pfeuty1970}. It has also been the most widely used in the simulations of fermionic system on quantum computers, see, e.g., Refs.~\cite{Barends2015,Arute2020,Chiaro2022,Hemery2023}. This is particularly the case for the variational quantum eigensolver~\cite{Peruzzo2014,Kandala2017} algorithm in the context of quantum chemistry.

To improve on the unfavourable non-locality of the Jordan-Wigner mapping, several other approaches have been put forward. An alternative that achieves a logarithmic scaling in system size is the Bravyi-Kitaev transformation~\cite{Bravyi2002}. While conceptually the approach is similar to the use of Jordan-Wigner strings, the Bravyi-Kitaev approach uses a combination of both occupation information (like Jordan-Wigner) and parity information~\cite{Tranter2015}. The result is a dramatic improvement in the asymptotic scaling of the transformation, but at the cost of potentially significant overhead. The benefits of using the Bravyi-Kitaev mapping only emerge for larger systems sizes, but already for those that are well within experimental reach~\cite{Tranter2018}. 

An alternative approach is to introduce auxiliary degrees of freedom in order to achieve a strictly local mapping of pairwise-fermionic operators. The locality of these approaches come with two main costs. The first is that the mapping is restricted to pairwise-fermionic operators and not to the creation/annihilation operators themselves. However, this is not so restrictive, since it allows for any even combination of fermionic operators, and any odd combination of operators would be unphysical in a closed quantum system. In these mappings, single creation/annihilation operators can also still be implemented non-locally. Secondly, the Hilbert spaces are no longer in one-to-one correspondence and an extra condition is placed on the auxiliary d.o.f. in order to restrict to the physical subspace. This conditions can interpreted as being in the (degenerate) ground state space of a particular Hamiltonian. The non-local nature of the fermion mapping therefore reemerges as long-range correlations in the mapped states. The was first realised by Bravyi and Kitaev~\cite{Bravyi2002}, then a few years later, the Verstrate-Cirac encoding~\cite{Verstraete2005} provided a more general framework for local fermion mappings. Other notable examples are string-net models~\cite{Levin2005} or Kitaev's quantum double models~\cite{Kitaev2003}, which are bosonic spin models that support fermionic excitations, and more general anyonic excitations. This approach has also been generalised to higher-dimensions and arbitrary graphs in the case of fermions~\cite{Ball2005,OBrien2024}.

With the advent of Noisy Intermediate Scale Quantum (NISQ) computers, an increased focus has been placed on developing fermion-to-qubit mappings with minimal resource requirements. What the precise resource limitations are depends on the physical implementation of the quantum computer, but these typically include some combination of the circuit depth (runtime), the number of qubits, the implemented gate set, and qubit connectivity. These limitations have motivated a range of new proposals~\cite{Whitfield2016,Jiang2019,OBrien2024,Ballarin2024}. While some work in higher-dimensions or on arbitrary graphs, in two-dimensions these mappings have been shown to ultimately be connected to the $\mathbb{Z}_2$ topological order of the toric code~\cite{Chen2023b}. These mappings offer different balances between the number of auxiliary qubits required and the support of the mapped operators. 

In this paper we have chosen to focus on the Derby-Klassen compact fermion mapping~\cite{Derby2021}, and will only refer to this mapping from now on. Because of the large number of possible mappings, we have chosen to fix one for the purposes of providing a step-by-step recipe for implementing the mapping on real devices in order to simulate quantum many-body dynamics. Similar steps should be followed if a different mapping is preferred, or your purpose for simulation differs. We have specifically chosen the Derby-Klassen mapping due to its simplicity and its balanced circuit depth and qubit requirements.

\subsubsection{Details of the Derby-Klassen compact mapping}


Let us start by providing the necessary details of the Derby-Klassen compact fermion mapping~\cite{Derby2021}, on which we will focus in this paper. The mapping is most conveniently defined in terms of Majorana fermions:
\begin{equation}
\gamma_i = c_i + c_i^\dagger, \qquad \bar{\gamma}_i = i(c_i^\dagger - c_i).
\end{equation} 
The mapping then primarily deals with even products of fermion operators, all of which can be written as products of the fermionic edge and vertex operators:
\begin{equation}\label{eq:Edge_Vertx_Op}
    E_{jk}=-i\gamma_j\gamma_k,\quad V_j=-i\gamma_j\bar{\gamma}_j,
\end{equation}
where $j$ and $k$ are nearest-neighbour sites. For simplicity, we will consider only a square lattice, but details of how to generalise to other lattices can be found in Ref.~\cite{Derby2021b}. The edge operators are antisymmetric, $E_{jk} = -E_{kj}$, and all edge and vertex operators are hermitian, traceless, and self-inverse, that is,
\begin{equation}\label{eq:Edge_Vertx_Op_Basics}
 E_{jk}^\dagger=E_{jk},\quad V_{j}^\dagger=V_{j},\quad 
 E_{jk}^2=V_{j}^2=1.
\end{equation}
From the canonical fermion anti-commutation relations, we see they  satisfy the following (anti-)commutation relations:
\begin{equation}\label{eq:Edge_Vertx_Op_AntiCom}
    \{E_{jk}, V_j\} = 0,\quad \{E_{ij} , E_{jk}\} = 0,
\end{equation}
and for all $i\neq j \neq m \neq n$:
\begin{equation}\label{eq:Edge_Vertx_Op_Comm}
[V_i, V_j] = 0,\quad [E_{ij} , V_m] = 0, \quad [E_{ij} , E_{mn}] = 0.
\end{equation}
Finally, there exists one additional non-local relation in this system, that the product of any loop of edge operators must be proportional to the identity, namely,
\begin{equation}\label{eq:Edge_Op_Loop}
\prod_{j \in \text{loop}}\left( i E_{p_j,p_{j+1}} \right) = \mathds{1}.
\end{equation}
This will be important when defining the physical subspace of the enlarged qubit Hilbert space.

\begin{figure}[t]
    \begin{center}
    \begin{tikzpicture}[scale=1.25,>=stealth,thick]
\colorlet{darkgreen}{green!40!black}
\colorlet{darkred}{red!70!black}
\colorlet{Bpurple}{red!50!blue}

    \foreach \x in {0,1,2,3,4,5,6}{  
	\foreach \y in {0,1,2,3,4}{
		\node at (\x,\y)[circle,fill=NodeGray,scale=0.5]{};
    }	}
    \foreach \x in {0.5,2.5,4.5}{  
	\foreach \y in {0.5, 2.5}{
		\node at (\x,\y)[circle,fill=NodeGray,scale=0.5]{};
		\node at (\x+1.0,\y+1.0)[circle,fill=NodeGray,scale=0.5]{};
    }   }


    \foreach \x in {0,1,2,3,4,5,6}{
	\foreach \y in {0,1,2,3}{
		\draw[<-, draw=EdgeGray] (\x,{\y+0.5-pow(-1,\x)*0.35})--(\x,{\y+0.5+pow(-1,\x)*0.35});
    }   }
    \foreach \x in {0,1,2,3,4}{
	\foreach \y in {0,1,2,3,4,5}{
		\draw[<-, draw=EdgeGray] ({\y+0.5-pow(-1,\x)*0.35},\x)--({\y+0.5+pow(-1,\x)*0.35},\x);
    }   }	
    \foreach \x in {1,2,3,4,5,6,7}{
	\foreach \y in {1,2,3,4,5}{
        \pgfmathsetmacro\z{int( 7*(\y-1)+\x)}
        \node at (\x-1.16,-\y+5.16)[text=NodeGray, scale=0.63]{ $\z$};
    }   }
    
\begin{scope}[shift={(3,2)}, draw=SpinRed, text=SpinRed]
    \node at (0,0)[scale=.8,circle,draw,fill=none,line width=0.4mm]{};
    \node at (0.25,0.2)[scale=0.9]{$V_{_{18}}$};
    \node at (-.25,-.2)[scale=0.8]{$\boldsymbol{Z}$};
\end{scope}
    
\begin{scope}[shift={(1,2)},draw=SpinBlue, text=SpinBlue]
    \draw[line width=0.5mm,->] (0.15,1.0)--(0.85,1.0);
    \draw[line width=0.5mm,-] (0.5,1.0)--(0.5,1.4);
    \node at (0,1)[scale=.8,circle,draw, fill=none,line width=0.4mm]{};
    \node at (0.5,1.5)[scale=.8,circle,draw, fill=none,line width=0.4mm]{};
    \node at (1,1)[scale=.8,circle,draw, fill=none,line width=0.4mm]{};
    \node at (0.5,0.8)[scale=0.9]{$E_{_{9,10}}$};
    
    \node at (-0.2,0.8)[scale=0.8]{$X$};
    \node at (0.4,1.7)[scale=0.8]{$Y$};
    \node at (1.2,0.8)[scale=0.8]{$Y$};
\end{scope}

\begin{scope}[shift={(3,3)}, draw=orange, text=orange]
    \draw[line width=0.5mm,->] (0,0.15)--(0,0.85);
    \draw[line width=0.5mm,-] (0,.50)--(0.4,.5);
    \node at (0,0)[scale=.8,circle,draw,fill=none,line width=0.4mm]{};
    \node at (0.5,.5)[scale=.8,circle,draw,fill=none,line width=0.4mm]{};
    \node at (0,1)[scale=.8,circle,draw,fill=none,line width=0.4mm]{};
    \node at (0.6,0.8)[scale=.9]{$E_{_{11,4}}$};
    \node at (0.18,-.15)[scale=.8]{$X$};
    \node at (0.55,0.3)[scale=.8]{$-X$};
    \node at (0.18,1.15)[scale=.8]{$Y$};
\end{scope}
    
\begin{scope}[shift={(0,1)}, draw=GreenCell, double=GreenCell, text=GreenCell]
    \draw[line width=0.5mm,->] (0.15,0)--(0.85,0);
    \draw[line width=0.5mm,->] (1.15,0)--(1.85,0);
    \draw[line width=0.5mm,-] (0.5,-0.4)--(0.5,0.0);
    \draw[line width=0.5mm,-] (1.5,0)--(1.5,0.4);
    
    \node (a1) at (0,0)[scale=.8,circle,draw,fill=none,line width=0.4mm]{};
    \node (a2) at (0.5,-.5)[scale=.8,circle,draw,fill=none,line width=0.4mm]{};
    \node (a3) at (1,0)[scale=.8,circle,draw,fill=none,line width=0.4mm]{};
    \node (a4) at (1.5,0.5)[scale=.8,circle,draw,fill=none,line width=0.4mm]{};
    \node (a5) at (2,0)[scale=.8,circle,draw,fill=none,line width=0.4mm]{};
    
    \node at (-.2,-.2)[scale=.8]{$X$};
    \node at (0.3,-.7)[scale=.8]{$Y$};
    \node at (1.2,.2)[scale=.8]{$Z$};
    \node at (1.7,.7)[scale=.8]{$Y$};
    \node at (2.2,.2)[scale=.8]{$Y$};
    
    \node at (1.6, -.3)[scale=1.0]{$F_{_{22,24}}$};
    
\end{scope}

\begin{scope}[shift={(4,0)}, draw=EdgePurple, double=EdgePurple, text=EdgePurple]
    \draw[line width=0.5mm,<-] (0.15,0.0)--(0.85,0.0);
    \draw[line width=0.5mm,->] (1,0.15)--(1,0.85);
    \draw[line width=0.5mm,->] (1.,1.15)--(1.,1.85);
    \draw[line width=0.5mm,<-] (1.15,2)--(1.85,2);
    \draw[line width=0.5mm,-] (0.5,0)--(0.5,0.4);
    \draw[line width=0.5mm,-] (0.6,0.5)--(1,.5);
    \draw[line width=0.5mm,-] (1,1.5)--(1.4,1.5);
    \draw[line width=0.5mm,-] (1.5,1.6)--(1.5,2);
    
    
    \node at (0,0)[scale=.8,circle,draw,fill=none,line width=0.4mm]{};
    \node at (0.5,.5)[scale=.8,circle,draw,fill=none,line width=0.4mm]{};
    \node at (1,0)[scale=.8,circle,draw,fill=none,line width=0.4mm]{};
    \node at (1,1)[scale=.8,circle,draw,fill=none,line width=0.4mm]{};
    \node at (1.5,1.5)[scale=.8,circle,draw,fill=none,line width=0.4mm]{};
    \node at (1,2)[scale=.8,circle,draw,fill=none,line width=0.4mm]{};
    \node at (2,2)[scale=.8,circle,draw,fill=none,line width=0.4mm]{};
    
    \node at (0.6,1.6)[scale=1.0]{$F_{_{21,33}}$};
    \node at (-.2,-.2)[scale=0.8]{$Y$};
    \node at (0.35,.65)[scale=0.8]{$Z$};
    \node at (1.2,-.2)[scale=0.8]{$\mathds{1}$};
    \node at (1.2,0.8)[scale=0.8]{$Z$};
    \node at (1.65,1.35)[scale=0.8]{$Z$};
    \node at (1.17,2.2)[scale=0.8]{$\mathds{1}$};
    \node at (2.2,2.2)[scale=0.8]{$X$};
    \end{scope}

\end{tikzpicture}
    \end{center}
    \caption{Examples of the Majorana operators mapped to local qubit operators, see Eqs.~\eqref{eq:Compact_Fermoin_E_def},\eqref{eq:Compact_Fermion_V_def},\eqref{eq:Compact_Fermoin_F_def}. We show the Pauli operators that are involved, but for the correct phases please refer to the main text.}\label{fig:MajoranaOperators}
\end{figure}
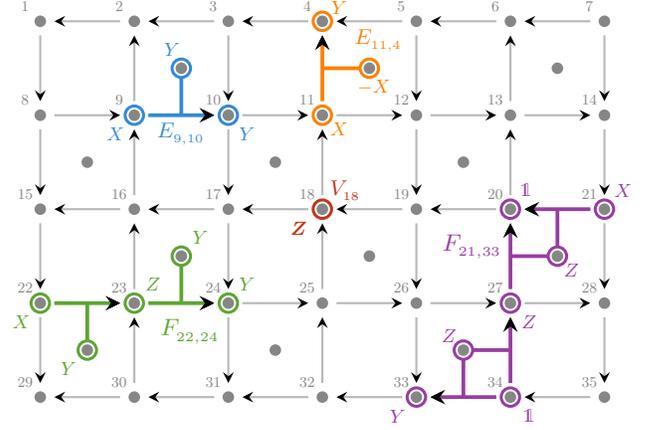

The task is then to find operators acting on qubits that have the same properties and satisfy the same (anti-) commutation relations. In order to achieve this with local operators, we need to add additional qubits to our system. For a square lattice, these new qubits can be added to some of the faces of the square lattice in a checkerboard pattern, as shown in Fig.~\ref{fig:MajoranaOperators}. Since the operators $E_{jk}$ are antisymmetric, we also need to assign a direction to each bond. We refer to the square lattice of qubits as \emph{primary} qubits, and to the additional face qubits as \emph{secondary}.
We will denote the transformed edge and vertex operators by tilde over-scripts, $\tilde{E}$ and $\tilde{V}$. Based on the reference \cite{Derby2021}, the compact fermionic edge operators, for every edge with $i$ pointing to $j$, are defined as:
\begin{equation}\label{eq:Compact_Fermoin_E_def}
    \tilde{E}_{jk}=
        \begin{cases}
        X_jY_kX_{f(j,k)} & (j,k)\text{ oriented downward}\\
        -X_jY_kX_{f(j,k)} & (j,k)\text{ oriented upward}\\
        X_jY_kY_{f(j,k)} & (j,k)\text{ horizontal}
        \end{cases}
\end{equation}
and $\tilde{E}_{ji}=-\tilde{E}_{ij}$. Here $f(j,k)$ denotes the additional secondary qubit that is in the face directly adjacent to the bond connecting $j$ and $k$, and the sign convention chosen in Eq.~\eqref{eq:Compact_Fermoin_E_def} is with respect to the directions assigned to each bond in Fig.~\ref{fig:MajoranaOperators}. For every vertex $j$, the corresponding mapped vertex operator is defined by
\begin{equation}\label{eq:Compact_Fermion_V_def}
    \tilde{V}_j=Z_j.
\end{equation}
These vertex and edge operators are illustrated in figure \ref{fig:MajoranaOperators}. These are now strictly local Pauli operators that satisfy the relations \eqref{eq:Edge_Vertx_Op_Basics}, \eqref{eq:Edge_Vertx_Op_AntiCom} and \eqref{eq:Edge_Vertx_Op_Comm}, as shown in Ref.~\cite{Derby2021}. 

When restricted to local qubit connectivity, we require longer range fermion hoppings.
To deal with these, we introduce a long-range edge operator, $F_{jk} = -i\gamma_j \gamma_k$, which is a product of edge operators connecting sites $j$ and $k$:
\begin{equation}\label{eq:Edge_Op_F}
F_{jk} = -i \prod_{\langle l, l' \rangle \in \text{path}(j,k)} \left(iE_{l l'} \right),
\end{equation}
where the path connecting $j$ and $k$ is an arbitrary sequence of nearest-neighbour sites. This operator is antisymmetric, $F_{jk} = -F_{kj}$, and satisfies the same (anti-) commutation relations as $E_{jk}$ and also commute with the stabilizers $\mathbb{J}_p$. The mapped operators
\begin{equation}\label{eq:Compact_Fermoin_F_def}
\tilde{F}_{jk} = -i \prod_{\langle l, l' \rangle \in \text{path}(j,k)} \left(i\tilde{E}_{l l'} \right),
\end{equation}
therefore automatically satisfy the correct (anti-) commutation relations. 

The hopping terms in the Hamiltonian can be written in terms of Majorana fermions and then as a combination of the edge and vertex operators, namely,
\begin{alignat}{2}
    c^{\dagger}_{j}c^{}_{k}+ c^{\dagger}_{k}c^{}_{j} &= -\frac{i}{2}(\bar{\gamma}_{j}\gamma_{k}- \gamma_{j}\bar{\gamma}^{}_{k})\label{eq:Ferm_Hopp_Majora}\\
    &=-\frac{i}{2}(V_jF_{jk}+ F_{jk}V_{k}),\label{eq:Ferm_Hopp_V_F}
\end{alignat}
and 
\begin{equation}\label{eq:density_correlator_operator}
n_j n_k = \frac{1}{4}(1-V_j)(1-V_k).
\end{equation}
These operators are then mapped to qubit operators by replacing all edge/vertex operators with their mapped counterparts. To be more explicit, a fermionic 
hopping from $j$ to $k$ will take the form
\begin{equation}\label{eq:first_vertical_hopping_qubit}
    c_{j}^\dagger c_{k}+c_{k}^\dagger c_{j}=\frac{1}{2}(X_{j}X_{k}+Y_{j}Y_{k})\prod_{\alpha}f_{\alpha}\sideset{}{'}\prod_{l \in \text{path}(j,k)} Z_{l},
\end{equation}
where $f_\alpha$ is Pauli operators for secondary qubits in the path from $j$ to $k$, and the prime on the product indicates that the end sites $j$ and $k$ are omitted. The interactions terms, are then simply
\begin{equation}\label{eq:mapped_interaction}
    n_{i,\uparrow} n_{i,\downarrow} = \frac{1}{4}(1-Z_j) (1-Z_k),
\end{equation}
where $j,k$ are the spinless fermion sites in Fig.~\ref{fig:Sq_Spinfull} corresponding to the up and down species on site $i$ of the spinful model.

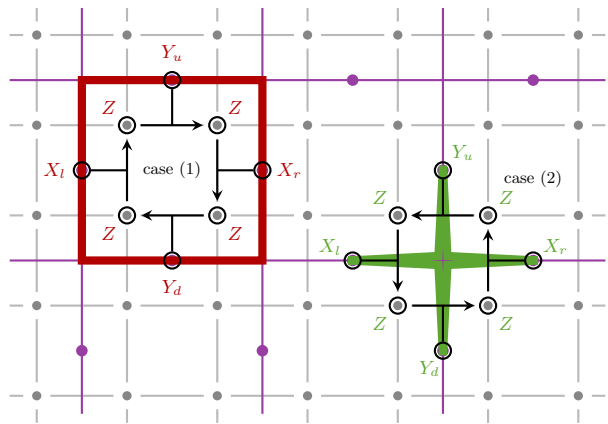
\begin{figure}[t]
    \begin{center}
    \begin{tikzpicture}[scale=1.2,>=stealth,thick]
\colorlet{darkgreen}{green!40!black}
\colorlet{SpinRed}{red!70!black}

    \foreach \x in {0,1,2,3,4,5,6}{
	\foreach \y in {2,3,4,5}{
		\draw[EdgeGray,-] (\x,{\y+0.5-pow(-1,\x)*0.35})--(\x,{\y+0.5+pow(-1,\x)*0.35});
    }   }
    \foreach \x in {2,3,4,5,6}{
	\foreach \y in {0,1,2,3,4,5}{
		\draw[EdgeGray,-] ({\y+0.5-pow(-1,\x)*0.35},\x)--({\y+0.5+pow(-1,\x)*0.35},\x);
    }   }	
\foreach \x in {0,1,2,3,4,5,6}{
	\draw[EdgeGray,-] (\x,1.85)--(\x,1.7);
	\draw[EdgeGray,-] (\x,6.15)--(\x,6.3);
    }
\foreach \y in {2,3,4,5,6}{
	\draw[EdgeGray,-] (-.15,\y)--(-.3,\y);
	\draw[EdgeGray,-] (6.15,\y)--(6.3,\y);
    }   

    \foreach \x in {0,1,2,3,4,5,6}{ 
	\foreach \y in {2,3,4,5,6}{
		\node at (\x,\y)[circle,fill=NodeGray,scale=0.4]{};
    }	}

    \foreach \x in {0.5,2.5,4.5}{
        \draw[EdgePurple,-] (\x,1.8)--(\x,6.3);         }
    \foreach \y in {3.5,5.5}{
        \draw[EdgePurple,-] (-0.3,\y)--(6.3,\y);        }

    \foreach \x in {0.5,2.5,4.5}{  
	\foreach \y in {2.5,4.5}{	
            \node at (\x,\y)[circle,fill=NodePurple,scale=0.5]{};
            \node at (\x+1,\y+1)[circle,fill=NodePurple,scale=0.5]{};
    }   }

\begin{scope}[shift={(0.5,3.5)}] 
    \draw[line width=1.09mm,SpinRed,opacity=0.65] (0,0) rectangle (2,2);
    
    
\end{scope}
    
\begin{scope}[shift={(1,4)}]  
    \foreach \x in {0,1}{
    \foreach \y in {0,1}{
        \node at (\x,\y)[scale=.7,circle,draw=black]{};
        }
    \foreach \y in {0}{
		\draw[black,->] (\x,{\y+0.5-pow(-1,\x)*0.34}) -- (\x,{\y+0.5+pow(-1,\x)*0.34});
            \draw[black,<-] ({\y+0.5-pow(-1,\x)*0.34},\x) -- ({\y+0.5+pow(-1,\x)*0.35},\x);
        }
    }
    
    \draw[black,-] (-.4,0.5)--(0,0.5);
    \node at (-.5,0.5)[scale=.7,circle,draw=black]{};
    \draw[black,-] (1,0.5)--(1.4,0.5);
    \node at (1.5,0.5)[scale=.7,circle,draw=black]{};
    \draw[black,-] (0.5,-.4)--(0.5,0);
    \node at (0.5,-.5)[scale=.7,circle,draw=black]{};    
    \draw[black,-] (0.5,1)--(0.5,1.4);
    \node at (0.5,1.5)[scale=.7,circle,draw=black]{};

    \node at (1.8,0.5)[scale=.8,SpinRed]{$X_r$};
    \node at (-.8,0.5)[scale=.8,SpinRed]{$X_l$};
    \node at (0.5,1.8)[scale=.8,SpinRed]{$Y_u$};
    \node at (0.5,-0.8)[scale=.8,SpinRed]{$Y_d$};

    \node at (-0.2,-0.2)[scale=.8,SpinRed]{$Z$};
    \node at (1.2,-.2)[scale=.8,SpinRed]{$Z$};
    \node at (-.2,1.2)[scale=.8,SpinRed]{$Z$};
    \node at (1.2,1.2)[scale=.8,SpinRed]{$Z$};

    \node at (.5,.5)[scale=0.7]{\text{case (1)}};
\end{scope}

\begin{scope}[shift={(4.5,3.5)}] 
    \draw[line width=1.09mm,GreenCell,opacity=0.65] (-1.05,0) -- (-.05,.05) -- (0,1.05) -- (+.05,.05) -- (+1.05,0) -- (+.05,-.05) -- (0,-1.05) -- (-.05,-.05) -- cycle;  
    
\end{scope}

\begin{scope}[shift={(4,3)},scale=1.]
    \foreach \x in {0,1}{
    \foreach \y in {0,1}{
        \node at (\x,\y)[scale=.7,circle,draw=black]{};
        }
    \foreach \y in {0}{
		\draw[black,<-] (\x,{\y+0.5-pow(-1,\x)*0.34}) -- (\x,{\y+0.5+pow(-1,\x)*0.34});
        \draw[black,->] ({\y+0.5-pow(-1,\x)*0.34},\x) -- ({\y+0.5+pow(-1,\x)*0.35},\x);
        }
    }
    
    \draw[black,-] (-.4,0.5)--(0,0.5);
    \node at (-.5,0.5)[scale=.7,circle,draw=black]{};
    \draw[black,-] (1,0.5)--(1.4,0.5);
    \node at (1.5,0.5)[scale=.7,circle,draw=black]{};
    \draw[black,-] (0.5,-.4)--(0.5,0);
    \node at (0.5,-.5)[scale=.7,circle,draw=black]{};    
    \draw[black,-] (0.5,1)--(0.5,1.4);
    \node at (0.5,1.5)[scale=.7,circle,draw=black]{};

    \node at (1.75,0.65)[GreenCell,scale=.8]{$X_r$};    
    \node at (-.75,0.65)[GreenCell,scale=.8]{$X_l$};
    \node at (0.72,1.7)[GreenCell,scale=.8]{$Y_u$};
    \node at (0.35,-0.7)[GreenCell,scale=.8]{$Y_d$};
    
    \node at (-0.2,-0.2)[scale=.8,GreenCell]{$Z$};
    \node at (1.2,-.2)[scale=.8,GreenCell]{$Z$};
    \node at (-.2,1.2)[scale=.8,GreenCell]{$Z$};
    \node at (1.2,1.2)[scale=.8,GreenCell]{$Z$};

    \node at (1.5,1.4)[scale=0.7]{\text{case (2)}};
\end{scope}

\end{tikzpicture}
    \end{center}
    \caption{Action of stabilizers $\mathbb{J}$, which have eigenvalue $+1$ in the physical sector. These operators are related to the toric code stabilizers. Primary qubits are shown as grey circles, and secondary qubits are purple. By convention, we can distinguish between plaquette (case 1) and star (case 2) operators based on the rotation direction of the bonds. Equivalently, we can choose to draw a lattice for the secondary qubits, as shown in purple, which makes this artificial distinction clearer.}\label{fig:Sq_Loop_Toric}
\end{figure}

In mapping the fermion operators to Pauli operators, we enlarged the Hilbert space. Necessarily, we have introduced new states of the qubits that do not correspond to any physical state of the fermions. In order to resolve the physical subspace, we use the extra constraint on the product of loop operators around the plaquettes in Eq.~\eqref{eq:Edge_Op_Loop}. On the qubit side of the mapping, these products take one of two forms, depending on whether the plaquette contains a secondary qubit or not, as shown in Fig.~\ref{fig:Sq_Loop_Toric}. For plaquettes with a secondary qubit, the product of edge operators around the plaquette is
\begin{equation}
i^4 \prod_{\circlearrowright_p}\tilde{E}_{jk} = \mathds{1},
\end{equation}
which matches the physical subspace.
For empty plaquettes, labelled by $p$, we get a non-trivial operators, which we denote by $\mathbb{J}_p$:
\begin{equation}\label{eq:plaquette_operators}
i^4 \prod_{\circlearrowright_p}\tilde{E}_{jk} = \left(\prod_{\circlearrowright_p} Z_i\right) X_a Y_b X_c Y_d  = \mathbb{J}_p.
\end{equation}
These plaquette operators are stabilizers, which are self-inverse, $\mathbb{J}^2_p = \mathbb{I}$ and commute $[ \mathbb{J}_p, \mathbb{J}_{p'}] = 0$. They also commute with all the edge and vertex operators. Working with even number of plaquettes, the number of stabilizers $\mathbb{J}_p$ is equal to the number of additional qubits. It means that the physical subspace is the unique $+1$ eigenspace of all the $\mathbb{J}_p$ operators. For the case where we have odd number of plaquettes, we can choose the arrangement of extra qubits so that we end up with a bigger Hilbert space, which can accommodate the full physical subspace. For simplicity, we will assume an even number of plaquettes in this paper, and see Ref.~\cite{Derby2021} for more details on the odd case. 

The stabilizers are closely related to the stabilizers in the toric code~\cite{Kitaev2003}. They take the form of the toric code star and plaquette operators coupled to diagonal operators on the primary qubits. With the convention used, they take a symmetric form that may be less familiar, which does not naturally distinguish between star and plaquette terms---although we can arbitrarily make this distinction based on the rotation directions (clockwise or anti-clockwise), as in Fig.~\ref{fig:Sq_Loop_Toric}. The ability to define strictly local operators that satisfy the correct algebra of the fermion operators, is facilitated by the $\mathbb{Z}_2$ topological order of the states in the physical subspace. This is a general feature of all known local fermion-to-qubit mappings in two dimensions~\cite{Chen2023}.

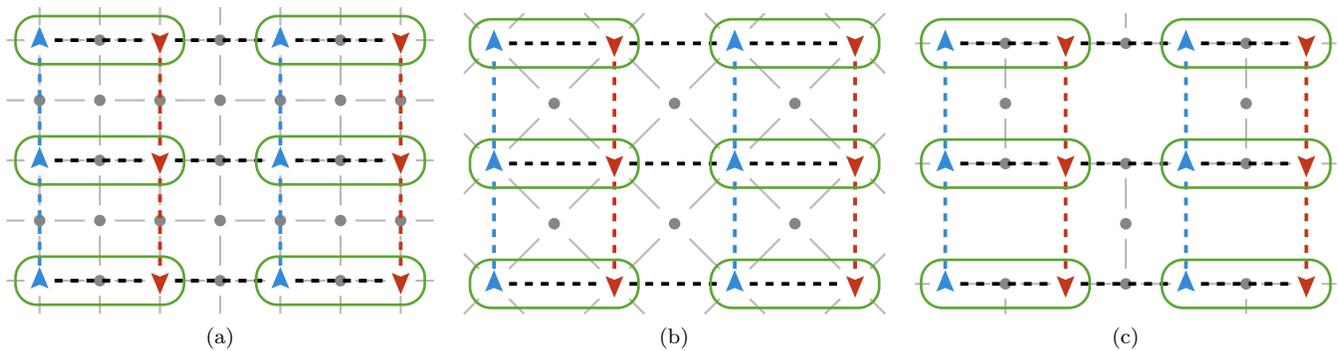
\begin{figure*}[t]
    \subfloat[]{%
      \usetikzlibrary{shapes.geometric}
\begin{tikzpicture}[scale=0.8,>=stealth,thick]

\foreach \x in {1,3,5}{  
\foreach \y in {0,1,2,3,4}{
    \node at (\x,\y)[shape=circle,fill=NodeGray,scale=0.5]{};
}}
\foreach \x in {0,2,4,6}{  
\foreach \y in {1,3}{
    \node at (\x,\y)[shape=circle,fill=NodeGray,scale=0.5]{};
}}

\def \padding {0.25}
\foreach \x in {0,1,2,3,4,5,6}{
\foreach \y in {0,1,2,3}{
		\draw[-, draw=EdgeGray] (\x,\y+\padding)--(\x,\y+1.0-\padding);      }}
\foreach \x in {0,1,2,3,4,5}{
\foreach \y in {0,1,2,3,4}{
		\draw[-, draw=EdgeGray] (\x+\padding,\y)--(\x+1.0-\padding, \y);     }}	
\foreach \x in {0,1,2,3,4,5,6}{
		\draw[-, draw=EdgeGray] (\x,0-\padding)--(\x,0.0-0.3-\padding);     
		\draw[-, draw=EdgeGray] (\x,4+\padding)--(\x,4.0+0.3+\padding);     
    }	
\foreach \y in {0,1,2,3,4}{
		\draw[-, draw=EdgeGray] (0.-\padding,\y)--(-.3-\padding,\y);     
		\draw[-, draw=EdgeGray] (6.+\padding,\y)--(6.3+\padding,\y);     
  }	
  
\foreach \y in {0,1}{
\foreach \x in {0,4}{
		\draw[dashed, draw=SpinBlue, opacity=1.0,line width=0.5mm] (\x,2*\y+\padding)--(\x,2*\y+2.0-\padding);}
\foreach \x in {2,6}{
		\draw[dashed, draw=SpinRed, opacity=1.0,line width=0.5mm] (\x,2*\y+\padding)--(\x,2*\y+2.0-\padding);}
    }
\foreach \y in {0,2,4}{
\foreach \x in {0,1,2}{
		\draw[dashed, draw=black,line width=0.5mm] (2*\x+\padding,\y)--(2*\x+2.0-\padding,\y);
}}	

\foreach \x in {0,4}{  
\foreach \y in {0,2,4}{
    \node at (\x,\y)[shape=dart,shape border rotate=90,fill=SpinBlue,scale=0.5]{};
}}
\foreach \x in {2,6}{  
\foreach \y in {0,2,4}{
    \node at (\x,\y)[shape=dart,shape border rotate=270,fill=SpinRed,scale=0.5]{};
}}

\foreach \x in {0,4}{
	\foreach \y in {0,2,4}{
    \draw[GreenCell,rounded corners=\RectRadius, line width=0.35mm] (\x-\RectSize,\y-\RectSize) rectangle (\x+2+\RectSize,\y+\RectSize);
    }   }

\end{tikzpicture}%
    }\hspace{0.5mm}
    \subfloat[]{%
      \usetikzlibrary{shapes.geometric}
\begin{tikzpicture}[scale=.80,>=stealth,thick]

\foreach \x in {1,3,5}{  
\foreach \y in {1,3}{
    \node at (\x,\y)[shape=circle,fill=NodeGray,scale=0.5]{};
}}

\def \z {0.25}

\foreach \x in {0,2,4}{
\foreach \y in {0,2}{
	\draw[-, draw=EdgeGray] (\x+\z,\y+\z)--(\x+1.0-\z,\y+1.-\z);}}
\foreach \x in {1,3,5}{
\foreach \y in {1,3}{
	\draw[-, draw=EdgeGray] (\x+\z,\y+\z)--(\x+1.0-\z,\y+1.-\z);}}
\foreach \x in {5,3,1}{
\foreach \y in {3,1}{
	\draw[-, draw=EdgeGray] (\x+\z,\y-\z)--(\x+1.0-\z,\y-1.+\z);}}
\foreach \x in {4,2,0}{
\foreach \y in {4,2}{
	\draw[-, draw=EdgeGray] (\x+\z,\y-\z)--(\x+1.0-\z,\y-1.+\z);}}
\foreach \x in {0,2,4,6}{
	\draw[-, draw=EdgeGray] (\x+\z,4+\z)--(\x+.5,4+.5);
	\draw[-, draw=EdgeGray] (\x-\z,4+\z)--(\x-.5,4+.5);
	\draw[-, draw=EdgeGray] (\x-.5,0-.5)--(\x-\z,0.0-\z);
	\draw[-, draw=EdgeGray] (\x+.5,0-.5)--(\x+\z,0.0-\z);
 }
\draw[-, draw=EdgeGray] (0-\z,0+\z)--(0-.5,0+.5);
\draw[-, draw=EdgeGray] (0-\z,2-\z)--(0-.5,2-.5);
\draw[-, draw=EdgeGray] (0-\z,2+\z)--(0-.5,2+.5);
\draw[-, draw=EdgeGray] (0-\z,4-\z)--(0-.5,4-.5);
\draw[-, draw=EdgeGray] (6+\z,0+\z)--(6.5,0+.5);
\draw[-, draw=EdgeGray] (6+\z,2-\z)--(6.5,2-.5);
\draw[-, draw=EdgeGray] (6+\z,2+\z)--(6.5,2+.5);
\draw[-, draw=EdgeGray] (6+\z,4-\z)--(6.5,4-.5);

\foreach \x in {0,2,4}{
\foreach \y in {0,2,4}{
		\draw[dashed, draw=black, opacity=1.0,line width=0.5mm] (\x+\z,\y)--(\x+2.0-\z,\y);
}}	


\foreach \y in {0,1}{
\foreach \x in {0,4}{
	\draw[dashed, draw=SpinBlue, line width=0.5mm] (\x,2*\y+\z)--(\x,2*\y+2-\z);}
\foreach \x in {2,6}{
	\draw[dashed, draw=SpinRed, opacity=1.0,line width=0.5mm] (\x,2*\y+\z)--(\x,2*\y+2-\z);}
}

\foreach \x in {0,4}{  
\foreach \y in {0,2,4}{
    \node at (\x,\y)[shape=dart,shape border rotate=90,fill=SpinBlue,scale=0.5]{};
}}
\foreach \x in {2,6}{  
\foreach \y in {0,2,4}{
    \node at (\x,\y)[shape=dart,shape border rotate=270,fill=SpinRed,scale=0.5]{};
}}


\foreach \x in {0,4}{
	\foreach \y in {0,2,4}{
    \draw[GreenCell,rounded corners=\RectRadius, line width=0.35mm] (\x-\RectSize,\y-\RectSize) rectangle (\x+2+\RectSize,\y+\RectSize);
    }   }

\end{tikzpicture}%
    }\hspace{0.5mm}
    \subfloat[]{%
      \usetikzlibrary{shapes.geometric}
\begin{tikzpicture}[scale=0.8,>=stealth,thick]

\foreach \x in {1,3,5}{  
\foreach \y in {0,2,4}{
    \node at (\x,\y)[shape=circle,fill=NodeGray,scale=0.5]{};
}}
\node at (3,1)[shape=circle,fill=NodeGray,scale=0.5]{};
\node at (1,3)[shape=circle,fill=NodeGray,scale=0.5]{};
\node at (5,3)[shape=circle,fill=NodeGray,scale=0.5]{};

\pgfmathsetmacro\z{0.25}

\foreach \x in {0,1,2,3,4,5}{
\foreach \y in {0,2,4}{
		\draw[-, draw=EdgeGray] (\x+\z,\y)--(\x+1-\z,\y);  }}
\foreach \x in {1,5}{
\foreach \y in {2,3}{
		\draw[-, draw=EdgeGray] (\x,\y+\z)--(\x,\y+1.0-\z);}}
\foreach \y in {0,1}{
		\draw[-, draw=EdgeGray] (3,\y+\z)--(3,\y+1.0-\z);}
\foreach \y in {0,2,4}{
\draw[-, draw=EdgeGray] (0-\z,\y)--(-0.5,\y);
\draw[-, draw=EdgeGray] (6+\z,\y)--(6.5,\y);
}    

\draw[-, draw=EdgeGray] (1,-\z)--(1,-.5);
\draw[-, draw=EdgeGray] (5,-\z)--(5,-.5);
\draw[-, draw=EdgeGray] (3,4+\z)--(3,4.5);
  
\foreach \y in {0,2}{
\foreach \x in {0,4}{
		\draw[dashed, draw=SpinBlue, opacity=1.0,line width=0.5mm] (\x,\y+\z)--(\x,\y+2-\z);
		\draw[dashed, draw=SpinRed, opacity=1.0,line width=0.5mm] (\x+2.,\y+\z)--(\x+2,\y+2.0-\z);
  }}
    
\foreach \x in {0,2,4}{
\foreach \y in {0,1,2}{
		\draw[dashed, draw=black, opacity=1.0,line width=0.5mm] (\x+\z,2*\y)--(\x+2-\z,2*\y);}}	

\foreach \x in {0,4}{  
\foreach \y in {0,2,4}{
    \node at (\x,\y)[shape=dart,shape border rotate=90,fill=SpinBlue,scale=0.5]{};
}}
\foreach \x in {2,6}{  
\foreach \y in {0,2,4}{
    \node at (\x,\y)[shape=dart,shape border rotate=270,fill=SpinRed,scale=0.5]{};
}}

\foreach \x in {0,4}{
	\foreach \y in {0,2,4}{
    \draw[GreenCell,rounded corners=\RectRadius, line width=0.35mm] (\x-\RectSize,\y-\RectSize) rectangle (\x+2+\RectSize,\y+\RectSize);
    }   }

\end{tikzpicture}%
    }
    \caption{Possible embeddings of the spinful fermion lattice from Fig.~\ref{fig:Sq_Spinfull} on different local geometries: (a) Square, (b) Diamond, (c) Heavy-honeycomb. The grey sites and bonds indicated the physical qubits and connections on the device. The coloured sites and dashed lines indicate the sites and coupling of the desired lattice. }\label{fig:Embeddings}
\end{figure*}

\subsection{Decomposition of Unitary Operators}\label{sec:decomposition unitary}

The main building blocks of any algorithm we wish to implement on a quantum computer are the unitary operators. These unitary operators will be used both for state preparation, and for simulating dynamics. A minimal set of unitary operators that we will need is 
\begin{equation}
\left\{ e^{-i\theta V_j}, \; e^{-i\phi V_j V_k}, \; e^{-i\lambda F_{jl}} \right\},
\end{equation}
where $j$ and $k$ are nearest-neighbour sites, and $j$ and $l$ are at most next-nearest-neighbour sites. Since, the Fermi-Hubbard model in Eq.~\eqref{eq:Hubbard model} contains only hopping terms and no pair creation/annihiliation, it will also be convenient to consider the operators
\begin{equation}
    \begin{aligned}
    e^{-i2\alpha(c^\dagger_n c^{}_{m}+c^\dagger_{m}c^{}_n)} &=e^{\alpha({V}_n {F}_{nm}+ {F}_{nm} {V}_{m})} \\
    &=e^{\alpha {V}_n {F}_{nm}}e^{ \alpha{F}_{nm} {V}_{m}},
    \end{aligned}
\end{equation}
where the final equality follows since the exponents commute, i.e. $[{V}_n {F}_{nm}, {F}_{nm} {V}_{m}] = -[\bar{\gamma}_i\gamma_j,\gamma_i\bar{\gamma}_j] = 0$. More precisely, we will need the mapped qubit analogues of these operators:
\begin{subequations}
\begin{align}
    e^{-i\theta V_j} &\mapsto e^{-i\theta Z_j} \\
    e^{-i\phi V_j V_k} &\mapsto e^{-i\phi Z_j Z_k} \\
    e^{\alpha {V}_n {F}_{nm}} &\mapsto e^{i\alpha Y_{j}Y_{k}Z_{l} \prod_{\alpha}f_{\alpha} } \\
    e^{\alpha {F}_{nm}V_{m}} &\mapsto e^{i\alpha X_{j}X_{k}Z_{l}\prod_{\alpha}f_{\alpha} },
\end{align}
\end{subequations}\label{eq:majorana_gates}
where $l$ is the site in between the next-nearest-neighbour sites $n$ and $m$. If $n$ and $m$ are nearest neighbours then we have $e^{\alpha {V}_n {F}_{nm}} \mapsto e^{i\alpha Y_{j}Y_{k}f_{\alpha} }$, and $e^{\alpha {F}_{nm}V_{m}} \mapsto e^{i\alpha X_{j}X_{k}f_{\alpha} }$.
In all cases, the mapped operator is the exponential of a Pauli-string, which can be efficiently decomposed into a sequence of local two-qubit gates~\cite{Nielsen_Chuang_2010}. We include details of how to decompose these unitary gates into specific gate sets with local connectivity in Appendix~\ref{app:Diamond Connectivity}.

\subsection{Embedding on a physical device}

With this mapping in hand, a significant practical consideration is how to realize or embed this mapping on a real device. Physical devices will have their own connectivity, which may not match the lattice used in our mapping. Different choices may lead to different trade-offs between the number of qubits and the depth of the resulting circuits.

In some physical realizations of quantum computers, we may have access to long-range or all-to-all couplings, or the ability to physically move any pair of qubits to interact, such as is typical for trapped ion quantum computers~\cite{Bruzewicz2019,Pino2021,Moses2023,Chen2023c}. In this case, we can directly implement the lattice from Fig.~\ref{fig:MajoranaOperators}. However, in the majority of cases, the connectivity in the device will be local and fixed. For instance, in superconducting circuits, the qubits are arranged in the 2D array and pairs of qubits are explicitly coupled. In Fig.~\ref{fig:Embeddings} we show two possible embeddings for a square lattice (employed by e.g. Google~\cite{Arute2019}), and one for a heavy-honeycomb lattice (employed by e.g. IBM~\cite{Kim2023}). This mapping has also been considered in the context of quantum annealers~\cite{Levy2022}, which we do not consider here.

This explicit connectivity will now have an impact on how the multi-qubit gates arising from the mapping are most efficiently implemented. This will have to be done on a case-by-case basis and may depend on the trade-off between number of qubits and number of gates. In Appendix~\ref{app: explicit example}, we consider some explicit decompositions of unitary operators for different geometries.

\section{The Recipe}

With our ingredients laid out and prepped, we are ready to cook (that is, simulate quantum many-body physics). In this section we will provide a step-by-step recipe for simulating the two dimensional Fermi-Hubbard model on a quantum computer. We will start by preparing the initial state, then simulate dynamics under unitary time evolution, and finally measure the observables of interest. 

 The reader should note that in providing this recipe we must walk a fine line between providing a general framework for simulating quantum many-body physics, and providing concrete steps for a particular model and setup. For example, we do not go into detail about the properties of the ground state, nor do we consider open quantum dynamics described by non-unitary channels. The setting we have in mind is relevant for global quantum quench protocols or measuring spectral properties in experiment. By preparing either a simple initial state or approximation of the ground state, simulating the time evolution, and measuring the expectation values of the observables as well as Green's functions, provides a lot of access to the properties of the many-body quantum system.


\subsection{State Preparation}

In this section we will provide a guide for the first step in any quantum simulation: state preparation. We will start by preparing the fermionic vacuum state under the Derby-Klassen fermion-to-qubit mapping, and then show how to prepare simple fermionic density patterns. Finally, we will discuss how to prepare the ground state of the Fermi-Hubbard model by reviewing the range of approaches that have been recently developed. 


\subsubsection{Preparing the Vacuum: the Toric Code}\label{sec:vacuum}

The first step in creating an initial fermionic state is to prepare the vacuum state in the physical sub-sector. This vacuum state $|\text{vac}\rangle$ is defined as the state, absent of any fermions, and in the $+1$ eigenspace of the $\mathbb{J}_p$ plaquette operators defined in Eq.~\eqref{eq:plaquette_operators}. That is, the vacuum state is defined by
\begin{equation}
    Z_j |\text{vac}\rangle = |\text{vac}\rangle, \qquad \mathbb{J}_p |\text{vac}\rangle = |\text{vac}\rangle,
\end{equation}
for all sites $j$ and plaquettes $p$. In this vacuum state the primary qubits decouple from the secondary qubits and are in a product state polarized along the $Z$-axis. The secondary qubits on the other hand are in the Toric code ground state: the simultaneous $+1$ eigenstate of the commuting operators $\mathbb{J}_p$. Note that depending on the boundary conditions (periodic vs open, and rough vs smooth), the vacuum state may not be unique---corresponding to the topological ground state degeneracy of the toric code~\cite{Kitaev2003}.

Efficient quantum circuits have been found for preparing the ground state of the toric code. Using only unitary gates, this state can be prepared in a circuit depth that scale with the shortest linear dimension of the system~\cite{Satzinger2021,Liu2022}, inspired by isometric Tensor Networks~\cite{Zaletel2020,Soejima2020,Wei2022}. This construction proceeds in a sequential manner where ``columns" of the system are entangled in parallel, and explicit constructions can be found in Appendix~\ref{app: explicit example}. 
Alternatively, by using measurements and classical feedback, the toric code state can be prepared in finite depth~\cite{Piroli2021,Tantivasadakarn2023}. By preparing a simple product state, followed by projective measurements of the star (or plaquette) operators, a state with $\mathbb{Z}_2$ topological order is realized with a random configuration of static excitations. Using the results of the measurements, these excitations can be removed by a single layer of local unitary gates, leaving the clean toric code ground state. Again the trade-off between circuit depth and the cost of performing measurements will depend on the specific quantum computing platform.

\subsubsection{Simple Fermion Density Patterns}

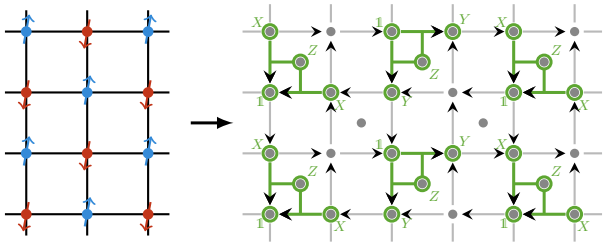
\begin{figure}[t]
    \begin{center}
    \begin{tikzpicture}[scale=0.81,>=stealth,thick]

\tikzset{middlearrow/.style={
        decoration={markings,
            mark= at position 0.45 with {\arrow{#1}} ,
        },
        postaction={decorate}
    }
}

\begin{scope}[shift={(0,0)}]
\foreach \x in {0,1,2,3,4,5}{  
\foreach \y in {0,1,2,3}{
    \node at (\x,\y)[circle,fill=NodeGray,scale=0.4,opacity=1.0]{}; }}

\foreach \x in {0.5,2.5,4.5}{  
\foreach \y in {0.5, 2.5}{
    \node at (\x,\y)[circle,fill=NodeGray, scale=0.4,opacity=1.]{}; }}
    
\foreach \x in {1.5,3.5}{  
\foreach \y in {1.5}{
    \node at (\x,\y)[circle,fill=NodeGray, scale=0.4,opacity=1.0]{}; }}

\foreach \x in {0,1,2,3,4,5}{
\foreach \y in {0,1,2}{
    \draw[<-, draw=EdgeGray, opacity=1.] (\x,{\y+0.5-pow(-1,\x)*0.35})--(\x,{\y+0.5+pow(-1,\x)*0.35}); }}
\foreach \x in {0,1,2,3}{
\foreach \y in {0,1,2,3,4}{
    \draw[<-, draw=EdgeGray, opacity=1.] ({\y+0.5-pow(-1,\x)*0.35},\x)--({\y+0.5+pow(-1,\x)*0.35},\x);}}

\foreach \y in {0,1,2,3}{
    \draw[-,draw=EdgeGray, opacity=1.] (5.15,\y)--(5.45,\y); 
    \draw[-,draw=EdgeGray, opacity=1.] (-.45,\y)--(-.15,\y); 
    }
\foreach \x in {0,1,2,3,4,5}{
    \draw[-,draw=EdgeGray, opacity=1.] (\x,3.15)--(\x,3.45); 
    \draw[-,draw=EdgeGray, opacity=1.] (\x,-.45)--(\x,-.15); 
    }
\foreach \x in {0,4}{
\foreach \y in {0,2}{
    \begin{scope}[shift={(\x,\y)}, draw=GreenCell, double=GreenCell, text=GreenCell]
        \draw[line width=0.4mm, <-] (0,0.15)--(0,0.85);
        \draw[line width=0.4mm, <-] (0.15,0)--(0.85,0);
        \draw[line width=0.4mm,-] (.5,0)--(0.5,.4);
        \draw[line width=0.4mm,-] (0,0.5)--(0.4,0.5);
        
        \node at (0,0)[scale=.6,circle,draw,fill=none,line width=0.4mm]{};
        \node at (.5,.5)[scale=.6,circle,draw,fill=none,line width=0.4mm]{};
        \node at (0,1)[scale=.6,circle,draw,fill=none,line width=0.4mm]{};
        \node at (1,0)[scale=.6,circle,draw,fill=none,line width=0.4mm]{};
        
        \node at (-.15,-.15)[scale=.6]{$\mathds{1}$};
        \node at (.7,.7)[scale=.6]{$Z$};
        \node at (1.15,-.2)[scale=.6]{$X$};
        \node at (-.2,1.15)[scale=.6]{$X$};
    \end{scope}
    }   }

\foreach \y in {0,2}{
    \begin{scope}[shift={(2,\y)}, draw=GreenCell, double=GreenCell, text=GreenCell]
        \draw[line width=0.4mm,<-] (0,0.15)--(0,0.85);
        \draw[line width=0.4mm,->] (0.15,1)--(0.85,1);
        \draw[line width=0.4mm,-] (.5,.6)--(0.5,1.);
        \draw[line width=0.4mm,-] (0,0.5)--(0.4,0.5);
        
        \node at (0,0)[scale=.6,circle,draw,fill=none,line width=0.4mm]{};
        \node at (.5,.5)[scale=.6,circle,draw,fill=none,line width=0.4mm]{};
        \node at (0,1)[scale=.6,circle,draw,fill=none,line width=0.4mm]{};
        \node at (1,1)[scale=.6,circle,draw,fill=none,line width=0.4mm]{};
        
        \node at (.25,-.15)[scale=.6]{$Y$};
        \node at (.7,.3)[scale=.6]{$Z$};
        \node at (-.2,1.15)[scale=.6]{$\mathds{1}$};
        \node at (1.2,1.2)[scale=.6]{$Y$};
    \end{scope}
    }
\end{scope}

\begin{scope}[shift={(-4,0)}]

\foreach \y in {0,1,2,3}{ 
    \draw[-, draw=black] (-.35,\y)--(2.35,\y); } 
\foreach \x in {0,1,2}{ 
    \draw[-, draw=black] (\x,-.35)--(\x,3.35); } 

\foreach \x in {0,2}{ 
    \foreach \y in {1,3}{ 
        \draw[SpinBlue ,thick,-{Classical TikZ Rightarrow}] (\x-0.04,\y-0.2) -- (\x+0.05,\y+0.28);
        \fill[SpinBlue] (\x,\y) circle (2.5pt);} 
    \foreach \y in {0,2}{ 
        \draw[SpinRed ,thick,-{Classical TikZ Rightarrow}] (\x+0.04,\y+0.2) -- (\x-0.05,\y-0.28);
        \fill[SpinRed] (\x,\y) circle (2.5pt);}    
    }	

\foreach \y in {1,3}{ 
    \draw[SpinRed ,thick,-{Classical TikZ Rightarrow}] (1+0.04,\y+0.2) -- (1-0.05,\y-0.28);
    \fill[SpinRed] (1,\y) circle (2.5pt);}	

\foreach \y in {0,2}{ 
    \draw[SpinBlue ,thick,-{Classical TikZ Rightarrow}] (1-0.04,\y-0.2) -- (1+0.05,\y+0.28);
    \fill[SpinBlue] (1,\y) circle (2.5pt);}	
\end{scope}

\begin{scope}[shift={(-1.3,1.5)}, draw=black, fill=black, text=black]
    \draw[line width=.4mm,-latex, scale=1.] (0.,0.) -- (.7,0.);
\end{scope}

\end{tikzpicture}
    \end{center}
    \caption{ Example of a half-filled spin density wave configuration for the fermions with spins in a checkerboard pattern, along with the corresponding Pauli operators to create this state on top of the vacuum state.
    }\label{fig:density_pattern}
\end{figure}

Once we have prepared the vacuum state for the model, the simplest fermions states to prepare are simple density patterns. Such states are relevant for global quantum quench protocols~\cite{Polkovnikov2011}. These are states where each site has a definite fermion number, and are created by the action of the fermion pairwise creation operators on the vacuum state:
\begin{equation}\label{eq:pair-creation-strip}
    c_j^\dagger c_{k}^\dagger = \frac{i}{4}(1-V_j)F_{jk}(1+V_{k}).
\end{equation}
In fact, since the vacuum state is the $+1$ eigenstate of the $V_k$ operators, we can simply apply the pairwise Majorana operators $F_{jk}$ to create a pair of fermions at sites $j$ and $k$. To create a density pattern that contains an even number of fermions, we simply pair the filled sites in the pattern and apply the pairwise Majorana operators to each pair. Figure~\ref{fig:density_pattern} shows an example of a spin density wave, consisting of a half-filled pattern with spins arranging in a checkerboard pattern. The corresponding Pauli operators to create this state on top of the vacuum state are shown in the right of the figure. Note that we can additionally use the fact that any $Z$ Pauli operator on the physical sites will simply add a global phase, and so can be ignored. Alternatively, if idling qubits are problematic for the quantum computing architecture, we can trivially add $Z$ operators to any idle physical qubits. Note that the $Z$ operators acting on the secondary qubits have a non-trivial action and cannot be ignored---indeed, these imprint the pattern of fermions onto the background toric code state.

\subsubsection{Fermionic Ground State Preparation}\label{sec:ground state}

It may also be of interest to prepare the ground state of some fermionic Hamiltonian, or indeed the ground state of the Fermi-Hubbard model itself. Finding the ground state may be our end goal, or it could be the initial state for dynamics, including global quantum quenches, or the computation of spectral functions. Ground state preparation is a large topic in itself with substantial challenges. We will not attempt to cover the full topic here, but we will provide a brief overview of the main approaches. 

One approach for approximating the ground state is to consider a mean-field solution to Fermi-Hubbard model, for example, using the Hartree-Fock method~\cite{Bruus2004}. Using efficient classical numerical methods, we can use a single Slater determinant to self-consistently find the Hartree-Fock solution. Since the resulting state is Gaussian, it can be efficiently prepared as a quantum circuit~\cite{Jozsa2008}. Furthermore, shallower circuit approximations can be found on classical computers, which can then be prepared on the quantum computer. This approach has the downside that we only have an approximate mean-field solution but it might be a good starting point more involved variational approaches.

In the near-term (and potentially beyond) it is likely that the most efficient way to find ground states, is to classically optimize the quantum circuit within a given class of circuits. Ground states are observed to obey an area law scaling of entanglement entropy, which means that the corresponding quantum circuits can be efficiently captured using tensor network methods. While these types of tensor network simulations can be difficult in two dimensions, classical computing resources are currently dramatically cheaper and more plentiful than their quantum counterparts. More concretely, to perform this optimization classically, we can use the approach introduced by Evenbly and Vidal~\cite{Evenbly2009}, which was utilized in the context of quantum circuits in Ref.~\cite{Lin2021}. The basic idea is to represent the given quantum circuit structure as a tensor network and update the gates (tensors) one by one to maximize the overlap with the target state, which can be obtained using standard tensor network methods. To maximize the overlap, we construct an environment tensor $E$, which is the contraction of the tensor network with the gate in question removed. From a singular value decomposition of $E = U S V^\dag$, the optimal gate is then given by $V U^\dag$. We can then iteratively sweep through the gates locally maximizing the overlap until convergence. For more details see the Appendix of Ref.~\cite{Lin2021}. Alternatively the gates could be optimized using a gradient-based optimization algorithm.

Using tensor networks to approximate ground states is a well-studied problem, there are many efficient algorithms available algorithms, and they are known to converge quickly. While brute force classical optimisation of the quantum circuits is a viable option, the question of how to prepare tensor networks states most efficiently on a quantum computer is still an open question, but there has been a lot of recent progress, particularly in the context of matrix product states (MPS)~\cite{Haghshenas2022,Wei2023,Malz2024,Tantivasadakarn2023,Smith2023,Smith2024,Sahay2024a} and certain subclasses of PEPS~\cite{Wei2022,Sahay2024b}. For instance, while it is known that certain PEPS cannot be efficiently prepared by unitary quantum circuits~\cite{Schuch2007}, there are also no-go results for the tensor networks representation of chiral topological phases of matter~\cite{Dubail2015}.

With an eye on the future development of quantum computing, there is some belief that variational quantum algorithms may ultimately be a more powerful or efficient approach to finding certain quantum many-body ground states~\cite{Cerezo2021} (and also in the context of classical optimization problems~\cite{Fahri2014}).
The variational quantum eigensolver (VQE)~\cite{Peruzzo2014,Wecker2015,Kandala2017,Tilly2022,Consiglio2022,Li2023} is a hybrid quantum-classical algorithm, where a quantum computer is used to prepare a trial (or ansatz) wavefunction, and a classical computer is used to optimize the parameters of the trial wavefunction. Note that the optimisation of variational quantum circuits is currently an open and subtle problem due to the complex optimisation landscapes for generic circuits. Here we can construct a trial wavefunctions as a parametrized circuit consisting of fermionic unitary gates introduced in Section~\ref{sec:decomposition unitary}, applied to the fermionic vacuum state introduced in Section~\ref{sec:vacuum}. This ensures that the trial wavefunction is in the physical subspace. Furthermore, since the product and sums of the pairwise Majorana operators spans the space of even fermionic operators, this type of circuit can in principle be used to approximate the ground state to arbitrary accuracy.

An alternative approach could be the use of imaginary time evolution. For instance, the QITE or QLanczos algorithms could be used to prepare the ground state using purely unitary circuits~\cite{McArdle2019,Motta2020}. Through the use of a variational ansatz and projective measurements, a non-unitary analogue to TEBD in MPS was also introduced in Ref.~\cite{Lin2021} in 1D, but could be generalised to 2D. Other alternatives that make use of designed Master equation dynamics are the \emph{Dissipative Quantum Eigensolver}~\cite{Cubitt2023} and the \emph{Thermal Gradient Descent}~\cite{Chen2023}.

\subsection{Time Evolution}

While the preparation of the ground state is valuable in itself, the initial states can serve as the starting point for the simulation of unitary dynamics. For instance, local quenches can probe the elementary excitations of the model and are relevant for the experimental investigation of materials properties, through linear response. Global quenches are also becoming increasingly relevant due to the existence of quantum simulators and quantum computers. The resulting far-from-equilibrium dynamics reveals how quantum many-body systems relax or thermalise~\cite{Vasseur2016,Essler2016}, and can be used to study dynamical phase transitions~\cite{Heyl2018}, the emergence of hydrodynamics~\cite{Nahum2018,vonKeyserlingk2018,Rakovszky2018}, and the scrambling of quantum information~\cite{Maldacena2016,Hosur2016,Roberts2017}. Quantum many-body dynamics is particularly challenging for classical computers due to the rapid growth of entanglement, which severely limits the applicability of classical algorithms based on tensor networks. Quantum computers, on the other hand, provide a natural platform for simulating quantum dynamics, where the cost of the simulation (in principle) grows polynomially in both the number of qubits and the simulation time. Here we focus on the most direct method to simulate quantum dynamics on gate-based quantum computers: the Trotter decomposition.


\subsubsection{The Trotter Decomposition}

For local Hamiltonians, such as the Fermi-Hubbard model, the most natural approach to simulating unitary dynamics is direct application of the Trotter or Suzuki-Trotter decompositions~\cite{Trotter1959,Hatano2005}. The idea is to discretize time into small steps $dt$ and to approximate the time evolution operator $e^{-iH dt}$ as a product of local unitary operators, each of which can be efficiently implemented on a quantum computer~\cite{Lloyd1996}. The simplest version of the Trotter decomposition is the first-order approximation, 
\begin{equation}\label{eq:Trotter}
e^{-i\sum_i h_i dt} = \prod_i e^{-i h_i dt} + O(dt^2),
\end{equation}
where $h_i$ are the local terms in the Hamiltonian $H = \sum_i h_i$, and $dt$ is the time step. If the total time $t = n dt$, then the total error in the evolution is $O(n dt^2) = O(t dt)$. The error can be reduced further by using a smaller time step and/or by using a higher order Suzuki-Trotter decomposition~\cite{Hatano2005}. Choosing the optimal timestep is a problem that goes beyond this work (see e.g. Ref.~\cite{Heyl2019}) but this is often done through numerical experimentation. Importantly, the Trotter decomposition requires a quantum circuit that is polynomial in the time you want to reach.

In Eq.~\eqref{eq:Trotter}, we did not specify the order in the product. While this doesn't affect the asymptotic scaling of the expansion, there are practical considerations for choosing specific orderings. By grouping terms that don't overlap, these local unitary gates can be applied in parallel so that a single time step can be implemented by a finite depth local unitary circuit, and each local term is of the form introduced in Sec.~\ref{sec:decomposition unitary}. In two dimensions, there is freedom to choose the ordering of the terms while achieving the same minimal depth. This ordering can potentially have an impact on the prefactor of the error term~\cite{Tranter2019}. By classically optimizing the unitary gates appearing in the Trotter decomposition, it is possible to achieve an improved prefactor and even improved asymptotic scaling~\cite{Causer2023,McKeever2023,Maurits2023,Mansuroglu2023,Kotil2024}, although this is yet to be investigated thoroughly in two-dimensions.

\subsection{Measurement}

Now that we have given the recipe for initial state preparation and for simulating quantum dynamics, all that is left is to measure the observables of interest. These come in various forms with various levels of complexity to extract from the quantum computer. In this section we will give a brief overview of the types of observables that can be measured, and discuss the challenges that arise in measuring them.

\begin{figure}[t]
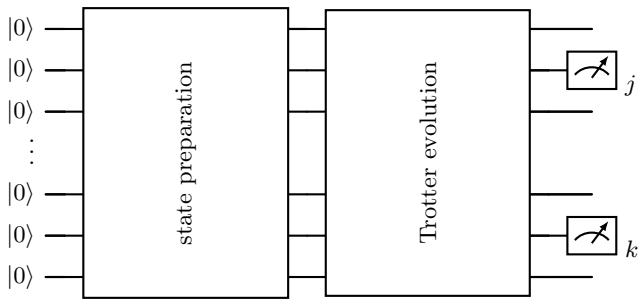

    \begin{center}
    \include{Figures/Tikz/LocalMeasurement}
    \end{center}
    \caption{Schematic of the quantum circuit to measure local observables and equal-time correlation functions. The circuits consist of a state preparation stage, then time evolution implemented using a Trotter decomposition, and finally local measurements. The measurements on sites $j$ and $k$ could be performed in parallel for local observables such as the local fermion density. Alternatively, sites $j$ and $k$ could correspond to two-sites in an equal time correlation function, such as the density-density correlator. 
    }\label{fig:LocalMeasurement}
\end{figure}

\subsubsection{Local observables}

The simplest observables to measure are local observables. These are observables that have a finite support and are geometrically local, for instance, the local density of fermions. These can be measured by simply measuring the Pauli operators that correspond to the local fermionic operators. For example, the local density of fermions at time $t$ is given by
\begin{equation}
\begin{aligned}
\langle n_j(t) \rangle &= \frac{1}{2}\left(1-\langle \psi | U(t)^\dag Z_j U(t) |\psi\rangle  \right).
\end{aligned}
\end{equation}
Here we simply need to measure the state $|\psi(t)\rangle = U(t)|\psi\rangle$ on site $j$ in the $Z$-basis, as illustrated in Fig.~\ref{fig:LocalMeasurement}. The local density is then given by the probability of measuring the $-1$ eigenstate, i.e. $p(|1\rangle)$. In practice, we measure in the computational basis $|0\rangle, |1\rangle$ and estimate the probabilities $p(|0\rangle)$ and $p(|1\rangle)$ by running the circuit many times and averaging over the observed measurement outcomes, as we discuss in Sec.~\ref{sec:measurement}. We can also easily measure sums of local observables, such as the total fermion number $N = \sum_j n_j$. In this case, since the operators in the sum have distinct support, we can measure all terms in parallel by measuring all the qubits in the $Z$ basis. Whether measuring in parallel or running separate circuits for each term is more efficient will depend on the specific hardware implementation since measurement cross-talk can be a significant source of error.

\subsubsection{Equal-time correlation functions}

The same procedure can be used to measure equal-time correlations functions, which may be non-local. For the Fermi-Hubbard model, a natural observable to consider is the density-density correlator given by
\begin{equation}
\langle n_j (t) n_k (t) \rangle = \frac{1}{4} \langle \psi | (1-Z_j(t))(1-Z_k(t))  |\psi \rangle.
\end{equation}
This can be measured in a similar way to the local density, by the site $j$ and $k$ in the $Z$ basis, as illustrated in Fig.~\ref{fig:LocalMeasurement}. The expectation value of the observable is then proportional to the probability of measuring both qubits in the $|1\rangle$ state after evolution time $t$.

\subsubsection{Unequal-time correlation functions}\label{sec:unequal}

It is also possible to extract unequal-time correlation functions. One of the simplest examples of which is the unequal-time density-density correlator
\begin{equation}
\langle n_j (t) n_k(0) \rangle = \langle \psi | e^{iHt} n_j e^{-iHt} n_k | \psi \rangle.
\end{equation}
In contrast to the previous section, the Pauli operators are inserted at different times, meaning that the densities cannot simply be measured at the end of the time evolution. 

There are multiple ways to compute such a quantity. If we only need access to the absolute value, the most direct way is to perform the forward and backwards time evolution and to compute the overlap with the initial state. More explicitly, if $U_\psi$ is the unitary that creates the initial state, i.e. $|\psi\rangle = U_\psi |0\rangle$, then the correlator can be written as
\begin{equation}
\frac{1}{4}\langle 0| U^\dagger_\psi e^{iHt} (1-Z_j) e^{-iHt} (1-Z_k) U_\psi| 0 \rangle.
\end{equation}
We can interpret this as the sum of four terms, where in each we apply the sequence of unitary gates to the initial $|0\rangle$ state, and then at the end measure the overlap with the state $|0\rangle$. More precisely we would have access to the probability of measuring the $|0\rangle$ state at the end of the process. This may not be desired if there is significant measurement error, as it will compound exponentially with the length of the bit string set my the system size.

An alternative method, which would also give us access to both the real and imaginary part of the correlator, would be to use a generalized Hadamard test~\cite{Cleve1998,Somma2002} (as shown in bottom of Fig.~\ref{fig:Measure_Interfero_Two}). The four correlators we want to compute are all of the form $\langle \psi | A(t) B |\psi\rangle$, where $A$ and $B$ are Pauli strings, in this case, just $\mathds{1}$ or $Z$. We choose to use two ancilla, one each for the operators at $j$ and $k$, but it is also possible to use just one~\cite{Somma2002}. If we have all-to-all connectivity, or we are able to have ancilla next to the site $j$ and $k$, then the controlled operations in Fig.~\ref{fig:Measure_Interfero_Two} would simply be controlled-Z gates. Otherwise, we may need to perform a sequence of swap gates. The measurement results on the ancilla qubits (denoted by $a$ and $b$) would be:
\begin{equation}\label{eq: ancilla measurements Greens function}
    \langle (X_a -iY_a)(X_b-iY_b)\rangle = \langle A(t)B^{} \rangle
\end{equation}
In Appendix~\ref{app: hadamard test} we provide a derivation of this protocol, and it is possible to perform the same measurement using a single ancilla qubit, as explained in Ref.~\cite{Somma2002}.

\subsubsection{Green's functions}\label{subsubsec: Greensfunction}

Green's functions---and the related spectral functions---are of particular interest for the study of quantum many-body systems since they probe the linear response of the system, and reveal the elementary excitations. In the context of the Fermi-Hubbard model, the Green's functions take the form
\begin{equation}\label{eq: fermion Greens function}
\begin{aligned}
G_{ij}(t) &= i\langle c^\dag_j(t) c^{}_k(0) \rangle \\
&= \frac{i}{4} \big[ \langle \gamma_j(t) \gamma_k \rangle + \langle \bar{\gamma}_j(t) \bar{\gamma}_k \rangle \\ 
& \qquad \quad {} + i\langle \gamma_j(t) \bar{\gamma}_k \rangle -i \langle \bar{\gamma}_j(t) \gamma_k \rangle  \big]
\end{aligned}
\end{equation}
These quantities are also examples of unequal time correlators. However, they introduce a new difficulty: they contain individual fermionic creation/annihilation operators, which cannot be written as products of the edge and vertex operators, $F_{jk}$ and $V_j$, and are necessarily non-local as qubit operators.

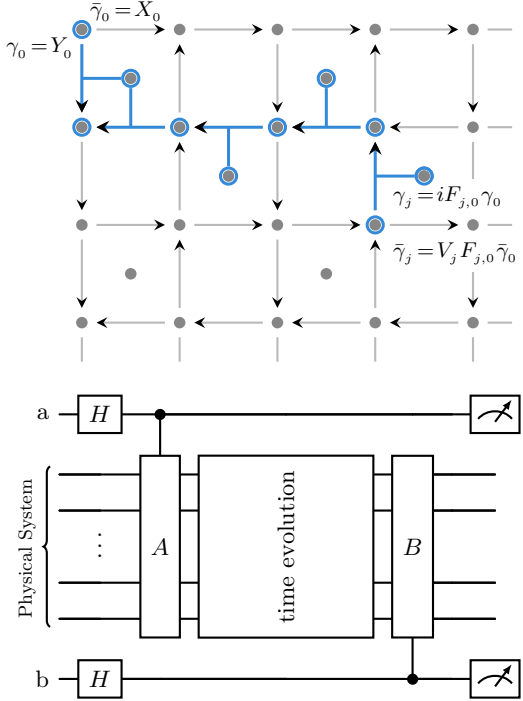
\begin{figure}[t]
    \begin{center}
    \begin{tikzpicture}[scale=1.3,>=stealth,thick]

\tikzset{middlearrow/.style={
    decoration={markings,
        mark= at position 0.45 with {\arrow{#1}},
    },
    postaction={decorate}
    }   
}

\begin{scope}[shift={(0,0)}, local bounding box = scope1]
\foreach \x in {0,1,2,3,4}{  
\foreach \y in {0,1,2,3}{
    \node at (\x,\y)[circle,fill=NodeGray,scale=0.5,opacity=1.0]{}; }}

\foreach \x in {0.5,2.5}{  
\foreach \y in {0.5, 2.5}{
    \node at (\x,\y)[circle,fill=NodeGray, scale=0.5,opacity=1.]{}; }}
    
\foreach \x in {1.5,3.5}{  
\foreach \y in {1.5}{
    \node at (\x,\y)[circle,fill=NodeGray, scale=0.5,opacity=1.0]{}; }}

\foreach \x in {0,1,2,3,4}{
\foreach \y in {0,1,2}{
    \draw[<-, draw=EdgeGray, opacity=1.] (\x,{\y+0.5-pow(-1,\x)*0.35})--(\x,{\y+0.5+pow(-1,\x)*0.35}); }}
\foreach \x in {0,1,2,3}{
\foreach \y in {0,1,2,3}{
    \draw[<-, draw=EdgeGray, opacity=1.] ({\y+0.5-pow(-1,\x)*0.35},\x)--({\y+0.5+pow(-1,\x)*0.35},\x);}}

\foreach \y in {0,1,2,3}{
    \draw[-,draw=EdgeGray, opacity=1.] (4.15,\y)--(4.4,\y); 
    }
\foreach \x in {0,1,2,3,4}{
    \draw[-,draw=EdgeGray, opacity=1.] (\x,-.4)--(\x,-.15); 
}
\node (n0) at (0,3)[scale=.7,circle,draw=SpinBlue,fill=none,line width=0.4mm]{};
\node (n1) at (0,2)[scale=.7,circle,draw=SpinBlue,fill=none,line width=0.4mm]{};
\node (n2) at (1,2)[scale=.7,circle,draw=SpinBlue,fill=none,line width=0.4mm]{};
\node (n3) at (2,2)[scale=.7,circle,draw=SpinBlue,fill=none,line width=0.4mm]{};
\node (n4) at (3,2)[scale=.7,circle,draw=SpinBlue,fill=none,line width=0.4mm]{};
\node (n5) at (3,1)[scale=.7,circle,draw=SpinBlue,fill=none,line width=0.4mm]{};
\node at (0.5,2.5)[scale=.7,circle,draw=SpinBlue,fill=none,line width=0.4mm]{};
\node at (2.5,2.5)[scale=.7,circle,draw=SpinBlue,fill=none,line width=0.4mm]{};
\node at (1.5,1.5)[scale=.7,circle,draw=SpinBlue,fill=none,line width=0.4mm]{};
\node at (3.5,1.5)[scale=.7,circle,draw=SpinBlue,fill=none,line width=0.4mm]{};
\def\ss{0.1}    
\draw[line width=0.4mm, ->, draw=SpinBlue] (0,2.85)--(0,2.15);
\draw[line width=0.4mm, <-, draw=SpinBlue] (0.15,2)--(0.85,2);
\draw[line width=0.4mm, <-, draw=SpinBlue] (1.15,2)--(1.85,2);
\draw[line width=0.4mm, <-, draw=SpinBlue] (2.15,2)--(2.85,2);
\draw[line width=0.4mm, <-, draw=SpinBlue] (3,1.85)--(3,1.15);

\draw[line width=0.4mm,-, draw=SpinBlue] (.5,2)--(0.5,2.4);
\draw[line width=0.4mm,-, draw=SpinBlue] (2.5,2)--(2.5,2.4);
\draw[line width=0.4mm,-, draw=SpinBlue] (1.5,1.6)--(1.5,2.0);
\draw[line width=0.4mm,-, draw=SpinBlue] (0,2.5)--(0.4,2.5);
\draw[line width=0.4mm,-, draw=SpinBlue] (3,1.5)--(3.4,1.5);

\node at (0.0,3.0)[scale=.9, anchor=south west]{$\bar{\gamma}_{_{0}}\! =\! X_{_{0}}$};
\node at (0.0,3.0)[scale=.9, anchor=north east]{$\gamma_{_{0}}\! =\! Y_{_{0}}$};

\node at (3.1,0.91)[scale=.9, anchor=north west, fill=white,opacity=.7,text opacity=1]{$\bar{\gamma}_{_{j}} \!=\! V_{_{j}} F_{_{j,0}} \bar{\gamma}_{_{0}}$};
\node at (3.1,1.08)[scale=.9, anchor=south west, fill=white,opacity=.7,text opacity=1]{$\gamma_{_{j}} \!=\! iF_{_{j,0}} \gamma_{_{0}} $};

\node at (3.5,1.5)[circle,fill=NodeGray,scale=0.5,opacity=1.0]{};
\node at (3.5,1.5)[scale=.7,circle,draw=SpinBlue,fill=none,line width=0.4mm]{};

\end{scope}


\node [below=-2mm of scope1.south] {
\def\rsep{-3.2}
\begin{quantikz}[row sep={0.8cm,between origins}, column sep=.25cm]
\lstick{a} & \gate{H} & \ctrl{1} &\qw & \qw &\qw &\meter{}\\
\lstick[wires=5, label style={black,rotate=90, anchor=south west, label distance=13mm, scale=0.8}]{Physical System} &\qw & \gate[wires=5]{A} & \gate[wires=5, label style={black,rotate=90}]{\text{time evolution}} & \gate[wires=5]{B} &\qw & \qw \\[\rsep mm,between origins]
\lstick{} &\qw & & & &\qw &\qw \\[\rsep mm,between origins]
\lstick{} & \wireoverride{n}\rvdots & \wireoverride{n} & \wireoverride{n} & \wireoverride{n} & \wireoverride{n} & \wireoverride{n} \\[\rsep mm,between origins]
\lstick{} &\qw & & & &\qw &\qw \\[\rsep mm,between origins]
\lstick{} &\qw & & & &\qw &\qw \\ 
\lstick{b} &\gate{H} & \qw &\qw & \ctrl{-1} &\qw &\meter{} \\ [-5.0cm]
\end{quantikz}
};
\end{tikzpicture}
    \end{center}
    \caption{(Top) Implementation of individual Majorana fermion operators. By placing the secondary qubits in the top left plaquette, these Majorana operators can be realized locally. When acting on sites in the rest of the system, these operators are mapped to non-local qubit operators, which are simple Pauli-strings. (Bottom) The quantum circuit to measure the unequal-time correlators of the form $\langle B(t) A\rangle$ needed to compute the Green's function.
    }\label{fig:Measure_Interfero_Two}
\end{figure}

In order to implement the creation/annihilation operators, or equivalently the single Majorana operators $\gamma_i, \bar{\gamma}_i$, we can always make a strategic choice of where to place the secondary qubits. If we ensure that there is a secondary qubit in one of the corner plaquettes of the system (c.f. Fig.~\ref{fig:Measure_Interfero_Two}(top)), then we have access to the full fermionic Hilbert space, including the odd parity sector, because we have qubits that are not involved in the stabilizer plauette operators $\mathbb{J}_p$. If we label this corner site by index $0$, then we can then implement the mapping of the Majorana Operators
\begin{equation}\label{eq:single majorana operators}
\gamma_0 \mapsto Y_0, \qquad \bar{\gamma}_0 \mapsto X_0.
\end{equation}
It can easily be checked that these operators have the correct (anti-) commutation relations with $\tilde{E}_{0j}$ and $\tilde{V}_0$, as we show in Appendix~\ref{app: single operators}. For any other site on the lattice, we can then use the $F$  operators to move the operators, that is 
\begin{equation}
\gamma_j = iF_{j0} \gamma_0, \qquad \bar{\gamma}_j = -V_j F_{0j} \gamma_0.
\end{equation}
So while these operators are now non-local, they correspond to simple Pauli-strings, see Fig.~\ref{fig:Measure_Interfero_Two}. 

With the Pauli string representation of the Majorana operators, we can then compute the unequal time correlator in a similar way to the previous Section.~\ref{sec:unequal}. That is, we can use the generalized Hadamard test shown in Fig.~\ref{fig:Measure_Interfero_Two}. Due to the non-local nature of the Majorana operators, these would now correspond in general to controlled Pauli operators with linear length. For a device with local connectivity, these operators can be implemented using a unitary circuit with depth that scales linearly with the length of the Pauli string (see Appendix~\ref{sec:controlled_pauli} for explicit circuit). If, however, we have all-to-all connectivity these controlled operations can be implemented with a logarithmic depth circuit, as shown in Appendix~\ref{sec:controlled_pauli}. Furthermore, the use of mid-circuit measurements could enable a constant depth implementation~\cite{Piroli2021,Baumer2024}.



\section{Finishing Touches}

In principle, we now have all the ingredients and steps required to perform an end-to-end simulation of dynamics in the Fermi-Hubbard model. We finish by discussing some of the practical considerations that arise when implementing these steps on a real quantum computer. In particular, we will discuss the impact of noise on the results, and the various error mitigation techniques that can be used to improve the accuracy of the results.

\subsection{Measurement Statistics}\label{sec:measurement}


When working with quantum computers, we typically do not have direct access to the observables of interest. Instead, we must estimate these by performing a series of individual experiments at the end of which we perform a projective measurement (a shot). To estimate simple observables, such as expectation values of Pauli operators, we must measure in the appropriate basis and then average over many shots. While this is quite a basic fact of interacting with quantum computers, it is important factor in the resource cost and the resulting statistical errors when performing experiments.

The statistical error in the estimate of expectation values scales as $1/\sqrt{N}$, where $N$ is the number of shots. Even in a theoretical noise-free device, this can put a limit on the attainable accuracy for variational approaches such as VQE~\cite{Peruzzo2014}. However, by using more sophisticated techniques inspired by Baysian optimisation, it may be possible to achieve improved performance with limited numbers of shots~\cite{Sauvage2020}.



\subsection{Error Mitigation}

The success of many near-term quantum algorithms before fault-tolerance is reliant on error mitigation techniques to improve the accuracy or scope~\cite{Cai2023}. In contrast to error correction, error mitigation methods are not able to systematically reduce the effective error rates of the quantum computation. Instead, these techniques generally involve simplified modelling of the noise channels in the device in order to account for their impact on the computation where possible. While in many cases this will involve an explicit model for the quantum computer, alternative approaches are model independent and instead involve the extrapolation of noisy results where the noise channels have been controllably boosted, or where results are averaged over multiple circuits that are theoretically equivalent.


\subsubsection{Measurement Error Mitigation}

One form of error common to all devices is readout error. The process of projective measurement and readout is a destructive process that involves strong coupling to an auxiliary device and can take a significant amount of time. As a result, this process is prone to bit flip errors that happen at the end of the desired computation. These forms of errors can be mitigated relatively simply by preparing a full set of basis states on the qubits you wish to measure and measuring the probability distribution for the measured states. This provides a matrix relating the probability distributions of the prepared and measured states. By using applying the (pseudo-)inverse of this matrix, it is possible to correct for the readout errors~\cite{Maciejewski2020}.

While conceptually simple, this mitigation technique naturally has a few pitfalls. Firstly, like many error mitigation techniques, it may result in unphysical results, such as expectation values outside the range $[-1,1]$. Secondly, this technique does not differentiate between errors in the initial state preparation and those that occur during measurement. This approach assumes that the latter are dominant. Finally, this technique scales exponentially with the number of qubits. For local operators, this may necessitate the need to measure each operator with a separate set of experiments, whereas for non-local operators is may be prohibitively costly. 

\subsubsection{Post-selection}

When simulating quantum many-body systems, it is often the case that there are conservation laws in the model that are not respected by the quantum computer. This provides a physically motivated mechanism for mitigating errors. Namely, we can measure these conserved quantities and only use the results that satisfy the conservation laws. This is known as post-selection. Conservation laws can come in various forms, such as the conservation of particle number or parity, the presence of stabilizers, or gauge degrees of freedom in lattice gauge theories. The use of ancilla qubits may also introduce additional conservation laws that can be exploited, for examples, that they should remain in zero state after the simulation.

In our case, we know that the plaquette stabilizers are conserved quantities and that our physical subspace is the one where all of these stabilizers are in the $+1$ eigenstate. By measuring the plaquette stabilizers at the end of the computation, we can discard the results that do not satisfy this condition. We additionally have that the total fermion number is conserved, and so we may be able to post-select with respect to this as well. Whether we are able to postselect is dependent on being able to simultaneously measure the conserved quantities and the observables of interest. When the observable commutes with some of the conserved quantities, there may still be significant overhead to measure them simultaneously. This approach also scales exponentially with the number of conserved quantities---due to the reduction in the useable measurement outcomes---so may be impractical for larger systems.


\subsubsection{Depolarizing Error Model}

Another simple approach to error mitigation tackles the non-unitary aspect of quantum circuits when run on a real quantum computer. The simplest model for this is the global depolarizing channel, which is a completely positive trace-preserving map, $\mathcal{E}$, that takes the density matrix $\rho$ to the form 
\begin{equation}
    \mathcal{E}(\rho) = (1-p)\rho + p \frac{\mathbb{I}}{2^n},
\end{equation}
where $p$ is the depolarizing probability and $n$ is the number of qubits~\cite{Nielsen_Chuang_2010}. The expected density matrix of the noise-free quantum computation is a pure state, $\rho = |\psi\rangle \langle \psi|$, and the global depolarizing error channel uniformly moves this density matrix towards the fully mixed state $\rho = \mathbb{I}$. Despite its simplicity, this model can effectively capture the behaviour of real quantum devices~\cite{Vovrosh2021,Urbanek2021}.

The power of this approach is that the effect on the expectation values of an observable can be easily calculated. For instance, the expectation value of a Pauli operator $P \neq \mathbb{I}$ is given by
\begin{equation}
    \langle P \rangle = \text{Tr}(\mathcal{E}(\rho) P) = (1-p)\langle P \rangle_{\text{ideal}}.
\end{equation}
This equation can be inverted so that errors can be mitigated by simply rescaling the measured expectation values. All that remain is to estimate the depolarizing probability $p$ for a particular quantum circuit on a given device. This can be done by running a circuit with the same structure, but with parameters chosen such that the result is known or can be efficiently computed~\cite{Vovrosh2021,Urbanek2021}. The depolarizing probability can then be estimated by comparing the ideal and noisy results. These techniques can also be extended to the measurement of entanglement entropies~\cite{Vovrosh2021}.

\subsubsection{Randomized Compiling}

In many cases, coherent gate errors can be more detrimental to quantum computations or simulations than decoherence. Randomized compiling is a technique that can be used to mitigate these errors~\cite{Wallman2016}. It was originally introduced in the context of error correction as a way of improving error thresholds~\cite{Knill2004,Kern2005}, but has since been utilized in quantum simulation on gate-based quantum computers.

The basic idea is to average over a set of equivalent circuits that differ in their implementation. This results in an effective channel where the coherent errors have been replaced by stochastic Pauli noise~\cite{Wallman2016}. It is based on the idea of Pauli-twirling~\cite{Bennet1996,Knill2008,Geller2013}, where two-qubit Clifford gates (e.g. CNOT or CZ gates), can be dressed by single qubit Clifford gates without changing the operation of the gate. The added single qubit gates can then be incorporated into existing single qubit gates in the circuit. 

A variant on randomized compiling is to exploit a gauge freedom in the quantum circuit. For instance, Ref.~\cite{Lin2021} studied circuits inspired by matrix product states, which have a gauge freedom on the virtual bonds. On the level of quantum circuits, this amounts to inserting random unitary gates and their inverse in certain point in the circuit and then incorporating them separately into existing gates.

\subsubsection{Dynamical Decoupling}

Many types of error in quantum computers ultimately arise due to unintended couplings, either to other qubits, or to the external environment. However, it was realized that strongly driving your system, can effectively suppress other couplings--an effect referred to as dynamical decoupling~\cite{Viola1998,Viola1999,Duan1999,Ezzell2023}. This can be exploited in a quantum computer by realising that any idling qubits not currently involved in the computation can be driven to suppress the effect of noise. Dynamical decoupling can be achieved by implementing a kind of spin echo process~\cite{Hahn1950,Maudsley1986}, where a sequence of pulses (or gates) are applied that ultimately have a trivial effect in the absence of errors, and is now routinely implemented in experiments on superconducting qubits.~\cite{Biercuk2009}. For more details on dynamical decoupling, see Ref.~\cite{Ezzell2023}.

\subsubsection{Zero-noise extrapolation and probabilistic error cancellation}

The final error mitigation techniques that we will mention are zero-noise extrapolation~\cite{Li2017,Temme2017} and the related approach of probabilistic error cancellation~\cite{Temme2017}. The basis of these techniques is to run multiple experiments where the noise is artificially, but controllably, increased. For example, this could be achieved by stretching the pulse sequences for the gates, making them take longer to implement, and thus increasing the exposure noise channels~\cite{Kandala2019}. The errors are then mitigated by fitting the data and extrapolating the results to the limit of zero noise. Probabilistic error cancellation, in particular, considers a probabilistic error model, which can be amplified by stochastically adding the corresponding gates to the circuit and averaging the results~\cite{Temme2017}.


\section{Explicit Example: Resource Estimation}\label{sec: explicit example}

The above recipe allows us to simulate many different scenarios in the context of the two-dimensional Fermi-Hubbard model. To offer some concreteness, it is useful to consider a specific example. This also enables us to estimate the required resources to realistically implement these simulations on near-term devices. In our discussion we assume that entangling operations (here assumed to be CNOT gates) are significantly more costly and prone to errors that single qubit gates, and so don't keep count of the latter. For this explicit example, the number of shots required scales as $(1/\epsilon)^2$, where $\epsilon$ is the standard error of the mean for the observable.

The explicit example we will consider is global quantum quench under the Fermi-Hubbard model from a product state, followed by the measurement of equal-time density-density correlations for a $6\times 8$ square lattice of spinful fermions. This is one of the simplest setups, but also one of the most difficult to simulate using classical numerical methods. It therefore provides an estimate of the quantum resources required to achieve a potential quantum advantage in a physically relevant problem. While this type of simulation may currently be more suitable for analog quantum simulators, our goal is not to compare these current technologies, and indeed due to the flexibility and potential for error correction it is still likely to be a practical application for future digital quantum computers. For this example we consider two possibilities for the connectivity of the qubits: All-to-all and nearest-neighbour with diamond connectivity (see Fig.~\ref{fig:Embeddings}(b)). The former is typical for quantum computing architectures where the qubits can be physically rearranged, such as in trapped ions, and the latter is typically of the nearest-neighbour coupling for fixed qubits, e.g. in superconducting circuits. The resources required for these two cases are summarised in Table~\ref{tab:Resources}, and the detailed circuits are provided in Appendix~\ref{app: explicit example}. We also include the total circuit depth (number of entangling CNOT layers) to simulate using 10 second-order Trotter steps.

\begin{table}[h!]
\centering
\begin{tabular}{ |p{0.31\linewidth}||p{0.31\linewidth}|p{0.31\linewidth}| } 
 \hline
 Architecture & All-to-All & Nearest-neighbour (diamond) \\
 \hline
 Qubits & 132 & 203\\
 \hline
 State Prep. & 3 & 9 \\
 \hline
 Trotter ($1^\text{st}$) & 26 & 46 \\
 \hline
 Trotter ($2^\text{nd}$) & 40 (+6) &  72 (+4) \\
 \hline\hline
 Total Depth\newline (State Prep. + 10 $\times$ $2^\text{nd}$-Trotter) & 409 & 733 \\
 \hline
\end{tabular}
\caption{Resource requirements to simulate a global quantum quench from a product state for a $6\times 8$ square lattice of spinful fermions. The two columns show the numbers for the All-to-all and the nearest-neighbour (diamond) connectivity of the qubits. We include the number of qubits required as well as the number of entangling CNOT layers. At the bottom we also include the total number of entangling layers needed for a full simulation including the state preparation and ten second-order trotter steps. The numbers in brackets are the additional cost for the first trotter step only.}
\label{tab:Resources}
\end{table}

For the all-to-all connectivity, we are able to implement the Derby-Klassen mapping directly, and so we require the minimum possible number of qubits to implement the Derby-Klassen mapping. Furthermore, we can implement the mapping separately for the up and down species of fermion, since all operators have even combinations of these operators. Because of the all-to-all coupling there is no issue with long-range hopping or interactions, see Appendix~\ref{app: explicit example} for more details. The state preparation then consists of constructing the vacuum states, which is equivalent to preparing the toric code ground state on the secondary qubits, which can be done in 3 layers of CNOTs using the sequential preparation from Ref.~\cite{Satzinger2021,Liu2022}. The trotterized evolution operator can then be decomposed into 26 CNOT layers for the first-order decomposition, and 40 for second-order with an extra 6 layers for the first step. The density-density correlator can then be extracted by measuring all the physical qubits in the $Z$ basis.

When we have nearest-neighbour connectivity (on the diamond lattice), we have to embed the mapped qubits into the device qubits, and so we end up with additional ancilla qubits. These additional qubits may be used in the circuit decomposition as long as they are returned to the $|0\rangle$ state. Here we also need to interleave the spin up and down species so that the unit cells are local. This results in a total of 203 qubits to simulate the $6\times 8$ spinful fermions. This restricted connectivity also leads to deeper circuits for the state preparation and both orders of trotterization, as shown in Table~\ref{tab:Resources}. For first-order we need 46 CNOT layers, and 72 layers for second-order, with an addition 4 for the first layer.

If we consider a simulation with 10 second-order trotter steps, then we find that we need 409 CNOT layers for the all-to-all connectivity and 733 for the nearest-neighbour connectivity. The restricted connectivity has resulted in a circuit almost twice as deep, and using around $50\%$ more qubits. Whatever, the quantum computer architecture, simulations on this scales require 100+ qubits, with around 1000 CNOT layers and 10,000 CNOT gates in total. While existing devices have a comparable number of qubits, a new generation of devices will be needed with improved gate fidelity and coherence times in order to convincingly probe non-equilibrium dynamics beyond the reach of our best classical numerical methods.

\section{Outlook}

In this paper, we have provided a step-by-step recipe for simulating the quantum many-body dynamics of the two-dimensional Fermi-Hubbard model on a quantum computer. There has now been a wide range of works that have tackled various aspects of such a simulation, from the mapping of fermionic operators to qubits, the preparation of initial states, the simulation of dynamics, and the measurement of observables. We have provided a comprehensive overview of the current state-of-the-art in these areas. By focussing on a concrete fermion-to-qubit mapping, we have been able to provide a detailed recipe covering all the steps in such a simulation. We have also provided a detailed worked example in Appendix~\ref{app: explicit example}, which has allowed us to estimate realistic resource requirements to access physics beyond classical numerical methods. Different choices may be made at each step, depending on the problem of interest and the hardware implementation, and this recipe can also provide a template in these cases.

The Fermi-Hubbard model is a particularly interesting model to study on a quantum computer as it is a paradigmatic model for strongly correlated systems, and is relevant for the study of metals, Mott insulators, magnetically ordered systems, and high-temperature superconductivity. The model in two-dimensions has also proven challenging for existing analytical and numerical approaches. Fermi-gases in optical lattices~\cite{Omran2015,Cheuk2015,Haller2015,Parsons2015,Edge2015} have proven to be a powerful setting to study the Fermi-Hubbard model in one, two and three dimensions~\cite{Cheuk2016,Boll2016,Cocchi2016,Mazurenko2017,Sompet2022,Xu2023, kohl2005}. While these so-called quantum simulators are currently much larger and higher fidelity than the available quantum computers, the flexibility of quantum computers, and the wider range of observables that can be readily accessed, make them a very promising tool for studying strongly-correlated quantum many-body physics. While we have focussed on the simplest version of the Fermi-Hubbard model on a square lattice, the same mapping allows us to study the model on other lattices, such as the triangular or Kagome lattices, or to include next-nearest neighbour hopping. These can all be incorporated into the same construction. Additionally, by further enlarging the unit cell, we can include more orbitals in the model. 

For the study of fermions on quantum computers, we are now spoilt for choice of what mapping to use between fermionic degrees of freedom and qubits. This choice allows us to balance the different resources available to us for our problem of interest, such as the connectivity of the quantum computer, the number of qubits available, and accessible circuit depths. The Jordan-Wigner transformation, while attractively simple, is particularly problematic for dynamics due to multi-qubit, long-range unitary gates that are required. By allowing for additional qubits in the mapping, it has been shown that a purely local mapping of operators can be achieved, at the cost of a long-range entangled vacuum state related to $\mathbb{Z}_2$ topological order. In this paper we focussed on the Derby-Klassen mapping~\cite{Derby2021}, which is particularly compact in the context of simulation the Fermi-Hubbard model and its dynamics. Some local fermion-to-qubit mappings discussed in Sec.~\ref{sec:Fermion_to_Qubit_Mappings} can also extend this recipe to study three-dimensional systems.

There have also been a range of algorithmic advances, for near-terms quantum simulation. These involved variational and non-variational algorithms for finding ground states~\cite{Peruzzo2014,Kandala2017,Tilly2022,McArdle2019,Motta2020,Lin2021,Cubitt2023,Chen2023}. By restricting the available qubit operators to those that correspond to the physical fermionic operators, these algorithms can be implemented with a minimal number of qubits and a minimal circuit depth. For simulating time evolution, variational algorithms have also been proposed as ways to improve upon simple trotterised evolution by taking inspiration from tensor network approaches and storing the current state in a compressed form as a parametrised quantum circuit~\cite{Lin2021,Barratt2021}. While most of these approaches have so far only been tested in one-dimension, they are promising for the future of quantum simulation in two-dimensions. 

An aspect of this quantum simulation that has not been a focus of this paper has been the experimental developments in quantum computers. We have taken the view that we have access to imperfect quantum computers and our goal is to maximize the utility of these devices. However, in parallel, experimental progress in quantum computers continues to roll on at a rapid and accelerating pace. There are now a collection of devices on the market, using a range of different technologies that present their own unique advantages and challenges. These quantum computers are now reaching a scale and a quality such that the community's collective goal of a practical quantum advantage seems tantalisingly close. As can be seen by the resource estimates we provided in Section~\ref{sec: explicit example}, quantum many-body simulations that challenge the limits of classical numerical methods may be possible with the next generation of devices. The quantum simulation of the Fermi-Hubbard model is a particularly promising application for these devices, and it appears to only be a matter of time before experimental advances allow us to follow a recipe of this kind to access new physics.





\section*{Acknowledgements}

We thank Henrik Dreyer and Daniel Malz for useful discussions and for comments on drafts of this manuscript.
A.G-S. and A.J. acknowledge support from the Royal Commission for the Exhibition of 1851.
A.G-S. 
was supported by the UK Research and Innovation (UKRI) under the UK government’s Horizon Europe funding guarantee [grant number EP/Y036069/1]. F.P.~acknowledges support by the European Research Council (ERC) under the European Union’s Horizon 2020 research and innovation program under Grant Agreement No. 771537. F.P.~acknowledges the support of the Deutsche Forschungsgemeinschaft (DFG, German Research Foundation) under Germany’s Excellence Strategy EXC-2111-390814868,  TRR 360 - 492547816, and FOR 5522 (project-id 499180199). This research is part of the Munich Quantum Valley, which is supported by the Bavarian state government with funds from the Hightech Agenda Bayern Plus.

\appendix

\section{Circuit details for explicit example}\label{app: explicit example}

In this appendix we work through the explicit circuit details for realising a concrete example of a quantum simulation, discussed in Sec.~\ref{sec: explicit example}. We consider simulating a quantum quench for a $6 \times 8$ system. Specifically, we start from a spin density wave initial state, then simulate dynamics under the Fermi-Hubbard model and measure the density-density correlation functions. We provide efficient quantum circuits for implementing this setup on two different geometries: (1) All-to-All connectivity; (2) Nearest-Neighbour connectivity in a diamond layout, shown in Fig.~\ref{fig:Embeddings}(b). In both cases, we must consider: the lattice mapping, the vacuum preparation, initial state preparation, and then the implementation of the vertical and horizontal hopping terms and the interactions in the trotterised evolution. In the following we count the number of entangling layers and assume that the layers of single qubits are effectively free, both in the gate fidelity and the gate duration in comparison to the two-qubit gates.

\subsection{All-to-all connectivity}

Let us first consider an implementation where we have all-to-all connectivity. More specifically, we will consider access to arbitrary single qubit gates, and CNOT gates that can be implemented between any pair of qubits. This setup is typical of quantum computing platforms based on trapped atoms or ions~\cite{Bruzewicz2019,Pino2021,Moses2023,Chen2023c}.

\subsubsection{Lattice Mapping}

\begin{figure}[b]
    \centering
    \begin{tikzpicture}[scale=0.45,thick]


\begin{scope}[shift={(3,0)},fill=SpinBlue, draw=SpinBlue]
\foreach \x in {0,1,2,3,4,5}{  
\foreach \y in {0,1,2,3,4,5,6,7}{
    \fill (\x,\y) circle (3pt);
    }}
\foreach \x in {0.5,2.5,4.5}{  
\foreach \y in {0.5,2.5,4.5,6.5}{
    \fill[NodePurple] (\x,\y) circle (3pt);}}
\foreach \x in {1.5,3.5}{  
\foreach \y in {1.5,3.5,5.5}{
    \fill[NodePurple] (\x,\y) circle (3pt); }}
    
\end{scope}

\begin{scope}[shift={(9,0)},fill=SpinRed]
\foreach \x in {0,1,2,3,4,5}{  
\foreach \y in {0,1,2,3,4,5,6,7}{
    \fill (\x,\y) circle (3pt);
    }}

\foreach \x in {0.5,2.5,4.5}{  
\foreach \y in {0.5,2.5,4.5,6.5}{
    \fill[NodePurple] (\x,\y) circle (3pt);}}
\foreach \x in {1.5,3.5}{  
\foreach \y in {1.5,3.5,5.5}{
    \fill[NodePurple] (\x,\y) circle (3pt); }}

\end{scope}

\begin{scope}[shift={(-4,0)}, draw=black, fill=black]
\foreach \y in {0,1,2,3,4,5,6,7}{ 
    \draw[-, opacity=1.] (0,\y)--(5,\y); } 
\foreach \x in {0,1,2,3,4,5}{ 
    \draw[-, opacity=1.] (\x,0)--(\x,7); } 
\foreach \x in {0, 1, 2, 3, 4, 5}{  
\foreach \y in {0, 1, 2, 3, 4, 5, 6, 7}{
    \fill (\x,\y) circle (3pt);
    }}
\end{scope}

\begin{scope}[shift={(1.5,3.5)}, draw=black, fill=black, text=black]
    \draw[line width=0.2mm,-latex, scale=0.6] (0.,0.) -- (1.7,0.);
\end{scope}

\end{tikzpicture}
    \caption{Mapping of a $6 \times 8$ square lattice of fermions when we have all-to-all connectivity. Due to the form of the Hamiltonian, the spin up (blue) and down (red) degrees of freedom can be mapped separately. The purple points correspond to the secondary qubits required by the Derby-Klassen mapping.}
    \label{fig:LatticeMappingAllToAll}
\end{figure}
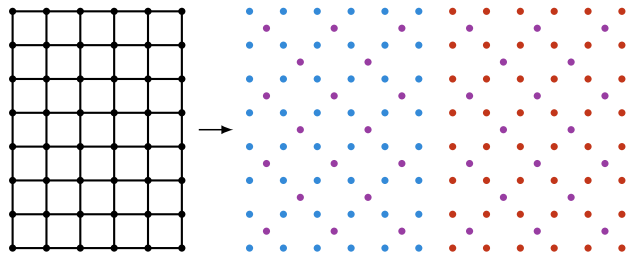

When we have all-to-all connectivity we are able to realise the Derby-Klassen mapping directly. Furthermore, since the Fermi-Hubbard model has only terms that are even both in spin up and down operators, we can separately implement the Derby-Klassen mapping for each species. The corresponding mapping for a $6 \times 8$ lattice of spinful fermions is shown in Fig.~\ref{fig:LatticeMappingAllToAll}. This consists of 132 qubits, 96 of which correspond to the primary qubits, and the remaining 36 are the secondary qubits from the Derby-Klassen mapping. The number of qubits required scales approximately as $3N$, where $N$ is the number of spinful lattice sites. The exact expression (given in Ref.~\cite{Derby2021}) for the number of qubits is not particularly insightful, and it is more practical to draw the setup to work out how many secondary qubits are required.

\subsubsection{Vacuum Preparation}

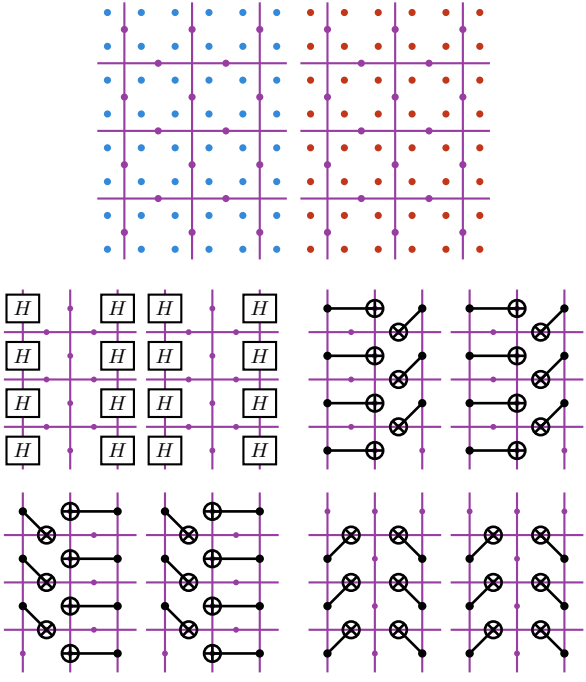
\begin{figure}[t!]
    \centering
    \begin{tikzpicture}[shift={(0,0)},scale=0.45,thick]

\begin{scope}[shift={(0,0)},local bounding box = scope1]
\foreach \x in {0.5,2.5,4.5,6.5,8.5,10.5}{
    \draw[-,EdgePurple, opacity=1.0] (\x,-.3)--(\x,7.3);
}  
\foreach \y in {1.5,3.5,5.5}{
    \draw[-,EdgePurple, opacity=1.0] (-.3,\y) -- (5.3,\y);
    \draw[-,EdgePurple, opacity=1.0] (5.7,\y) -- (11.3,\y);}
\foreach \x in {0,1,2,3,4,5}{  
\foreach \y in {0,1,2,3,4,5,6,7}{
    \fill[SpinBlue] (\x,\y) circle (3pt);   }}
\foreach \x in {6,7,8,9,10,11}{  
\foreach \y in {0,1,2,3,4,5,6,7}{
    \fill[SpinRed] (\x,\y) circle (3pt);   }}
\foreach \x in {0.5,2.5,4.5,6.5,8.5,10.5}{  
\foreach \y in {0.5,2.5,4.5,6.5}{
    \fill[NodePurple] (\x,\y) circle (3pt);  }}
\foreach \x in {1.5,3.5,7.5,9.5}{  
\foreach \y in {1.5,3.5,5.5}{
    \fill[NodePurple] (\x,\y) circle (3pt);  }}


\end{scope}

\coordinate (f11) at ($(scope1.south) + (-8.35,-6.)$);
\coordinate (f12) at ($(scope1.south) + (0.65,-6)$);
\coordinate (f21) at ($(scope1.south) + (-8.35,-12)$);
\coordinate (f22) at ($(scope1.south) + (0.65,-12)$);

\begin{scope}[shift={(f11)}, scale=0.7]
\foreach \x in {0.5,2.5,4.5,6.5,8.5,10.5}{
    \draw[-,EdgePurple] (\x,-.3)--(\x,7.3);}  
\foreach \y in {1.5,3.5,5.5}{
    \draw[-,EdgePurple] (-.3,\y) -- (5.3,\y);
    \draw[-,EdgePurple] (5.7,\y) -- (11.3,\y);}
\foreach \x in {0,2,4,6,8,10}{  
\foreach \y in {0,2,4,6}{
    \fill[NodePurple] (\x+0.5,\y+.5) circle (3.5pt);  }}
\foreach \x in {1,3,7,9}{  
\foreach \y in {1,3,5}{
    \fill[NodePurple] (\x+0.5,\y+0.5) circle (3.5pt);  }}
\foreach \x in {0,4,6,10}{  
\foreach \y in {0,2,4,6}{
    \draw (\x+0.5,\y+0.5) node[minimum size=2.mm,draw,scale=0.85, fill=white] {$H$};}}
    
\end{scope}

\begin{scope}[shift={(f12)}, scale=0.7]
\foreach \x in {0,2,4,6,8,10}{  
\foreach \y in {0,2,4,6}{
    \fill[NodePurple] (\x+0.5,\y+.5) circle (3.5pt);  }}
\foreach \x in {1,3,7,9}{  
\foreach \y in {1,3,5}{
    \fill[NodePurple] (\x+0.5,\y+0.5) circle (3.5pt);  }}
\foreach \x in {0.5,2.5,4.5,6.5,8.5,10.5}{
    \draw[-,EdgePurple] (\x,-.3)--(\x,7.3);}  
\foreach \y in {1.5,3.5,5.5}{
    \draw[-,EdgePurple] (-.3,\y) -- (5.3,\y);
    \draw[-,EdgePurple] (5.7,\y) -- (11.3,\y);}
\foreach \x in {0,6}{  
\foreach \y in {0,2,4,6}{
    \cnot{\x+0.5}{\y+0.5}{\x+2.5}{\y+0.5}   
    }}

\foreach \x in {4,10}{  
\foreach \y in {2,4,6}{
    \dcnot{\x+0.5}{\y+0.5}{\x-0.5}{\y-0.5}   
    }}

\end{scope}

\begin{scope}[shift={(f21)}, scale=0.7]
  
\foreach \x in {0.5,2.5,4.5,6.5,8.5,10.5}{
    \draw[-,EdgePurple] (\x,-.3)--(\x,7.3);}  
\foreach \y in {1.5,3.5,5.5}{
    \draw[-,EdgePurple] (-.3,\y) -- (5.3,\y);
    \draw[-,EdgePurple] (5.7,\y) -- (11.3,\y);}
\foreach \x in {0,2,4,6,8,10}{  
\foreach \y in {0,2,4,6}{
    \fill[NodePurple] (\x+0.5,\y+.5) circle (3.5pt);  }}
\foreach \x in {1,3,7,9}{  
\foreach \y in {1,3,5}{
    \fill[NodePurple] (\x+0.5,\y+0.5) circle (3.5pt);  }}
\foreach \x in {2,8}{  
\foreach \y in {0,2,4,6}{
    \cnot{\x+2.5}{\y+0.5}{\x+0.5}{\y+0.5}
    }}
\foreach \x in {0,6}{  
\foreach \y in {2,4,6}{
    \dcnot{\x+0.5}{\y+0.5}{\x+1.5}{\y-0.5}
    }}
    
\end{scope}

\begin{scope}[shift={(f22)}, scale=0.7]
    
\foreach \x in {0.5,2.5,4.5,6.5,8.5,10.5}{
    \draw[-,EdgePurple] (\x,-.3)--(\x,7.3);}  
\foreach \y in {1.5,3.5,5.5}{
    \draw[-,EdgePurple] (-.3,\y) -- (5.3,\y);
    \draw[-,EdgePurple] (5.7,\y) -- (11.3,\y);}
\foreach \x in {0,2,4,6,8,10}{  
\foreach \y in {0,2,4,6}{
    \fill[NodePurple] (\x+0.5,\y+.5) circle (3.5pt);  }}
\foreach \x in {1,3,7,9}{  
\foreach \y in {1,3,5}{
    \fill[NodePurple] (\x+0.5,\y+0.5) circle (3.5pt);  }}
\foreach \x in {3.5,9.5}{  
\foreach \y in {2,4,6}{
    \dcnot{\x+1}{\y-1.5}{\x}{\y-0.5}
    }}
\foreach \x in {0,6}{  
\foreach \y in {0,2,4}{
    \dcnot{\x+0.5}{\y+0.5}{\x+1.5}{\y+1.5}
    }}
    
\end{scope}


\end{tikzpicture}
    \caption{Preparation of the vacuum state for all-to-all connectivity. This type of connectivity allows us to apply operations only on the secondary (purple) qubits. Top left shows the hypothetical lattice for the secondary qubits that is used for the state preparation. In the bottom the gate sequence proceeds from top-left to bottom-right.}
    \label{fig:VacuumPrepAllToAll}
\end{figure}

For the vacuum preparation we can take advantage of the all-to-all connectivity, which allows us to act only on the secondary qubits. To create the toric code ground state using a unitary circuit we use the linear-depth construction from Refs.~\cite{Satzinger2021,Liu2022}. With our choice to separate the species, the vaccum state is a product state of two toric code ground states, which can be prepared in parallel. The circuit sequence is shown in Fig.~\ref{fig:VacuumPrepAllToAll}. This circuit constructs the circuit in the more standard toric code basis and so we follow this by a layer of single qubit gates. This basis transformation is achieved by applying a Hadamard gate on all secondary qubits, then applying a $R_X(-\pi/2)$ rotation on secondary qubits that lie on horizontal bonds, as shown by the purple lattice in Fig.~\ref{fig:VacuumPrepAllToAll}. This vacuum preparation takes 3 CNOT layers.

\subsubsection{Hopping}

\begin{figure}[t!]
    \centering
    \begin{tikzpicture}[scale=0.35,thick]


\begin{scope}[shift={(0,0)}]
\foreach \x in {0,2,4,6,8,10}{  
\foreach \y in {0,2,4,6}{
    \draw [rounded corners=1.5mm,fill=HighGray,draw=none, scale=1.0 ] (\x-.2,\y-.4)--(\x+.9,\y+.5)--(\x-.2,\y+1.4)--cycle;  }}
\foreach \x in {1,3,7,9}{  
\foreach \y in {1,3,5}{
    \draw [rounded corners=1.5mm,fill=HighGray,draw=none, scale=1.0 ] (\x-.2,\y-.4)--(\x+.9,\y+.5)--(\x-.2,\y+1.4)--cycle;  }}

\foreach \x in {5,11}{  
\foreach \y in {1,3,5}{
    \draw [rounded corners=0.72mm,fill=HighGray,draw=none, scale=1.0 ] (\x-0.2,\y-0.2) rectangle (\x+0.2,\y+1.2){};  }}

\foreach \x in {0,1,2,3,4,5}{  
\foreach \y in {0,1,2,3,4,5,6,7}{
    \fill[fill=SpinBlue] (\x,\y) circle (3pt); }}
\foreach \x in {6,7,8,9,10,11}{  
\foreach \y in {0,1,2,3,4,5,6,7}{
    \fill[SpinRed] (\x,\y) circle (3pt); }}
\foreach \x in {0.5,2.5,4.5,6.5,8.5,10.5}{  
\foreach \y in {0.5,2.5,4.5,6.5}{
    \fill[NodePurple] (\x,\y) circle (3pt);}}
\foreach \x in {1.5,3.5,7.5,9.5}{  
\foreach \y in {1.5,3.5,5.5}{
    \fill[NodePurple] (\x,\y) circle (3pt); }}

\end{scope}

\begin{scope}[shift={(13,0)}]
\foreach \x in {0,2,4,6,8,10}{  
\foreach \y in {0,2,4,6}{
    \draw [rounded corners=1.5mm,fill=HighGray,draw=none, scale=1.0 ] (\x+.1,\y+.5)--(\x+1.2,\y-.4)--(\x+1.2,\y+1.4)--cycle;
}}
\foreach \y in {1,3,5}{
\foreach \x in {1,3,7,9}{  
    \draw [rounded corners=1.5mm,fill=HighGray,draw=none, scale=1.0 ] (\x+.1,\y+.5)--(\x+1.2,\y-.4)--(\x+1.2,\y+1.4)--cycle;}
\foreach \x in {0,6}{  
    \draw [rounded corners=.72mm,fill=HighGray,draw=none, scale=1.0 ] (\x-0.2,\y-0.2) rectangle (\x+0.2,\y+1.2) {};}
}
\foreach \x in {0,1,2,3,4,5}{  
\foreach \y in {0,1,2,3,4,5,6,7}{
    \fill[fill=SpinBlue] (\x,\y) circle (3pt); }}
\foreach \x in {6,7,8,9,10,11}{  
\foreach \y in {0,1,2,3,4,5,6,7}{
    \fill[SpinRed] (\x,\y) circle (3pt); }}
\foreach \x in {0.5,2.5,4.5,6.5,8.5,10.5}{  
\foreach \y in {0.5,2.5,4.5,6.5}{
    \fill[NodePurple] (\x,\y) circle (3pt);}}
\foreach \x in {1.5,3.5,7.5,9.5}{  
\foreach \y in {1.5,3.5,5.5}{
    \fill[NodePurple] (\x,\y) circle (3pt); }}
\end{scope}

\begin{scope}[shift={(0,-9)}]
\foreach \x in {0,2,4,6,8,10}{  
\foreach \y in {0,2,4,6}{
    \draw [rounded corners=1.5mm,fill=HighGray,draw=none, scale=1.0 ] (\x+.5,\y+.1)--(\x-.4,\y+1.2)--(\x+1.4,\y+1.2)--cycle;
}}
\foreach \x in {1,3,7,9}{  
\foreach \y in {1,3,5}{
    \draw [rounded corners=1.5mm,fill=HighGray,draw=none, scale=1.0 ] (\x+.5,\y+.1)--(\x-.4,\y+1.2)--(\x+1.4,\y+1.2)--cycle;
}}
\foreach \x in {1,3,7,9}{  
\foreach \y in {0}{
    \draw [rounded corners=0.72mm,fill=HighGray,draw=none, scale=1.0 ] (\x-0.2,\y-0.2) rectangle (\x+1.2,\y+0.2) {};
}}
\foreach \x in {0,1,2,3,4,5}{  
\foreach \y in {0,1,2,3,4,5,6,7}{
    \fill[fill=SpinBlue] (\x,\y) circle (3pt); }}
\foreach \x in {6,7,8,9,10,11}{  
\foreach \y in {0,1,2,3,4,5,6,7}{
    \fill[SpinRed] (\x,\y) circle (3pt); }}
\foreach \x in {0.5,2.5,4.5,6.5,8.5,10.5}{  
\foreach \y in {0.5,2.5,4.5,6.5}{
    \fill[NodePurple] (\x,\y) circle (3pt);}}
\foreach \x in {1.5,3.5,7.5,9.5}{  
\foreach \y in {1.5,3.5,5.5}{
    \fill[NodePurple] (\x,\y) circle (3pt); }}

\end{scope}

\begin{scope}[shift={(13,-9)}]
\foreach \x in {0,2,4,6,8,10}{  
\foreach \y in {0,2,4,6}{
    \draw [rounded corners=1.5mm,fill=HighGray,draw=none, scale=1.0 ] (\x+.5,\y+.9)--(\x-.4,\y-.2)--(\x+1.4,\y-.2)--cycle;
}}
\foreach \x in {1,3,7,9}{  
\foreach \y in {1,3,5}{
    \draw [rounded corners=1.5mm,fill=HighGray,draw=none, scale=1.0 ] (\x+.5,\y+.9)--(\x-.4,\y-.2)--(\x+1.4,\y-.2)--cycle;}
\foreach \y in {7}{
    \draw [rounded corners=0.72mm,fill=HighGray,draw=none, scale=1.0 ] (\x-0.2,\y-0.2) rectangle (\x+1.2,\y+0.2) {};}
}

\foreach \x in {0,1,2,3,4,5}{  
\foreach \y in {0,1,2,3,4,5,6,7}{
    \fill[fill=SpinBlue] (\x,\y) circle (3pt); }}
\foreach \x in {6,7,8,9,10,11}{  
\foreach \y in {0,1,2,3,4,5,6,7}{
    \fill[SpinRed] (\x,\y) circle (3pt); }}
\foreach \x in {0.5,2.5,4.5,6.5,8.5,10.5}{  
\foreach \y in {0.5,2.5,4.5,6.5}{
    \fill[NodePurple] (\x,\y) circle (3pt);}}
\foreach \x in {1.5,3.5,7.5,9.5}{  
\foreach \y in {1.5,3.5,5.5}{
    \fill[NodePurple] (\x,\y) circle (3pt); }}

\end{scope}


\end{tikzpicture}
    \caption{A sequence for implementing the hopping terms in parallel for all-to-all connectivity to reduce the overall circuit depth. }
    \label{fig:VerticalParallelAllToAll}
\end{figure}
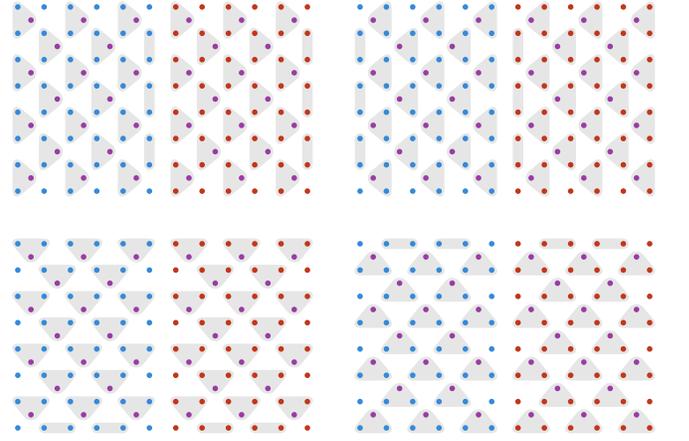

For the vertical hopping, we want to realise the unitary operators
\begin{equation}
    \exp\{ i\alpha (X_1 X_2 X_3 + Y_1 X_2 Y_3)\},
\end{equation}
and for the horizontal hopping
\begin{equation}
    \exp\{ i\alpha (X_1 Y_2 X_3 + Y_1 Y_2 Y_3)\},
\end{equation}
where qubits $1$ and $3$ are the primary qubits and qubit $2$ is the secondary qubit on the adjacent face, and where $\alpha = \frac{1}{2}J$. To implement this using our set of gates we can use the decomposition by Vatan and Williams~\cite{Vatan2004}:
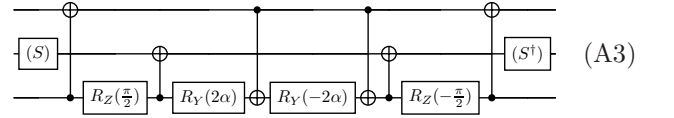
\begin{equation}
\begin{adjustbox}{width=0.85\columnwidth}
\begin{quantikz}[row sep={0.9cm,between origins}, column sep=0.1cm]
\qw & \qw & \targ{} & \qw & \qw & \qw & \ctrl{2} & \qw & \ctrl{2} & \qw & \qw & \targ{} & \qw & \qw \\
\qw & \gate{(S)} & \qw & \qw & \targ{} & \qw & \qw & \qw & \qw & \targ{} & \qw & \qw & \gate{(S^\dag)} & \qw \\
\qw & \qw & \ctrl{-2} & \gate{R_Z(\frac{\pi}{2})} & \ctrl{-1} & \gate{R_Y(2\alpha)} & \targ{} & \gate{R_Y(-2\alpha)} & \targ{} & \ctrl{-1} & \gate{R_Z(-\frac{\pi}{2})} & \ctrl{-2} & \qw & \qw
\end{quantikz}
\end{adjustbox}
\end{equation}
where the gates in brackets are only applied for the horizontal hopping.
This circuit has 6 layers of CNOT gates. By applying these operators on bonds in an alternating fashion, as illustrated in Fig.~\ref{fig:VerticalParallelAllToAll}, we can apply the operator to all bonds in four rounds, resulting in a total depth of 24 CNOT layers.

There are some bonds on the boundary of the system that do not involve the secondary qubits in the vertical hopping. For these bonds we can instead apply $\exp\{i\alpha(X_1 X_2 + Y_1 Y_2)\}$, implemented by the circuit
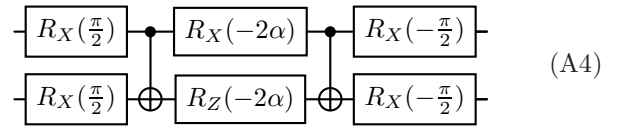
\begin{equation}
\begin{quantikz}[row sep={0.9cm,between origins}, column sep=0.15cm]
\qw & \gate{R_X(\frac{\pi}{2})} & \ctrl{1} & \gate{R_X(-2\alpha)} & \ctrl{1} & \gate{R_X(-\frac{\pi}{2})} & \qw \\
\qw & \gate{R_X(\frac{\pi}{2})} & \targ{} & \gate{R_Z(-2\alpha)} & \targ{} & \gate{R_X(-\frac{\pi}{2})} &\qw 
\end{quantikz}
\end{equation}
which can be applied in parallel with the rest of the bonds, and so doesn't affect the total depth.

\subsubsection{Interaction}

The interaction term is the simplest to implement since it only involves two qubits. The interactions are of the form
\begin{equation}
    \exp\{i\beta(Z_1 Z_2)\} \exp\{-i\beta Z_1\} \exp\{-i\beta Z_2\},
\end{equation}
where $\beta = -\frac{1}{4}U$.
The last two terms are simple $R_Z$ rotation gates, and the two-qubit term can be decomposed into the circuit
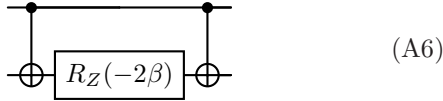
\begin{equation}
\begin{quantikz}[row sep={0.9cm,between origins}, column sep=0.15cm]
\qw & \ctrl{1} & \qw & \ctrl{1} & \qw \\
\qw & \targ{} & \gate{R_Z(-2\beta)} & \targ{} & \qw
\end{quantikz}
\end{equation}
which requires only 2 layers of CNOT gates.

\subsubsection{Final gate count}

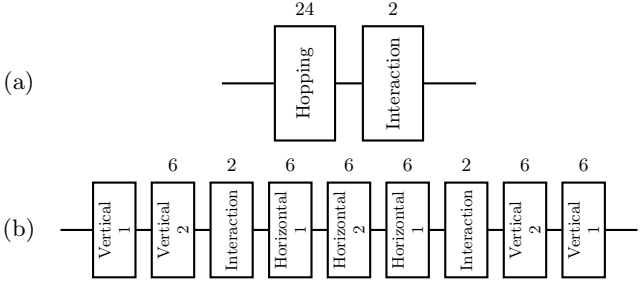
\begin{figure}[t!]
    \centering
\begin{tikzpicture}[scale=0.65,align=center, label distance=.5mm,every label/.style={scale=0.8, black}, tight background, thick]

\node at (-.75,0.0) [rotate=0, scale=1.0] {(b)};
\def \dist{1.2}
\node (a0) at (1*\dist,0) [draw=black, rotate=90, minimum height=8mm, minimum width=18mm, scale=0.7, label=right:${}$] {Vertical \\ 1};
\node (a1) at (2*\dist,0) [draw=black, rotate=90, minimum height=8mm, minimum width=18mm, scale=0.7, label=right:$6$] {Vertical \\ 2};
\node (a2) at (3*\dist,0) [draw=black, rotate=90, minimum height=8mm, minimum width=18mm, scale=0.7, label=right:$2$] {Interaction};
\node (a3) at (4*\dist,0) [draw=black, rotate=90, minimum height=8mm, minimum width=18mm, scale=0.7, label=right:$6$] {Horizontal \\ 1};
\node (a4) at (5*\dist,0) [draw=black, rotate=90, minimum height=8mm, minimum width=18mm, scale=0.7, label=right:$6$] {Horizontal \\ 2};
\node (a5) at (6*\dist,0) [draw=black, rotate=90, minimum height=8mm, minimum width=18mm, scale=0.7, label=right:$6$] {Horizontal \\ 1};
\node (a6) at (7*\dist,0) [draw=black, rotate=90, minimum height=8mm, minimum width=18mm, scale=0.7, label=right:$2$] {Interaction};
\node (a7) at (8*\dist,0) [draw=black, rotate=90, minimum height=8mm, minimum width=18mm, scale=0.7, label=right:$6$] {Vertical \\ 2};
\node (a8) at (9*\dist,0) [draw=black, rotate=90, minimum height=8mm, minimum width=18mm, scale=0.7, label=right:$6$] {Vertical \\ 1};

\coordinate (start) at (0.1,0);
\coordinate (end) at (10*\dist-0.1,0);
\foreach \from/\to in { start/a0,a0/a1,a1/a2,a2/a3,a3/a4,a4/a5,a5/a6,a6/a7,a7/a8,a8/end}{
    \draw [-] (\from) -- (\to);
}


\node at (-.75,3.0) [rotate=0, scale=1.0] {(a)};

\def \DIST {1.8}
\def \sh{3.3}

\node (b0) at (1*\DIST+\sh,3) [draw=black, rotate=90, minimum height=10mm, minimum width=19mm, scale=0.8, label=right:$24$] {Hopping};
\node (b1) at (2*\DIST+\sh,3) [draw=black, rotate=90, minimum height=10mm, minimum width=19mm, scale=0.8, label=right:$2$] {Interaction};

\coordinate (START) at (\sh+0.1,3);
\coordinate (END) at (3*\DIST-0.1+\sh,3);
\foreach \from/\to in { START/b0,b0/b1,b1/END}{
    \draw [-] (\from) -- (\to);
}

  
\end{tikzpicture}
    \caption{Gate count for the first-order (a) and second-order (b) trotter decompositions for $6\times 8$ lattice using all-to-all connectivity.}
    \label{fig:TrotterAllToAll}
\end{figure}

Combining all of these steps we can figure out the total circuit depth per layer of trotterization for both a first and second order decomposition, as shown in Fig.~\ref{fig:TrotterAllToAll}. The first order decomposition requires just the hopping and interaction steps to be applied alternating leading to 26 layers of entangling gates per trotter step. The second order decomposition requires that we apply the gate sequence in a time reversal symmetric fashion. Therefore we need to break up the hopping in to two vertical steps, and the two horizontal steps. By applying them in the order shown in Fig.~\ref{fig:TrotterAllToAll}, we get a count of 40 entangling layers per trotter step plus a constant 6 layers from the first trotter step.

\subsubsection{General Majorana Coupling}

Above we have explained efficient and parallel implementations of all of the operators required for the nearest-neighbour Fermi-Hubbard model. However, the mapping can be applied more generally, in which case we would require the general set of Majorana interactions in Eq.~\eqref{eq:majorana_gates}. All of these gates are of the form of an exponential of a Pauli string. By performing the correct basis transformation, we can therefore focus on the implementation of 
\begin{equation}
\exp\{ i \alpha Z_1 Z_2 \cdots Z_{N-1} Z_N \},
\end{equation}
for a string of $N$ Pauli-$Z$ operators.

When we have all-to-all connectivity, this unitary can be implemented with a logarithmic depth circuit, as we demonstrate for the case $N=8$ below in Eq.~\eqref{eq:log_exp_1}.
\begin{equation}
\begin{quantikz}[row sep={0.5cm,between origins}, column sep=0.15cm]
\qw & \ctrl{1} & \qw & \qw & \qw & \qw & \qw & \ctrl{1} & \qw \\
\qw & \targ{} & \ctrl{2} & \qw & \qw & \qw & \ctrl{2} & \targ{} & \qw \\
\qw & \ctrl{1} & \qw & \qw & \qw & \qw & \qw & \ctrl{1} & \qw \\
\qw & \targ{} & \targ{} & \ctrl{4} & \qw & \ctrl{4} & \targ{} & \targ{} & \qw \\
\qw & \ctrl{1} & \qw & \qw & \qw & \qw & \qw & \ctrl{1} & \qw \\
\qw & \targ{} & \ctrl{2} & \qw & \qw & \qw & \ctrl{2} & \targ{} & \qw \\
\qw & \ctrl{1} & \qw & \qw & \qw & \qw & \qw & \ctrl{1} & \qw \\
\qw & \targ{} & \targ{} & \targ{} & \gate{R_Z(-2\alpha)} & \targ{} & \targ{} & \targ{} & \qw \\
\end{quantikz} \label{eq:log_exp_1}
\end{equation}
This circuit is constructed by pairing up all of the qubits and applying CNOT gates between them. We then apply CNOTs between those qubits that were the target in the previous layer, and repeat until there is a single CNOT. We can then apply the rotation to the final target qubit, before doing the reverse sequence of CNOT gates. The basic idea of this decomposition is to use the pull-through relation
\begin{equation}
\begin{quantikz}[row sep={0.9cm,between origins}, column sep=0.2cm]
\qw & \targ{} & \gate{Z} & \qw \\
\qw & \ctrl{-1} & \qw & \qw
\end{quantikz}
=
\begin{quantikz}[row sep={0.9cm,between origins}, column sep=0.2cm]
\qw & \gate{Z} & \targ{} & \qw \\
\qw & \gate{Z} & \ctrl{-1} & \qw
\end{quantikz}\label{eq:CNOT_pull_through}
\end{equation}
This sequence of CNOTS transforms the Pauli string to a single $Z_N$, on which we can apply the rotation.

When the circuit contains a number of qubits that is not a power of two, we must pair up the (target) qubits in each layer as much as we can. Any qubits that were not paired up in the previous layer, should then be paired up next. This is demonstrated for the case $N=7$ below,
\begin{equation}
\begin{quantikz}[row sep={0.5cm,between origins}, column sep=0.15cm]
\qw & \ctrl{1} & \qw & \qw & \qw & \qw & \qw & \ctrl{1} & \qw \\
\qw & \targ{} & \ctrl{2} & \qw & \qw & \qw & \ctrl{2} & \targ{} & \qw \\
\qw & \ctrl{1} & \qw & \qw & \qw & \qw & \qw & \ctrl{1} & \qw \\
\qw & \targ{} & \targ{} & \ctrl{3} & \qw & \ctrl{3} & \targ{} & \targ{} & \qw \\
\qw & \ctrl{1} & \qw & \qw & \qw & \qw & \qw & \ctrl{1} & \qw \\
\qw & \targ{} & \ctrl{1} & \qw & \qw & \qw & \ctrl{1} & \targ{} & \qw \\
\qw & \qw & \targ{} & \targ{} & \gate{R_Z(-2\alpha)} & \targ{} & \targ{} & \qw & \qw \\
\end{quantikz}
\end{equation}

In summary, the circuit requires $2\ceil{\log_2 N}$ CNOT layers with a total of $2(N-1)$ CNOT gates. As we have shown above, there may be efficient ways to perform these gates (partially) in parallel. However, this would have to be determined on a case-by-case basis.

\begin{figure}[b!]
    \centering
    \begin{tikzpicture}[scale=0.45,thick]


\begin{scope}[shift={(3,0)}]
\foreach \x in {0,2,4,6,8,10}{  
\foreach \y in {0,1,2,3,4,5,6,7}{
    \fill[SpinBlue] (\x,\y) circle (3pt);
}}
\foreach \x in {1,3,5,7,9,11}{  
\foreach \y in {0,1,2,3,4,5,6,7}{
    \fill[SpinRed] (\x,\y) circle (3pt);
}}
    
\foreach \x in {0.5,2.5,4.5,6.5,8.5,10.5}{  
\foreach \y in {0.5,2.5,4.5,6.5}{
    \fill[NodePurple] (\x,\y) circle (3pt);}}
\foreach \x in {1.5,3.5,5.5,7.5,9.5}{  
\foreach \y in {1.5,3.5,5.5}{
    \fill[NodePurple] (\x,\y) circle (3pt); }}
    
\foreach \x in {0.5,2.5,4.5,6.5,8.5,10.5}{  
\foreach \y in {-0.5,1.5,3.5,5.5,7.5}{
    \fill[NodeGray] (\x,\y) circle (3pt);}}
\foreach \x in {-.5,1.5,3.5,5.5,7.5,9.5,11.5}{  
\foreach \y in {.5,2.5,4.5,6.5}{
    \fill[NodeGray] (\x,\y) circle (3pt); }}
\foreach \x in {1.5,3.5,5.5,7.5, 9.5}{
    \fill[NodeGray] (\x,7.5) circle (3pt); 
    \fill[NodeGray] (\x,-.5) circle (3pt);  }
\foreach \x in {0,1,2,3,4,5,6,7,8,9,10}{  
\foreach \y in {0,1,2,3,4,5,6,7}{
    \draw[-,EdgeGray, opacity=1.] (\x+.65,\y+.35)--(\x+.85,\y+.15);
    \draw[-,EdgeGray, opacity=1.] (\x+.65,\y-.35)--(\x+.85,\y-.15);
    \draw[-,EdgeGray, opacity=1.] (\x+.15,\y+.15)--(\x+.35,\y+.35);
    \draw[-,EdgeGray, opacity=1.] (\x+.15,\y-.15)--(\x+.35,\y-.35);  }}
\foreach \y in {0.5,2.5,4.5,6.5}{
    \draw[-,EdgeGray, opacity=1.] (-.35,\y-.15)--(-.15,\y-.35);
    \draw[-,EdgeGray, opacity=1.] (-.35,\y+.15)--(-.15,\y+.35);  
    \draw[-,EdgeGray, opacity=1.] (11.15,\y+.35)--(11.35,\y+.15);
    \draw[-,EdgeGray, opacity=1.] (11.15,\y-.35)--(11.35,\y-.15);  }
\end{scope}

\begin{scope}[shift={(-4,0)}, draw=black, fill=black]
\foreach \y in {0,1,2,3,4,5,6,7}{ 
    \draw[-, opacity=1.] (0,\y)--(5,\y); } 
\foreach \x in {0,1,2,3,4,5}{ 
    \draw[-, opacity=1.] (\x,0)--(\x,7); } 
\foreach \x in {0, 1, 2, 3, 4, 5}{  
\foreach \y in {0, 1, 2, 3, 4, 5, 6, 7}{
    \fill (\x,\y) circle (3pt); }}
\end{scope}

\begin{scope}[shift={(1.25,3.5)}, draw=black, fill=black, text=black]
    \draw[line width=0.3mm,-latex, scale=0.56] (0.,0.) -- (1.6,0.);
\end{scope}

\end{tikzpicture}
    \caption{Mapping from a $6 \times 8$ square lattice of spinful fermions to qubits for the diamond connectivity (shown by grey connecting lines). The blue points correspond to spin up and red to spin down. The purple points are the secondary qubits, and the grey points are ancilla qubits.}
    \label{fig:LatticeMappingGoogle}
\end{figure}
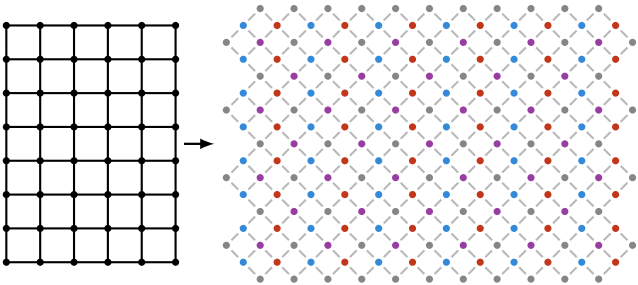

\subsection{Diamond connectivity}\label{app:Diamond Connectivity}

For our second explicit example we will consider the case where we have restricted local connectivity. For concreteness we consider the diamond connectivity from Fig.~\ref{fig:Embeddings}. This type of local connectivity is typical for superconducting qubits. We will again assume that we have access to arbitrary single qubit gates, but now we will only have entangling CNOT operations between connected neighbouring qubits.

\subsubsection{Lattice Mapping}

\begin{figure}[b!]
    \centering
    \begin{tikzpicture}[scale=0.62,thick]

\newcommand{\CNOT}[2]{
    \def\csize{2.1} 
    \def\tsize{3.6} 
    \def\lw{0.4} 
    \coordinate (c) at (#1);
    \coordinate (t) at (#2);
    \draw[{Circle[length=\csize pt, sep=-1*\csize*.5pt]}-{Circle[open,length=\tsize pt, line width=\lw pt, sep=-1*\tsize*.5pt]},line width=\lw pt] (c)--(t);
    \draw[-,line width=\lw pt] (t) --+(\tsize*.56pt,\tsize*.56pt) --+ (-\tsize*.56pt,-\tsize*.56pt);
    \draw[-,line width=\lw pt] (t) --+(\tsize*.56pt,-\tsize*.56pt) --+ (-\tsize*.56pt,\tsize*.56pt);    
}


\begin{scope}[shift={(3.051,-1)}, scale=0.7]
\foreach \y in {0.5,2.5,4.5,6.5}{
    \draw[-,EdgePurple] (-.3,\y) -- (11.3,\y);    }
\foreach \x in {1.5,3.5,5.5,7.5,9.5}{
    \draw[-,EdgePurple] (\x,-.3)--(\x,7.3);       }  
\foreach \x in {0,2,4,6,8,10}{  
\foreach \y in {0,1,2,3,4,5,6,7}{
    \fill[SpinBlue] (\x,\y) circle (3pt);                  }}
\foreach \x in {1,3,5,7,9,11}{  
\foreach \y in {0,1,2,3,4,5,6,7}{
    \fill[SpinRed] (\x,\y) circle (3pt);
}}
    
\foreach \x in {0.5,2.5,4.5,6.5,8.5,10.5}{  
\foreach \y in {0.5,2.5,4.5,6.5}{
    \fill[NodePurple] (\x,\y) circle (3pt);}}
\foreach \x in {1.5,3.5,5.5,7.5,9.5}{  
\foreach \y in {1.5,3.5,5.5}{
    \fill[NodePurple] (\x,\y) circle (3pt); }}
    
\foreach \x in {0.5,2.5,4.5,6.5,8.5,10.5}{  
\foreach \y in {1.5,3.5,5.5}{
    \fill[NodeGray] (\x,\y) circle (3pt);     }}
\foreach \x in {1.5,3.5,5.5,7.5,9.5}{  
\foreach \y in {.5,2.5,4.5,6.5}{
    \fill[NodeGray] (\x,\y) circle (3pt);     }}

\foreach \x in {0,1,2,3,4,5,6,7,8,9,10}{  
\foreach \y in {0,1,2,3,4,5,6}{
    \draw[-,EdgeGray] (\x+.65,\y+.35)--(\x+.85,\y+.15);
    \draw[-,EdgeGray] (\x+.15,\y+.15)--(\x+.35,\y+.35);  }}
\foreach \x in {0,1,2,3,4,5,6,7,8,9,10}{  
\foreach \y in {1,2,3,4,5,6,7}{
    \draw[-,EdgeGray] (\x+.65,\y-.35)--(\x+.85,\y-.15);
    \draw[-,EdgeGray] (\x+.15,\y-.15)--(\x+.35,\y-.35);  }}

\node (a) at (-.5,3.5) [anchor=east, scale=1.0]{(a)};
\end{scope}

\begin{scope}[shift={(0,-3)}]
\pgfmathsetmacro{\qsize}{1.5}
\pgfmathsetmacro{\csize}{2.0}
\def\tsize{3.5}
\foreach \N in {0,1,2,3,4,5}{
    \pgfmathsetmacro{\x}{\N*1.9 + 1.2}
    \draw[-,EdgePurple] (\x-.7,+0.5) -- (\x+0.7,+0.5);    
    \draw[-,EdgePurple] (\x-.7,-0.5) -- (\x+0.7,-0.5);    
    \draw[-,EdgePurple] (\x+.5,-0.7) -- (\x+0.5,+0.7);    
    \draw[-,EdgePurple] (\x-.5,-0.7) -- (\x-0.5,+0.7);
    \fill[NodePurple] (\x,0.5) coordinate (pt\N) circle (2 pt);
    \fill[NodePurple] (\x,-.5) coordinate (pb\N) circle (2 pt);
    \fill[NodePurple] (\x-0.5,0) coordinate (pl\N) circle (2 pt);
    \fill[NodePurple] (\x+0.5,0) coordinate (pr\N) circle (2 pt);
    
    \fill[NodeGray] (\x,0) coordinate (gc\N) circle (\qsize pt);
    \fill[NodeGray] (\x+.25,0+.25) coordinate (gtr\N) circle (\qsize pt);
    \fill[NodeGray] (\x-.25,0-.25) coordinate (gbl\N) circle (\qsize pt);
    \fill[NodeGray] (\x+.25,0-.25) coordinate (gbr\N) circle (\qsize pt);
    \fill[NodeGray] (\x-.25,0+.25) coordinate (gtl\N) circle (\qsize pt);    
}
    
\draw (gc0) node[minimum size=.2mm,draw,scale=0.7,line width=0.1mm, fill=white, inner sep=1.5pt] {$H$};

\draw[{Circle[length=\csize pt, sep=-1*\csize*.5pt]}-{Circle[open,length=\tsize pt, line width=0.3pt, sep=-\tsize*.5pt]},line width=0.3pt] (gc1)--(gtl1);
\draw[-,line width=0.3pt] (gtl1) --+(\tsize*.4pt,\tsize*.4pt) --+ (-\tsize*.4pt,-\tsize*.4pt);
\draw[-,line width=0.3pt] (gtl1) --+(\tsize*.4pt,-\tsize*.4pt) --+ (-\tsize*.4pt,\tsize*.4pt);

\CNOT{gtl2}{pt2}
\CNOT{gc2}{gbr2}

\CNOT{gbr3}{pb3}
\CNOT{gtl3}{pl3}

\CNOT{gtl4}{gc4}
\CNOT{gbr4}{pr4}

\CNOT{pt5}{gtl5}
\CNOT{pb5}{gbr5}

\foreach \N in {0,1,2,3,4,5,6}{
    \pgfmathsetmacro{\x}{\N*1.9 + 1.2}
    \draw[-,EdgePurple] (\x-.7,+0.5-2) -- (\x+0.7,+0.5-2);    
    \draw[-,EdgePurple] (\x-.7,-0.5-2) -- (\x+0.7,-0.5-2);    
    \draw[-,EdgePurple] (\x+.5,-0.7-2) -- (\x+0.5,+0.7-2);    
    \draw[-,EdgePurple] (\x-.5,-0.7-2) -- (\x-0.5,+0.7-2);
    \fill[NodePurple] (\x,0.5-2) coordinate (PT\N) circle (2 pt);
    \fill[NodePurple] (\x,-.5-2) coordinate (PB\N) circle (2 pt);
    \fill[NodePurple] (\x-0.5,0-2) coordinate (PL\N) circle (2 pt);
    \fill[NodePurple] (\x+0.5,0-2) coordinate (PR\N) circle (2 pt);
    
    \fill[NodeGray] (\x,0-2) coordinate (GC\N) circle (\qsize pt);
    \fill[NodeGray] (\x+.25,0+.25-2) coordinate (GTR\N) circle (\qsize pt);
    \fill[NodeGray] (\x-.25,0-.25-2) coordinate (GBL\N) circle (\qsize pt);
    \fill[NodeGray] (\x+.25,0-.25-2) coordinate (GBR\N) circle (\qsize pt);
    \fill[NodeGray] (\x-.25,0+.25-2) coordinate (GTL\N) circle (\qsize pt);    
}

\draw (GC0) node[minimum size=.2mm,draw,scale=0.7,line width=0.1mm, fill=white,inner sep=1.5pt] {$H$};
\CNOT{GC1}{GBR1}
\CNOT{GC2}{GTL2}
\CNOT{GBR2}{PR2}
\CNOT{GTL3}{PL3}
\CNOT{GBR3}{PB3}
\CNOT{GTL4}{PT4}
\CNOT{GC4}{GBR4}
\CNOT{GTL5}{GC5}
\CNOT{PT6}{GTL6}

\node (b) at (.5,0.) [anchor=east, scale=1.0]{(b)};
\node (c) at (.5,-2.) [anchor=east, scale=1.0]{(c)};

\end{scope}

\end{tikzpicture}
    \caption{(a) The embedded qubits superimposed with the lattice connectivity for the background toric code chosen for the vacuum state preparation. The purple qubits indicate the secondary qubits from the mapping and the purple lattice arbitrarily distinguished between the star and plaquette terms of the toric code. (b),(c) The main circuit motifs used in constructing the vacuum state for the central and edge rows in Fig.~\ref{fig:VacuumPrepGoogle}. }
    \label{fig:VacuumPrepIdeaGoogle}
\end{figure}
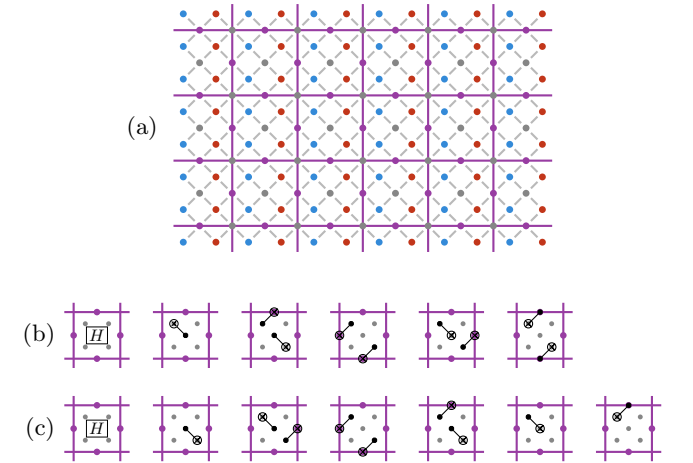

\begin{figure*}[t!]
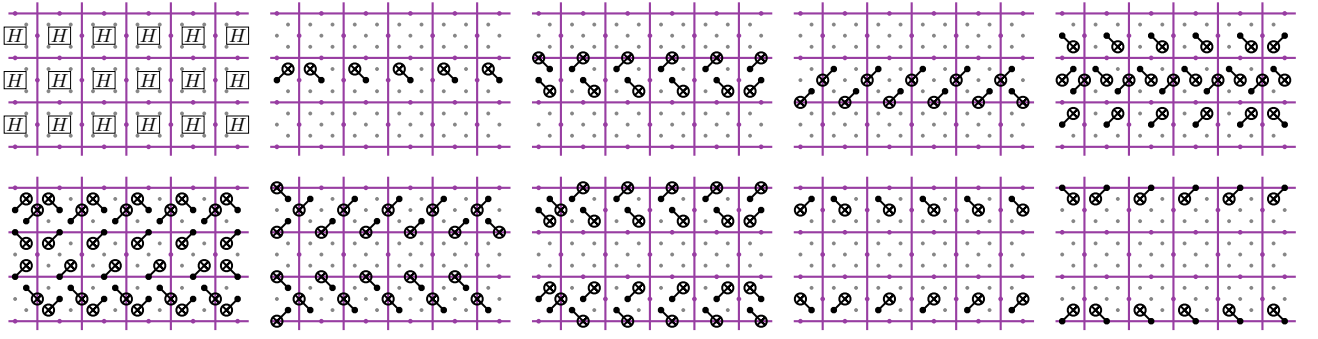

    \centering
    \include{Figures/Tikz/VacuumPrepGoogle}
    \caption{The explicit circuit for constructing the vacuum state for the $6 \times 8$ lattice of spinful fermions with the diamond connectivity of qubits.}
    \label{fig:VacuumPrepGoogle}
\end{figure*}

The mapping from the $6\times 8$ spinfull fermions on a square lattice to qubits is shown in Fig.~\ref{fig:LatticeMappingGoogle}. Because of the local connectivity, we have chosen to put qubits corresponding to the same unit cell next to each other in order to make both the hopping and interactions local. This will minimise the depth of the circuit to realise the hopping and the interaction terms. In this case we have secondary qubits from the Derby-Klassen mapping (shown in purple), as well as additional ancilla qubits (shown in grey) from the embedding. Furthermore, we have added ancilla qubits to the edge, which will aid in the parallelisation of the circuits in order to minimize the circuit depth. In this case we have a total of 203 qubits, 96 of which are physical, 39 are secondary qubits from the mapping, and 68 are ancilla qubits from the embedding which should be kept in the $|0\rangle$ state.

\subsubsection{Vacuum Preparation}

For the vacuum preparation with the diamond connectivity we are a bit more restricted than in the all-to-all connectivity case. Here we need to use intermediate ancilla qubits to implement the desired gates between the secondary qubits from the mapping. In Fig.~\ref{fig:VacuumPrepIdeaGoogle} we show the underlying lattice for the toric code used for the vacuum preparation and also the building blocks of the preparation circuit. Since all of the qubits start in the $|0\rangle$ state, we are able to use all other qubits, including the primary qubits as ancilla qubits for the preparation, as long as we return them to the $|0\rangle$ state. Fig.~\ref{fig:VacuumPrepGoogle} shows the explicit circuit for the full vacuum preparation for the $6 \times 8$ lattice of spinful fermions. Again we need to perform a basis transformation: applying a Hadamard gate on all secondary qubits, then applying a $R_X(-\pi/2)$ rotation on secondary qubits that lie on horizontal bonds, as shown by the purple lattice. This circuit takes 9 entangling layers. More generally, if the toric code background has an odd number $M$ of rows (or columns if less) then the depth would be $1 + 4(M-1)/2$, or $2 + 4 M/2$ if $M$ is even. Note that we have the freedom of how to choose the toric code lattice, we used in this case to have 3 rows rather than 4, allowing us to reduce the circuit depth by 1.

\subsubsection{Vertical Hopping}

\begin{figure}[b!]
    \centering
    \begin{tikzpicture}[scale=0.7,thick]


\begin{scope}[shift={(0,0)}]
\pgfmathsetmacro{\di}{0.15}
\draw[-,EdgeGray, opacity=1.] (1-\di,0-\di) -- (0+\di,-1+\di);    
\draw[-,EdgeGray, opacity=1.] (0-\di,-1+\di) -- (-1+\di,0-\di);    
\draw[-,EdgeGray, opacity=1.] (-1+\di,0+\di) -- (0-\di,1-\di);    
\draw[-,EdgeGray, opacity=1.] (0+\di,1-\di) -- (1-\di,0+\di);
\fill[SpinBlue] (0, 1) circle (4.5pt);
\fill[SpinBlue] (0,-1) circle (4.5pt);
\fill[NodeGray] (-1,0) circle (4.5pt);
\fill[NodePurple] (1,0) circle (4.5pt);

\pgfmathsetmacro{\nu}{0.35}
\node at (+1+\nu,0)[scale=0.91]{$2$};
\node at (-1+\nu,0)[scale=0.91]{$4$};
\node at (0+\nu,+1)[scale=0.91]{$1$};
\node at (0+\nu,-1)[scale=0.91]{$3$};

\end{scope}

\begin{scope}[shift={(5,0)}]
\pgfmathsetmacro{\di}{0.15}
\draw[-,EdgeGray, opacity=1.] (1-\di,0-\di) -- (0+\di,-1+\di);    
\draw[-,EdgeGray, opacity=1.] (0-\di,-1+\di) -- (-1+\di,0-\di);    
\draw[-,EdgeGray, opacity=1.] (-1+\di,0+\di) -- (0-\di,1-\di);    
\draw[-,EdgeGray, opacity=1.] (0+\di,1-\di) -- (1-\di,0+\di);
\fill[SpinBlue] (0, 1) circle (4.5pt);
\fill[SpinBlue] (0,-1) circle (4.5pt);
\fill[NodeGray] (1,0) circle (4.5pt);
\fill[NodePurple] (-1,0) circle (4.5pt);

\pgfmathsetmacro{\nu}{0.35}
\node at (+1+\nu,0)[scale=0.91]{$4$};
\node at (-1+\nu,0)[scale=0.91]{$2$};
\node at (0+\nu,+1)[scale=0.91]{$1$};
\node at (0+\nu,-1)[scale=0.91]{$3$};

\end{scope}
    
\end{tikzpicture}
    \vspace{8pt}
    \begin{tikzpicture}[scale=0.34,thick]


\begin{scope}[shift={(-6.5,0)}]
\foreach \x in {0,2,4,6,8,10}{  
\foreach \y in {0,2,4,6}{
    \draw [rounded corners=0.72mm,fill=HighGray, draw=none, scale=1.0] (\x,\y-.3)--(\x-.8,\y+.5)--(\x,\y+1.3)--(\x+.8,\y+.5)--cycle;    }}
\foreach \x in {1,3,5,7,9}{  
\foreach \y in {1,3,5}{
    \draw [rounded corners=0.72mm,fill=HighGray, draw=none, scale=1.0] (\x,\y-.3)--(\x-.8,\y+.5)--(\x,\y+1.3)--(\x+.8,\y+.5)--cycle;    }}
\foreach \x in {10}{  
\foreach \y in {1,3,5}{
    \draw [rounded corners=1.5mm,fill=HighGray, draw=none, scale=1.0] (\x+.1,\y+.5)--(\x+1.2,\y-.4)--(\x+1.2,\y+1.4)--cycle;   }}
\foreach \x in {0,2,4,6,8,10}{  
\foreach \y in {0,1,2,3,4,5,6,7}{
    \fill[SpinBlue] (\x,\y) circle (3pt);    }}
\foreach \x in {1,3,5,7,9,11}{  
\foreach \y in {0,1,2,3,4,5,6,7}{
    \fill[SpinRed] (\x,\y) circle (3pt);  }}
    
\foreach \x in {0.5,2.5,4.5,6.5,8.5,10.5}{  
\foreach \y in {0.5,2.5,4.5,6.5}{
    \fill[NodePurple] (\x,\y) circle (3pt);  }}
\foreach \x in {1.5,3.5,5.5,7.5,9.5}{  
\foreach \y in {1.5,3.5,5.5}{
    \fill[NodePurple] (\x,\y) circle (3pt);  }}
\foreach \x in {0.5,2.5,4.5,6.5,8.5,10.5}{  
\foreach \y in {1.5,3.5,5.5}{
    \fill[NodeGray] (\x,\y) circle (3pt);}}
\foreach \x in {-.5,1.5,3.5,5.5,7.5,9.5,11.5}{  
\foreach \y in {.5,2.5,4.5,6.5}{
    \fill[NodeGray] (\x,\y) circle (3pt); }}
\foreach \x in {0,1,2,3,4,5,6,7,8,9,10}{  
\foreach \y in {0,1,2,3,4,5,6}{
    \draw[-,EdgeGray,opacity=1.] (\x+.65,\y+.35)--(\x+.85,\y+.15);
    \draw[-,EdgeGray,opacity=1.] (\x+.65,\y-.35+1.0)--(\x+.85,\y-.15+1.0);
    \draw[-,EdgeGray,opacity=1.] (\x+.15,\y+.15)--(\x+.35,\y+.35);
    \draw[-,EdgeGray,opacity=1.] (\x+.15,\y-0.15+1.0)--(\x+.35,\y-0.35+1.0);  
}}
\foreach \y in {0.5,2.5,4.5,6.5}{
    \draw[-,EdgeGray,opacity=1.] (-.35,\y-.15)--(-.15,\y-.35);
    \draw[-,EdgeGray,opacity=1.] (-.35,\y+.15)--(-.15,\y+.35);  
    \draw[-,EdgeGray,opacity=1.] (11.15,\y+.35)--(11.35,\y+.15);
    \draw[-,EdgeGray,opacity=1.] (11.15,\y-.35)--(11.35,\y-.15);  }
\end{scope}

\begin{scope}[shift={(6.5,0)}]

\foreach \x in {1,3,5,7,9,11}{  
\foreach \y in {0,2,4,6}{
    \draw [rounded corners=1.mm, fill=HighGray,draw=none, scale=1.0, opacity=1.] (\x,\y-.3)--(\x-.8,\y+.5)--(\x,\y+1.3)--(\x+.8,\y+.5)--cycle;}}
\foreach \x in {2,4,6,8,10}{  
\foreach \y in {1,3,5}{
    \draw [rounded corners=1.mm, fill=HighGray,draw=none, scale=1.0, opacity=1.] (\x,\y-.3)--(\x-.8,\y+.5)--(\x,\y+1.3)--(\x+.8,\y+.5)--cycle;}}
\foreach \x in {0}{  
\foreach \y in {1,3,5}{
    \draw [rounded corners=2.0mm, fill=HighGray, draw=none, scale=1.0, opacity=1.] (\x-.2,\y-.4)--(\x+.9,\y+.5)--(\x-.2,\y+1.4)--cycle;  }}
\foreach \x in {0,2,4,6,8,10}{  
\foreach \y in {0,1,2,3,4,5,6,7}{
    \fill[SpinBlue] (\x,\y) circle (3pt);    }}
\foreach \x in {1,3,5,7,9,11}{  
\foreach \y in {0,1,2,3,4,5,6,7}{
    \fill[SpinRed] (\x,\y) circle (3pt);     }}
\foreach \x in {0.5,2.5,4.5,6.5,8.5,10.5}{  
\foreach \y in {0.5,2.5,4.5,6.5}{
    \fill[NodePurple] (\x,\y) circle (3pt);  }}
\foreach \x in {1.5,3.5,5.5,7.5,9.5}{  
\foreach \y in {1.5,3.5,5.5}{
    \fill[NodePurple] (\x,\y) circle (3pt);  }}
\foreach \x in {0.5,2.5,4.5,6.5,8.5,10.5}{  
\foreach \y in {1.5,3.5,5.5}{
    \fill[NodeGray] (\x,\y) circle (3pt);}}
\foreach \x in {-.5,1.5,3.5,5.5,7.5,9.5,11.5}{  
\foreach \y in {.5,2.5,4.5,6.5}{
    \fill[NodeGray] (\x,\y) circle (3pt); }}
\foreach \x in {0,1,2,3,4,5,6,7,8,9,10}{  
\foreach \y in {0,1,2,3,4,5,6}{
    \draw[-,EdgeGray,opacity=1.] (\x+.65,\y+.35)--(\x+.85,\y+.15);
    \draw[-,EdgeGray,opacity=1.] (\x+.65,\y-.35+1.0)--(\x+.85,\y-.15+1.0);
    \draw[-,EdgeGray,opacity=1.] (\x+.15,\y+.15)--(\x+.35,\y+.35);
    \draw[-,EdgeGray,opacity=1.] (\x+.15,\y-0.15+1.0)--(\x+.35,\y-0.35+1.0);  
}}
\foreach \y in {0.5,2.5,4.5,6.5}{
    \draw[-,EdgeGray,opacity=1.] (-.35,\y-.15)--(-.15,\y-.35);
    \draw[-,EdgeGray,opacity=1.] (-.35,\y+.15)--(-.15,\y+.35);  
    \draw[-,EdgeGray,opacity=1.] (11.15,\y+.35)--(11.35,\y+.15);
    \draw[-,EdgeGray,opacity=1.] (11.15,\y-.35)--(11.35,\y-.15);  }
\end{scope}

\end{tikzpicture}
    \caption{(Top) The two local connectivity configurations for vertical bonds in the diamond connectivity. Blue qubits indicate physics qubits, purple indicate secondary qubits from the mapping, and the grey qubits are ancilla qubits in the $|0\rangle$ state. (Bottom) A sequence for applying the vertical hopping terms in parallel to reduce the overall circuit depth.}
    \label{fig:VerticalParallelGoogle}
\end{figure}
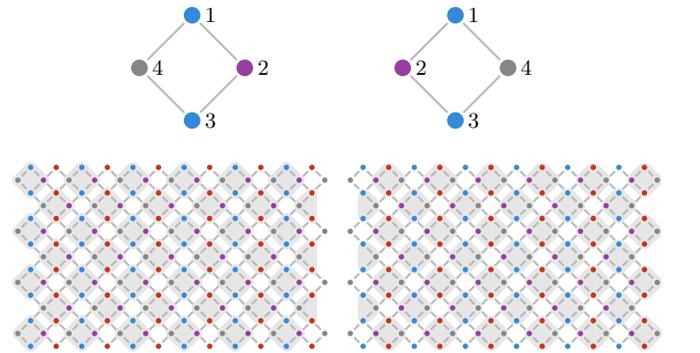

To implement the vertical hopping in the model, we need to implement the unitary operators
\begin{equation}
    \exp\{ i\alpha (X_1 X_2 X_3 + Y_1 X_2 Y_3)\},
\end{equation}
where the numbering is given in Fig.~\ref{fig:VerticalParallelGoogle}, and $\alpha = \frac{1}{2}J$. We can take advantage of the additional ancilla qubits at our disposal that are in the $|0\rangle $ state, as shown in Fig.~\ref{fig:VerticalParallelGoogle}. This allows us to use the following circuit decomposition,
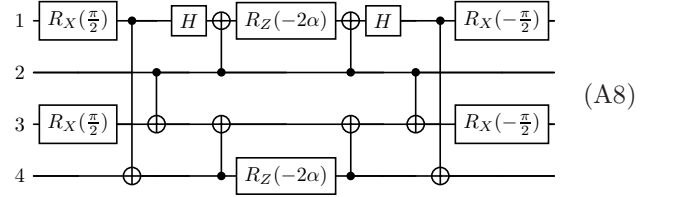
\begin{equation}
\begin{adjustbox}{width=0.85\columnwidth}
\begin{quantikz}[row sep={0.9cm,between origins}, column sep=0.1cm]
\lstick{1} & \gate{R_X(\frac{\pi}{2})} & \ctrl{3} & \qw & \gate{H} & \targ{} & \gate{R_Z(-2\alpha)} & \targ{} & \gate{H} & \qw & \ctrl{3} & \gate{R_X(-\frac{\pi}{2})} & \qw \\
\lstick{2} & \qw & \qw & \ctrl{1} & \qw & \ctrl{-1} & \qw & \ctrl{-1} & \qw & \ctrl{1} & \qw & \qw & \qw \\
\lstick{3} & \gate{R_X(\frac{\pi}{2})} & \qw & \targ{} & \qw & \targ{} & \qw & \targ{} & \qw & \targ{} & \qw & \gate{R_X(-\frac{\pi}{2})} & \qw \\
\lstick{4} & \qw & \targ{} & \qw & \qw & \ctrl{-1} & \gate{R_Z(-2\alpha)} & \ctrl{-1} & \qw & \qw & \targ{} & \qw & \qw
\end{quantikz}
\end{adjustbox}
\end{equation}
which has only 4 layers of entangling gates since the first and last pairs of CNOTs can be applied in parallel. These circuits can then be applied in parallel in two steps, as illustrated in Fig.~\ref{fig:VerticalParallelGoogle}, leading to a total depth of 8 entangling layers. At the boundary we additionally have terms that don't involve the secondary qubits. These terms correspond to 
\begin{equation}
    \exp\{ i\alpha (X_1 X_3 + Y_1 Y_3)\}
\end{equation}
and can be implemented by the circuit
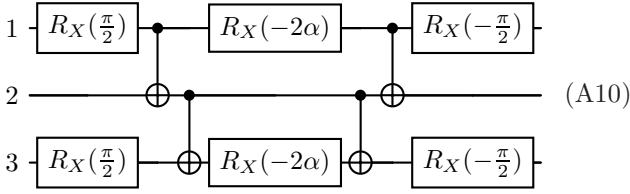
\begin{equation}
\begin{adjustbox}{width=0.85\columnwidth}
\begin{quantikz}[row sep={0.9cm,between origins}, column sep=0.1cm]
\lstick{1} & \gate{R_X(\frac{\pi}{2})} & \ctrl{1} & \qw & \gate{R_X(-2\alpha)} & \qw & \ctrl{1} & \gate{R_X(-\frac{\pi}{2})} & \qw \\
\lstick{2} & \qw & \targ{} & \ctrl{1} & \qw & \ctrl{1} & \targ{} & \qw & \qw \\
\lstick{3} & \gate{R_X(\frac{\pi}{2})} & \qw & \targ{} & \gate{R_X(-2\alpha)} & \targ{} & \qw & \gate{R_X(-\frac{\pi}{2})} & \qw
\end{quantikz}
\end{adjustbox}
\end{equation}
where the second qubit is an ancilla in the state $|0\rangle$.

\subsubsection{Interaction}

\begin{figure}[b!]
    \centering
    \begin{tikzpicture}[scale=0.7,thick]

\begin{scope}[shift={(0,0)}]
\pgfmathsetmacro{\di}{0.15}
\draw[-,NodeGray, opacity=0.85] (0+\di,0+\di) -- (1-\di,1-\di);    
\draw[-,NodeGray, opacity=0.85] (0-\di,0+\di) -- (-1+\di,1-\di);
\fill[SpinBlue] (-1, 1) circle (4.pt);
\fill[SpinRed] (+1, 1) circle (4.pt);
\fill[NodeGray] (0,0) circle (4.pt);

\pgfmathsetmacro{\nu}{0.35}
\node at (-1-\nu,+1)[scale=0.99]{1};
\node at (0,0-\nu)[scale=0.99]{$2$};
\node at (1+\nu,1)[scale=0.99]{$3$};

\end{scope}

\begin{scope}[shift={(5,0)}]
\pgfmathsetmacro{\di}{0.15}
\draw[-,EdgeGray] (0+\di,0+\di) -- (1-\di,1-\di);    
\draw[-,EdgeGray] (0-\di,0+\di) -- (-1+\di,1-\di);    
\fill[SpinBlue] (-1, 1) circle (4.pt);
\fill[SpinRed] (+1, 1) circle (4.pt);
\fill[NodePurple] (0,0) circle (4.pt);

\pgfmathsetmacro{\nu}{0.35}
\node at (-1-\nu,+1)[scale=0.99]{$1$};
\node at (0,0-\nu)[scale=0.99]{$2$};
\node at (1+\nu,1)[scale=0.99]{$3$};

\end{scope}
    
\end{tikzpicture}
    \caption{Two cases for the local connectivity for the interaction terms. The blue and red qubits correspond to the physical up and down states, the purple sites are the secondary qubits, and the grey are ancilla.}
    \label{fig:InteractionGoogle}
\end{figure}
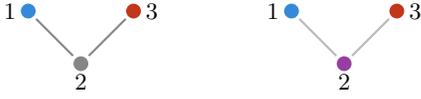
\begin{figure}[b!]
    \centering
    \begin{tikzpicture}[scale=0.5,thick]

\def\nsz{2.5}
\def\asz{2.5}

\newcommand{\CNOT}[2]{
    \def\csize{3.5} 
    \def\tsize{6}
    \def \lw{0.9}
    \coordinate (c) at (#1);
    \coordinate (t) at (#2);
    \draw[{Circle[length=\csize pt, sep=-1*\csize*.5pt]}-{Circle[open,length=\tsize pt, line width=\lw pt, sep=-1*\tsize*.5pt]},line width=\lw pt] (c)--(t);
    \draw[-,line width=\lw pt] (t) --+(\tsize*0.7pt,\tsize*0.7pt) --+ (-\tsize*0.7pt,-\tsize*0.7pt);
    \draw[-,line width=\lw pt] (t) --+(\tsize*0.7pt,-\tsize*0.7pt) --+ (-\tsize*0.7pt,\tsize*0.7pt);   
}


\begin{scope}[shift={(0,0)}]
    
\foreach \x in {0,2,4}{  
\foreach \y in {0,1,2,3}{
    \fill[SpinBlue] (\x,\y) circle (\nsz pt);   
    \fill[SpinRed] (\x+1,\y) circle (\nsz pt);       
    }}
\foreach \x in {0.5,2.5,4.5}{  
\foreach \y in {0.5,2.5}{
    \fill[NodePurple] (\x,\y) circle (\nsz pt);       
    \fill[NodeGray] (\x,\y-1.0) circle (\nsz pt);      
    }}
\foreach \x in {1.5,3.5}{  
\foreach \y in {-0.5,1.5}{
    \fill[NodePurple] (\x,\y) circle (\nsz pt);       
    \fill[NodeGray] (\x,\y+1.0) circle (\nsz pt);      
    }}
\foreach \x in {0,1,2,3,4}{  
\foreach \y in {-1,0,1,2}{
    \draw[-,EdgeGray] (\x+.65,\y-0.35+1.0)--(\x+.85,\y-.15+1.0);
    \draw[-,EdgeGray] (\x+.15,\y-0.15+1.0)--(\x+.35,\y-0.35+1.0);  }
\foreach \y in {0,1,2}{
    \draw[-,EdgeGray] (\x+.65,\y+.35)--(\x+.85,\y+.15);
    \draw[-,EdgeGray] (\x+.15,\y+.15)--(\x+.35,\y+.35);         }
}

\foreach \x in {0.5,2.5,4.5}{  
\foreach \y in {.5,2.5}{  
    \CNOT{\x,\y}{\x-0.5,\y+0.5}              
    }}
\end{scope}

\begin{scope}[shift={(6,0)}]
    
\foreach \x in {0,2,4}{  
\foreach \y in {0,1,2,3}{
    \fill[SpinBlue] (\x,\y) circle (\nsz pt);   
    \fill[SpinRed] (\x+1,\y) circle (\nsz pt);       
    }}
\foreach \x in {0.5,2.5,4.5}{  
\foreach \y in {0.5,2.5}{
    \fill[NodePurple] (\x,\y) circle (\nsz pt);       
    \fill[NodeGray] (\x,\y-1.0) circle (\nsz pt);      
    }}
\foreach \x in {1.5,3.5}{  
\foreach \y in {-0.5,1.5}{
    \fill[NodePurple] (\x,\y) circle (\nsz pt);       
    \fill[NodeGray] (\x,\y+1.0) circle (\nsz pt);      
    }}
\foreach \x in {0,1,2,3,4}{  
\foreach \y in {-1,0,1,2}{
    \draw[-,EdgeGray] (\x+.65,\y-0.35+1.0)--(\x+.85,\y-.15+1.0);
    \draw[-,EdgeGray] (\x+.15,\y-0.15+1.0)--(\x+.35,\y-0.35+1.0);  }
\foreach \y in {0,1,2}{
    \draw[-,EdgeGray] (\x+.65,\y+.35)--(\x+.85,\y+.15);
    \draw[-,EdgeGray] (\x+.15,\y+.15)--(\x+.35,\y+.35);         }
}
\foreach \x in {0,2,4}{  
\foreach \y in {0,1,2,3}{  
    \CNOT{\x,\y}{\x+0.5,\y-.5}              
    }}
\end{scope}

\begin{scope}[shift={(12,0)}]
    
\foreach \x in {0,2,4}{  
\foreach \y in {0,1,2,3}{
    \fill[SpinBlue] (\x,\y) circle (\nsz pt);   
    \fill[SpinRed] (\x+1,\y) circle (\nsz pt);       
    }}
\foreach \x in {0.5,2.5,4.5}{  
\foreach \y in {0.5,2.5}{
    \fill[NodePurple] (\x,\y) circle (\nsz pt);       
    \fill[NodeGray] (\x,\y-1.0) circle (\nsz pt);      
    }}
\foreach \x in {1.5,3.5}{  
\foreach \y in {-0.5,1.5}{
    \fill[NodePurple] (\x,\y) circle (\nsz pt);       
    \fill[NodeGray] (\x,\y+1.0) circle (\nsz pt);      
    }}
\foreach \x in {0,1,2,3,4}{  
\foreach \y in {-1,0,1,2}{
    \draw[-,EdgeGray] (\x+.65,\y-0.35+1.0)--(\x+.85,\y-.15+1.0);
    \draw[-,EdgeGray] (\x+.15,\y-0.15+1.0)--(\x+.35,\y-0.35+1.0);  }
\foreach \y in {0,1,2}{
    \draw[-,EdgeGray] (\x+.65,\y+.35)--(\x+.85,\y+.15);
    \draw[-,EdgeGray] (\x+.15,\y+.15)--(\x+.35,\y+.35);         }
}
\foreach \x in {1,3,5}{  
\foreach \y in {0,1,2,3}{  
    \CNOT{\x,\y}{\x-0.5,\y-0.5}              
    }}
\end{scope}

\begin{scope}[shift={(0,-4.5)}]
    
\foreach \x in {0,2,4}{  
\foreach \y in {0,1,2,3}{
    \fill[SpinBlue] (\x,\y) circle (\nsz pt);   
    \fill[SpinRed] (\x+1,\y) circle (\nsz pt);       
    }}
\foreach \x in {0.5,2.5,4.5}{  
\foreach \y in {0.5,2.5}{
    \fill[NodePurple] (\x,\y) circle (\nsz pt);       
    \fill[NodeGray] (\x,\y-1.0) circle (\nsz pt);      
    }}
\foreach \x in {1.5,3.5}{  
\foreach \y in {-0.5,1.5}{
    \fill[NodePurple] (\x,\y) circle (\nsz pt);       
    \fill[NodeGray] (\x,\y+1.0) circle (\nsz pt);      
    }}
\foreach \x in {0,1,2,3,4}{  
\foreach \y in {-1,0,1,2}{
    \draw[-,EdgeGray] (\x+.65,\y-0.35+1.0)--(\x+.85,\y-.15+1.0);
    \draw[-,EdgeGray] (\x+.15,\y-0.15+1.0)--(\x+.35,\y-0.35+1.0);  }
\foreach \y in {0,1,2}{
    \draw[-,EdgeGray] (\x+.65,\y+.35)--(\x+.85,\y+.15);
    \draw[-,EdgeGray] (\x+.15,\y+.15)--(\x+.35,\y+.35);         }}
\foreach \x in {.5,2.5,4.5}{  
\foreach \y in {-.5,.5,1.5,2.5}{  
\draw (\x,\y) node[minimum size=5mm,draw,scale=0.7,line width=0.1mm, fill=white,inner sep=1.8pt] {$R_z$};
    }}
\end{scope}

\begin{scope}[shift={(6,-4.5)}]
    
\foreach \x in {0,2,4}{  
\foreach \y in {0,1,2,3}{
    \fill[SpinBlue] (\x,\y) circle (\nsz pt);   
    \fill[SpinRed] (\x+1,\y) circle (\nsz pt);       
    }}
\foreach \x in {0.5,2.5,4.5}{  
\foreach \y in {0.5,2.5}{
    \fill[NodePurple] (\x,\y) circle (\nsz pt);       
    \fill[NodeGray] (\x,\y-1.0) circle (\nsz pt);      
    }}
\foreach \x in {1.5,3.5}{  
\foreach \y in {-0.5,1.5}{
    \fill[NodePurple] (\x,\y) circle (\nsz pt);       
    \fill[NodeGray] (\x,\y+1.0) circle (\nsz pt);      
    }}
\foreach \x in {0,1,2,3,4}{  
\foreach \y in {-1,0,1,2}{
    \draw[-,EdgeGray] (\x+.65,\y-0.35+1.0)--(\x+.85,\y-.15+1.0);
    \draw[-,EdgeGray] (\x+.15,\y-0.15+1.0)--(\x+.35,\y-0.35+1.0);  }
\foreach \y in {0,1,2}{
    \draw[-,EdgeGray] (\x+.65,\y+.35)--(\x+.85,\y+.15);
    \draw[-,EdgeGray] (\x+.15,\y+.15)--(\x+.35,\y+.35);         }
}
\foreach \x in {1,3,5}{  
\foreach \y in {0,1,2,3}{  
    \CNOT{\x,\y}{\x-0.5,\y-0.5}              
    }}
\end{scope}

\begin{scope}[shift={(12,-4.5)}]
    
\foreach \x in {0,2,4}{  
\foreach \y in {0,1,2,3}{
    \fill[SpinBlue] (\x,\y) circle (\nsz pt);   
    \fill[SpinRed] (\x+1,\y) circle (\nsz pt);       
    }}
\foreach \x in {0.5,2.5,4.5}{  
\foreach \y in {0.5,2.5}{
    \fill[NodePurple] (\x,\y) circle (\nsz pt);       
    \fill[NodeGray] (\x,\y-1.0) circle (\nsz pt);      
    }}
\foreach \x in {1.5,3.5}{  
\foreach \y in {-0.5,1.5}{
    \fill[NodePurple] (\x,\y) circle (\nsz pt);       
    \fill[NodeGray] (\x,\y+1.0) circle (\nsz pt);      
    }}
\foreach \x in {0,1,2,3,4}{  
\foreach \y in {-1,0,1,2}{
    \draw[-,EdgeGray] (\x+.65,\y-0.35+1.0)--(\x+.85,\y-.15+1.0);
    \draw[-,EdgeGray] (\x+.15,\y-0.15+1.0)--(\x+.35,\y-0.35+1.0);  }
\foreach \y in {0,1,2}{
    \draw[-,EdgeGray] (\x+.65,\y+.35)--(\x+.85,\y+.15);
    \draw[-,EdgeGray] (\x+.15,\y+.15)--(\x+.35,\y+.35);         }
}
\foreach \x in {0,2,4}{  
\foreach \y in {0,1,2,3}{  
    \CNOT{\x,\y}{\x+0.5,\y-0.5}              
    }}
\end{scope}

\begin{scope}[shift={(0,-9)}]
    
\foreach \x in {0,2,4}{  
\foreach \y in {0,1,2,3}{
    \fill[SpinBlue] (\x,\y) circle (\nsz pt);   
    \fill[SpinRed] (\x+1,\y) circle (\nsz pt);       
    }}
\foreach \x in {0.5,2.5,4.5}{  
\foreach \y in {0.5,2.5}{
    \fill[NodePurple] (\x,\y) circle (\nsz pt);       
    \fill[NodeGray] (\x,\y-1.0) circle (\nsz pt);      
    }}
\foreach \x in {1.5,3.5}{  
\foreach \y in {-0.5,1.5}{
    \fill[NodePurple] (\x,\y) circle (\nsz pt);       
    \fill[NodeGray] (\x,\y+1.0) circle (\nsz pt);      
    }}
\foreach \x in {0,1,2,3,4}{  
\foreach \y in {-1,0,1,2}{
    \draw[-,EdgeGray] (\x+.65,\y-0.35+1.0)--(\x+.85,\y-.15+1.0);
    \draw[-,EdgeGray] (\x+.15,\y-0.15+1.0)--(\x+.35,\y-0.35+1.0);  }
\foreach \y in {0,1,2}{
    \draw[-,EdgeGray] (\x+.65,\y+.35)--(\x+.85,\y+.15);
    \draw[-,EdgeGray] (\x+.15,\y+.15)--(\x+.35,\y+.35);         }
}
\foreach \x in {0.5,2.5,4.5}{  
\foreach \y in {0.5,2.5}{  
    \CNOT{\x,\y}{\x-0.5,\y+0.5}              
    }}
\end{scope}

\end{tikzpicture}
    \caption{Explicit circuit for the parallel application of all the interaction terms for the diamond connectivity. The lattice shown is just a portion of the full lattice.}
    \label{fig:InteractionParallelGoogle}
\end{figure}
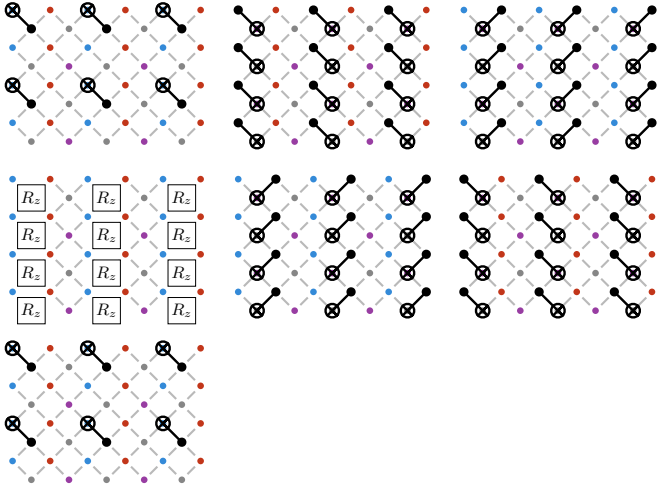

Dropping the single qubit rotations, the interactions are of the form 
\begin{equation}
    \exp\{i\beta(Z_1 Z_3)\},
\end{equation}
where the qubit numbering is shown in Fig.~\ref{fig:InteractionGoogle} and $\beta=-\frac{1}{4}U$. Due to the local connectivity, we need to apply this gate via an intermediate ancilla qubit. We have two possibilities, using an ancilla in the $|0\rangle$ state, or a secondary qubit from the mapping. The former can be decomposed into the circuit
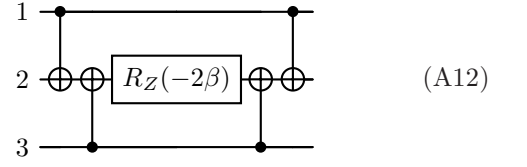
\begin{equation}
\begin{quantikz}[row sep={0.9cm,between origins}, column sep=0.1cm]
\lstick{1} & \ctrl{1} & \qw & \qw & \qw & \ctrl{1} & \qw \\
\lstick{2} & \targ{} & \targ{} & \gate{R_Z(-2\beta)} & \targ{} & \targ{} & \qw \\
\lstick{3} & \qw & \ctrl{-1} & \qw & \ctrl{-1} & \qw & \qw
\end{quantikz}
\end{equation}
where qubit 2 should start and end in the $|0\rangle$ state. For the decomposition including the secondary qubit from the mapping, the decomposition is
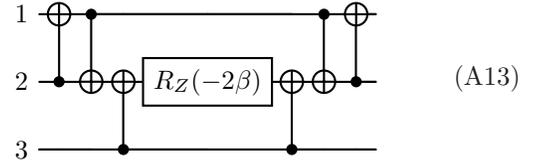
\begin{equation}
\begin{quantikz}[row sep={0.9cm,between origins}, column sep=0.1cm]
\lstick{1} & \targ{} & \ctrl{1} & \qw & \qw & \qw & \ctrl{1} & \targ{} & \qw \\
\lstick{2} & \ctrl{-1} & \targ{} & \targ{} & \gate{R_Z(-2\beta)} & \targ{} & \targ{} & \ctrl{-1} & \qw \\
\lstick{3} & \qw & \qw & \ctrl{-1} & \qw & \ctrl{-1} & \qw & \qw & \qw
\end{quantikz}
\end{equation}

Since the interaction terms all have distinct support, these two possible decompositions can be applied in parallel. The full explicit circuit for a subset of the lattice is shown in Fig.~\ref{fig:InteractionParallelGoogle}. In total this requires 6 layers of entangling gates.

\subsubsection{Horizontal Hopping}

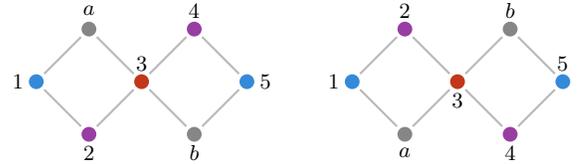
\begin{figure}[b!]
    \centering
    \begin{tikzpicture}[scale=0.7,thick]

\begin{scope}[shift={(-3,0)}]
\foreach \x in {0,2}{
    \pgfmathsetmacro{\di}{0.15}
    \draw[-,EdgeGray, opacity=1.] (\x+\di,0-\di) -- (\x+1-\di,-1+\di);    
    \draw[-,EdgeGray, opacity=1.] (\x+1+\di,-1+\di) -- (\x+2-\di,0-\di);    
    \draw[-,EdgeGray, opacity=1.] (\x+\di,0+\di) -- (\x+1-\di,1-\di);    
    \draw[-,EdgeGray, opacity=1.] (\x+1+\di,1-\di) -- (\x+2-\di,0+\di);
}
\pgfmathsetmacro{\nu}{0.35}
\fill[SpinBlue] (0,0) circle (4.pt);
\node at (0-\nu,0)[scale=0.91]{$1$};
\fill[SpinBlue] (4,0) circle (4.pt);
\node at (4+\nu,0)[scale=0.91]{$5$};
\fill[SpinRed] (2,0) circle (4.pt);
\node at (2,0+\nu)[scale=0.91]{$3$};
\fill[NodeGray] (1,1) circle (4.pt);
\node at (1,+1+\nu)[scale=0.91]{$a$};
\fill[NodeGray] (3,-1) circle (4.pt);
\node at (3,-1-\nu)[scale=0.91]{$b$};
\fill[NodePurple] (1,-1) circle (4.pt);
\node at (1,-1-\nu)[scale=0.91]{$2$};
\fill[NodePurple] (3,1) circle (4.pt);
\node at (3,1+\nu)[scale=0.91]{$4$};
\end{scope}

\begin{scope}[shift={(+3,0)}]
\foreach \x in {0,2}{
    \pgfmathsetmacro{\di}{0.15}
    \draw[-,EdgeGray, opacity=1.] (\x+\di,0-\di) -- (\x+1-\di,-1+\di);    
    \draw[-,EdgeGray, opacity=1.] (\x+1+\di,-1+\di) -- (\x+2-\di,0-\di);    
    \draw[-,EdgeGray, opacity=1.] (\x+\di,0+\di) -- (\x+1-\di,1-\di);    
    \draw[-,EdgeGray, opacity=1.] (\x+1+\di,1-\di) -- (\x+2-\di,0+\di);
}
\pgfmathsetmacro{\nu}{0.35}
\fill[SpinBlue] (0,0) circle (4.pt);
\node at (0-\nu,0)[scale=0.91]{$1$};
\fill[SpinBlue] (4,0) circle (4.pt);
\node at (4,+\nu,0)[scale=0.91]{$5$};
\fill[SpinRed] (2,0) circle (4.pt);
\node at (2,0-\nu)[scale=0.91]{$3$};
\fill[NodeGray] (1,-1) circle (4.pt);
\node at (1,-1-\nu)[scale=0.91]{$a$};
\fill[NodeGray] (3,+1) circle (4.pt);
\node at (3,+1+\nu)[scale=0.91]{$b$};
\fill[NodePurple] (1,1) circle (4.pt);
\node at (1,1+\nu)[scale=0.91]{$2$};
\fill[NodePurple] (3,-1) circle (4.pt);
\node at (3,-1-\nu)[scale=0.91]{$4$};
\end{scope}

\end{tikzpicture}
    \caption{Two local connectivity for horizontal bonds in the diamond connectivity. The blue, red and purple qubits correspond to the physical up state, down states and the secondary qubits as part of the mapping, respectively. The grey points are ancilla qubits.}
    \label{fig:HorizontalHoppingGoogle}
\end{figure}

\begin{figure}[t!]
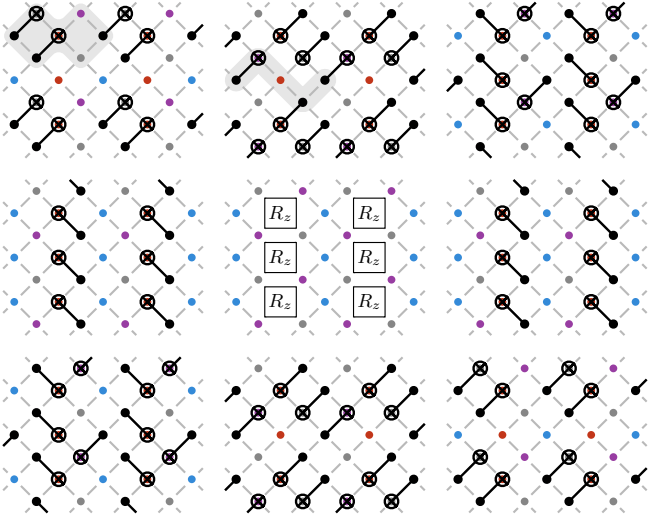

    \centering
    \include{Figures/Tikz/HorizontalParallelGoogle}
    \caption{Parallel application of the horizontal hopping terms with the diamond connectivity. Circuit is shown for a path of the full system and for the $X$-term of the up species. The other 3 terms can be implemented similarly.}
    \label{fig:HorizontalParallelGoogle}
\end{figure}

The horizontal hopping for the diamond connectivity is significantly more complicated since we need to implement next-nearest neighbour hopping. This now involves up to 7 qubits, as shown in Fig.~\ref{fig:HorizontalHoppingGoogle}. These hoppings correspond to 
\begin{equation}
\exp\{i\alpha X_1 Y_2 Z_3 Y_4 X_5\} \exp\{i\alpha Y_1 Y_2 Z_3 Y_4 Y_5\},
\end{equation}
with $\alpha=\frac{1}{2}J$.
The two terms are of the same form but are related by a basis transformation. 
For these hoppings, we will use two different circuits allowing us to implement the circuit in parallel. The most direct decomposition uses only the qubits involved in the mapping, and is
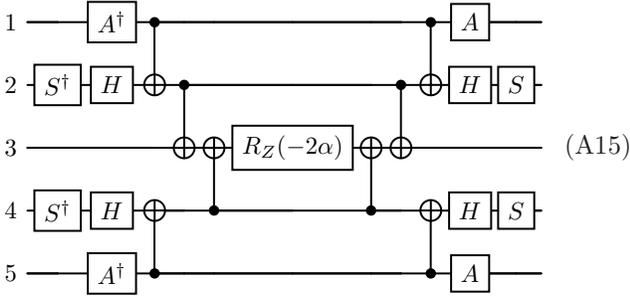
\begin{equation}
\begin{adjustbox}{width=0.85\columnwidth}
\begin{quantikz}[row sep={0.9cm,between origins}, column sep=0.1cm]
\lstick{1} & \qw & \gate{A^\dag} & \ctrl{1} & \qw & \qw & \qw & \qw & \qw & \ctrl{1} & \gate{A} & \qw & \qw \\
\lstick{2} & \gate{S^\dag} & \gate{H} & \targ{} & \ctrl{1} & \qw & \qw & \qw & \ctrl{1} & \targ{} & \gate{H} & \gate{S} & \qw \\
\lstick{3} & \qw & \qw & \qw & \targ{} & \targ{} & \gate{R_Z(-2\alpha)} & \targ{} & \targ{} & \qw & \qw & \qw & \qw \\
\lstick{4} & \gate{S^\dag} & \gate{H} & \targ{} & \qw & \ctrl{-1} & \qw & \ctrl{-1} & \qw & \targ{} & \gate{H} & \gate{S} & \qw \\
\lstick{5} & \qw & \gate{A^\dag} & \ctrl{-1} & \qw & \qw & \qw & \qw & \qw & \ctrl{-1} & \gate{A} & \qw & \qw 
\end{quantikz}
\end{adjustbox}
\end{equation}
where the rotations $A$ are given by $H$ for the $X$-terms and $S\cdot H$ for the $Y$-terms.
Alternatively, we can make use of the ancilla qubits in the $|0\rangle$ state to get the decomposition
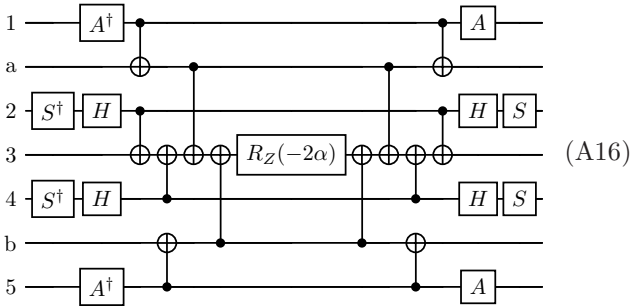
\begin{equation}
\begin{adjustbox}{width=0.85\columnwidth}
\begin{quantikz}[row sep={0.7cm,between origins}, column sep=0.1cm]
\lstick{1} & \qw & \gate{A^\dag} & \ctrl{1} & \qw & \qw & \qw & \qw & \qw & \qw & \qw & \ctrl{1} & \gate{A} & \qw & \qw \\
\lstick{a} & \qw & \qw & \targ{} & \qw & \ctrl{2} & \qw & \qw & \qw & \ctrl{2} & \qw & \targ{} & \qw & \qw & \qw \\
\lstick{2} & \gate{S^\dag} & \gate{H} & \ctrl{1} & \qw & \qw & \qw & \qw & \qw & \qw & \qw & \ctrl{1} & \gate{H} & \gate{S} & \qw \\
\lstick{3} & \qw & \qw & \targ{} & \targ{} & \targ{} & \targ{} & \gate{R_Z(-2\alpha)} & \targ{} & \targ{} & \targ{} & \targ{} & \qw & \qw & \qw \\
\lstick{4} & \gate{S^\dag} & \gate{H} & \qw & \ctrl{-1} & \qw & \qw & \qw & \qw & \qw & \ctrl{-1} & \qw & \gate{H} & \gate{S} & \qw \\
\lstick{b} & \qw & \qw & \qw & \targ{} & \qw & \ctrl{-2} & \qw & \ctrl{-2} & \qw & \targ{} & \qw & \qw & \qw & \qw \\
\lstick{5} & \qw & \gate{A^\dag} & \qw & \ctrl{-1} & \qw & \qw & \qw & \qw & \qw & \ctrl{-1} & \qw & \gate{A} & \qw & \qw
\end{quantikz}
\end{adjustbox}
\end{equation}
By using both decompositions, we are then able to apply all terms of a given type ($X$ or $Y$) for a given species (up or down) in parallel, as shown in Fig.~\ref{fig:HorizontalParallelGoogle}. This means that the total depth for all horizontal hopping terms is 32 entangling layers.

\subsubsection{Final Gate Count}

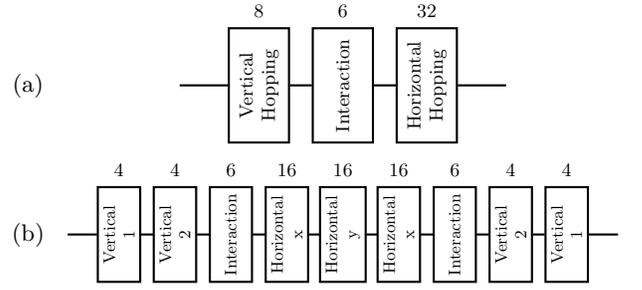
\begin{figure}[t!]
    \centering
\begin{tikzpicture}[scale=0.62, align=center, label distance=.5mm,every label/.style={scale=0.8, black}, thick, tight background]

\node at (-.75,0.0) [rotate=0, scale=1.0] {(b)};
\def \dist{1.2}
\node (a0) at (1*\dist,0) [draw=black, rotate=90, minimum height=8mm, minimum width=18mm, scale=0.7, label=right:$4$] {Vertical \\ 1};
\node (a1) at (2*\dist,0) [draw=black, rotate=90, minimum height=8mm, minimum width=18mm, scale=0.7, label=right:$4$] {Vertical \\ 2};
\node (a2) at (3*\dist,0) [draw=black, rotate=90, minimum height=8mm, minimum width=18mm, scale=0.7, label=right:$6$] {Interaction};
\node (a3) at (4*\dist,0) [draw=black, rotate=90, minimum height=8mm, minimum width=18mm, scale=0.7, label=right:$16$] {Horizontal \\ x};
\node (a4) at (5*\dist,0) [draw=black, rotate=90, minimum height=8mm, minimum width=18mm, scale=0.7, label=right:$16$] {Horizontal \\ y};
\node (a5) at (6*\dist,0) [draw=black, rotate=90, minimum height=8mm, minimum width=18mm, scale=0.7, label=right:$16$] {Horizontal \\ x};
\node (a6) at (7*\dist,0) [draw=black, rotate=90, minimum height=8mm, minimum width=18mm, scale=0.7, label=right:$6$] {Interaction};
\node (a7) at (8*\dist,0) [draw=black, rotate=90, minimum height=8mm, minimum width=18mm, scale=0.7, label=right:$4$] {Vertical \\ 2};
\node (a8) at (9*\dist,0) [draw=black, rotate=90, minimum height=8mm, minimum width=18mm, scale=0.7, label=right:$4$] {Vertical \\ 1};

\coordinate (start) at (0.1,0);
\coordinate (end) at (10*\dist-0.1,0);
\foreach \from/\to in { start/a0,a0/a1,a1/a2,a2/a3,a3/a4,a4/a5,a5/a6,a6/a7,a7/a8,a8/end}{
    \draw [-] (\from) -- (\to);
}

\node at (-.75,3.2) [rotate=0, scale=1.0] {(a)};

\def \DIST {1.8}
\def \sh{2.4}

\node (b0) at (1*\DIST+\sh,3.2) [draw=black, rotate=90, minimum height=10mm, minimum width=19mm, scale=0.8, label=right:$8$] {Vertical \\ Hopping};
\node (b1) at (2*\DIST+\sh,3.2) [draw=black, rotate=90, minimum height=10mm, minimum width=19mm, scale=0.8, label=right:$6$] {Interaction};
\node (b2) at (3*\DIST+\sh,3.2) [draw=black, rotate=90, minimum height=10mm, minimum width=19mm, scale=0.8, label=right:$32$] {Horizontal \\ Hopping};

\coordinate (START) at (\sh+0.1,3.2);
\coordinate (END) at (4*\DIST-0.1+\sh,3.2);
\foreach \from/\to in { START/b0,b0/b1,b1/b2,b2/END}{
    \draw [-] (\from) -- (\to);
}

  
\end{tikzpicture}
    \caption{(a) First-order trotter decomposition for the diamond connectivity with number of entangling layers shown. (b) The corresponding second-order decomposition.}
    \label{fig:TrotterGoogle}
\end{figure}

Combining all of these steps we can calculate the total circuit depth per layer of trotterization for both the first and second order decomposition for the diamond connectivity, as shown in Fig.~\ref{fig:TrotterGoogle}. The first-order decomposition requires a total of 46 layers, whereas the second-order decomposition requires 72 layers per step, plus a constant 4 layers from the first trotter step.

\subsubsection{General Majorana Coupling}

To implement the more general Majorana couplings with local connectivity, we require circuits that scale linearly with the lenght of the Pauli-string (without the use of measurements and feedforward). To implement the Pauli string exponential
\begin{equation}
\exp\{ i \alpha Z_1 Z_2 \cdots Z_{N-1} Z_N \},
\end{equation}
we can use a circuit of the form,
\begin{equation}
\begin{quantikz}[row sep={0.5cm,between origins}, column sep=0.15cm]
\qw & \ctrl{1} & \qw & \qw & \qw & \qw & \qw & \qw & \qw & \ctrl{1} & \qw \\
\qw & \targ{} & \ctrl{1} & \qw & \qw & \qw & \qw & \qw & \ctrl{1} & \targ{} & \qw \\
\qw & \qw & \targ{} & \ctrl{1} & \qw & \qw & \qw & \ctrl{1} & \targ{} & \qw & \qw \\
\qw & \qw & \qw & \targ{} & \targ{} & \gate{R_z(-2\alpha)} & \targ{} & \targ{} & \qw & \qw & \qw \\
\qw & \qw & \targ{} & \qw & \ctrl{-1} & \qw & \ctrl{-1} & \qw & \targ{} & \qw & \qw \\
\qw & \targ{} & \ctrl{-1} & \qw & \qw & \qw & \qw & \qw & \ctrl{-1} & \targ{} & \qw \\
\qw & \ctrl{-1} & \qw & \qw & \qw & \qw & \qw & \qw & \qw & \ctrl{-1} & \qw \\
\end{quantikz}
\end{equation}
shown for the explicit example of $N=7$. This circuit uses the same pull-through relation as in Eq.~\eqref{eq:CNOT_pull_through} but is now restricted by the connectivity. Note that the actual circuit may require additional swaps that depend on the details of the connectivity. There may also be ways to reduce the depth in specific cases, but this would have to be determined on a case-to-case basis, and would not improve the asymptotic scaling. This circuit would require $2\ceil{\frac{N}{2}}$ CNOT layers and a total of $2(N-1)$ CNOT gates.

\section{Derivation of the Hadamard test protocol}\label{app: hadamard test}

As discussed in section \ref{sec:measurement}, measurement difficulties amplifies when dealing with string order parameters at different times, such as Green's function, on the quantum computer. We take the advantage of the interferometry method, known as a Hadamard test in quantum computing, for measuring theses types of quantities on a quantum computer. 
This interferometry method relies on single qubit measurement which will reduce the measurement error, in contrast to the direct measurement method \cite{Smith2022}. 

The fermionic Green's function, introduced in Eq.~\eqref{eq: fermion Greens function} can be measured using the this method 
, as introduced in section~\ref{subsubsec: Greensfunction}. 
This alternative interferometry measurement protocol (see figure~\ref{fig:Measure_Interfero_Two}) uses two ancilla qubits to reduce both measurement error and gate error. The steps of this method can be explained as follow:

Acting with Hadamard gates on two ancilla qubits produces 
    \begin{equation}
        \frac{1}{2}(\ket{0}+\ket{1})\otimes\ket{\psi}\otimes(\ket{0}+\ket{1}).
    \end{equation}
The first controlled operation (which can be equivalent to creating and moving a single Majorana) would give us 
    \begin{equation}
    \begin{aligned}
        \frac{1}{2} \big( & \ket{0}\ket{\psi}\ket{0} +
        \ket{1}\hat{A}\ket{\psi}\ket{0}+
        \ket{0}\ket{\psi}\ket{1}  \\
        & +\ket{1}\hat{U}\ket{\psi}\ket{1} \big).
    \end{aligned}
    \end{equation}
Then system is evolved in time using operator $\hat{T}$:
    \begin{equation}
    \begin{aligned}
        \frac{1}{2}\big( & \ket{0}\hat{T}\ket{\psi}\ket{0} +
        \ket{1} \hat{T}\hat{A}\ket{\psi} \ket{0}+
        \ket{0} \hat{T}\ket{\psi} \ket{1} \\
        & +\ket{1} \hat{T}\hat{A}\ket{\psi} \ket{1}\big).
    \end{aligned}
    \end{equation}
The second controlled operation (which is equivalent to creating and moving a single Majorana) gives us: 
    \begin{equation}
    \begin{aligned}
        \frac{1}{2}\big( & \ket{0} \hat{T}\ket{\psi} \ket{0} +
        \ket{1} \hat{T}\hat{A}\ket{\psi} \ket{0}+
        \ket{0} \hat{B}\hat{T}\ket{\psi} \ket{1} \\
        & +\ket{1}\hat{B}\hat{T}\hat{A}\ket{\psi} \ket{1}\big).
    \end{aligned}
    \end{equation}\normalsize
Finally, we preform measurements in $X$ and $Y$ basis on ancilla qubits to construct the Green's function expectation, as written in Eq.~\eqref{eq: ancilla measurements Greens function}. 
Fermionic Green's function measurement relies on being able to create a single Majorana(s), see appendix~\ref{app: single operators} for detailed discussions. 

\section{Single creation/annihilation operators}\label{app: single operators}

In order to realise single Majorana fermion operators in the Derby-Klassen compact fermion mapping~\cite{Derby2021}, we require a secondary qubit in at least one of the corner faces.
A single Majorana mode created at site $0$ satisfies the fermionic algebra (commutation relations):
\begin{equation}
\begin{aligned}
    \{\gamma_0,E_{0k} \} = 0, \; [\bar{\gamma}_0, E_{0k} ] = 0, \\
    \{\gamma_0,V_{0}\}=\{\bar{\gamma}_0,V_{0}\}  = 0.
\end{aligned}
\end{equation}
A single Majorana can be mapped to $\hat{X}$ or $\hat{Y}$ operator acting on the corner qubit, as shown in Fig.~\ref{fig:Measure_Interfero_Two}
There are two possible choices for Majorana, $\gamma_0=\hat{X}_0$ if arrows are pointing toward the corner and $\gamma_0=\hat{Y}_0$ if arrows are pointing away from the corner. Once a single Majorana at a corner is chosen, the other Majorana type is also fixed, such as Eq.~\eqref{eq:single majorana operators}.

\section{Implementation of Controlled Pauli-Strings}\label{sec:controlled_pauli}

To realize the generalized Hadamard test in Fig.~\ref{fig:Measure_Interfero_Two} for the Green's function, we need to implement a multi-qubit controlled Pauli string operation. Since these are simple Pauli strings, we can always perform a basis transformation to a product of only $Z$ gates. To implement these gates we can use the pull-through relation in Eq.~\eqref{eq:CNOT_pull_through} to simplify the circuits to a single controlled-Z gate.

For all-to-all connectivity this can be implemented with logarithmic depth in the length $N$ of the Pauli string. For example,
\begin{equation}
\begin{quantikz}[row sep={0.5cm,between origins}, column sep=0.15cm]
\qw & \ctrl{1} & \qw \\
\qw & \gate[8]{\otimes^8 Z} & \qw \\
\qw & \qw & \qw \\
\qw & \qw & \qw \\
\qw & \qw & \qw \\
\qw & \qw & \qw \\
\qw & \qw & \qw \\
\qw & \qw & \qw \\
\qw & \qw & \qw 
\end{quantikz}
=
\begin{quantikz}[row sep={0.5cm,between origins}, column sep=0.15cm]
\qw & \qw & \qw & \qw & \ctrl{1} & \qw & \qw & \qw & \qw \\
\qw & \targ{} & \targ{} & \targ{} & \ctrl{0} & \targ{} & \targ{} & \targ{} & \qw \\
\qw & \ctrl{-1} & \qw & \qw & \qw & \qw & \qw & \ctrl{-1} & \qw \\
\qw & \targ{} & \ctrl{-2} & \qw & \qw & \qw & \ctrl{-2} & \targ{} & \qw \\
\qw & \ctrl{-1} & \qw & \qw & \qw & \qw & \qw & \ctrl{-1} & \qw \\
\qw & \targ{} & \targ{} & \ctrl{-4} & \qw & \ctrl{-4} & \targ{} & \targ{} & \qw \\
\qw & \ctrl{-1} & \qw & \qw & \qw & \qw & \qw & \ctrl{-1} & \qw \\
\qw & \targ{} & \ctrl{-2} & \qw & \qw & \qw & \ctrl{-2} & \targ{} & \qw \\
\qw & \ctrl{-1} & \qw & \qw & \ghost{H} & \qw & \qw & \ctrl{-1} & \qw 
\end{quantikz}
\end{equation}
for $N=8$.
More generally, it can be implemented with $2\ceil{\log_2 N}+1$ entangling layers and a total of $2N-1$ entangling gates.

For local connectivity, we can implement the gates with linear depth, as demonstrated for $N=5$:
\begin{equation}
\begin{quantikz}[row sep={0.5cm,between origins}, column sep=0.15cm]
\qw & \ctrl{1} & \qw \\
\qw & \gate[5]{\otimes^5 Z} & \qw \\
\qw & \qw & \qw \\
\qw & \qw & \qw \\
\qw & \qw & \qw \\
\qw & \qw & \qw 
\end{quantikz}
=
\begin{quantikz}[row sep={0.5cm,between origins}, column sep=0.15cm]
\qw & \qw & \qw & \qw & \qw & \ctrl{0} & \qw & \qw & \qw & \qw & \qw \\
\qw & \qw & \qw & \qw & \targ{} & \ctrl{-1} & \targ{} & \qw & \qw & \qw & \qw \\
\qw & \qw & \qw & \targ{} & \ctrl{-1} & \qw & \ctrl{-1} & \targ{} & \qw & \qw & \qw \\
\qw & \qw & \targ{} & \ctrl{-1} & \qw & \qw & \qw & \ctrl{-1} & \targ{} & \qw & \qw \\
\qw & \targ{} & \ctrl{-1} & \qw & \qw & \qw & \qw & \qw & \ctrl{-1} & \targ{} & \qw \\
\qw & \ctrl{-1} & \qw & \qw & \qw & \ghost{H} & \qw & \qw & \qw & \ctrl{-1} & \qw
\end{quantikz}
\end{equation}
In general this requires $2N-1$ entangling layers and the same number of gates. 

If measurements and feedforward are available, it may be possible and more efficient to implement this controlled operations in constant depth~\cite{Piroli2021,Baumer2024}.

\bibliographystyle{apsrev4-2}
\bibliography{references}

\end{document}